\documentclass{aa}
\usepackage[utf8]{inputenc}
\usepackage{graphicx}
\usepackage{txfonts}
\usepackage{xcolor}
\usepackage{float}
\usepackage{natbib}
\usepackage{enumitem}

\bibpunct{(}{)}{;}{a}{}{,} 
\newcommand{\dss} {$\delta$ Scuti stars}

\newcommand{\gd} {$\gamma$ Dor}

\newcommand{\cd} {\mathrm{d}^{-1}}

\newcommand{\vsini} {v\sin\,i}

\newcommand{\teff} {T_{\mathrm{eff}}}

\newcommand{\sigspec} {{\sc sigspec}}
\newcommand{\combi}{{\sc combine}}

\newcommand{\corot} {CoRoT}

\newcommand{\kepler} {\emph{Kepler}}

\newcommand{\miar} {MIARMA}

\newcommand\Tstrut{\rule{0pt}{2.6ex}}         
\newcommand\Bstrut{\rule[-0.9ex]{0pt}{0pt}}   

\begin{document} 
   \title{Impact of gaps in the asteroseismic characterization of pulsating stars. I. On the efficiency of pre-whitening}

	\titlerunning{I. On the efficiency of pre-whitening}
    
   \author{J. Pascual-Granado\inst{1}
          \and
          J.C. Su\'arez\inst{2,1}\fnmsep\thanks{Associate researcher at institute (1)}
          \and
          R. Garrido\inst{1}
          \and
          A. Moya\inst{2,3,4}    
          \and
          A. Garc\'ia Hern\'andez\inst{2}
          \and  
          J.R. Rod\'on\inst{1}
          \and
          M. Lares-Martiz\inst{1}         
          }

   \institute{Instituto de Astrof\'{\i}sica de Andaluc\'{\i}a, Glorieta de la Astronom\'{\i}a 
   			  s/n, 18008, Granada, Spain\\
              \email{javier@iaa.es}
         \and
             Dept. F\'{\i}sica Te\'orica y del Cosmos. Universidad de Granada. Campus de 
             Fuentenueva, 18071, Granada, Spain
	 	 \and
			 Departamento de Astrofísica, Centro de Astrobiología (CAB/INTA-CSIC), Spain
         \and
         	 School of Physics and Astronomy, University of Birmingham, Birmingham B15 2TT, UK
             }


 
  \abstract
  {It is known that the observed distribution of frequencies in \corot\ and \kepler\ \dss\ has no parallelism with any theoretical model. Pre-whitening is a widespread technique in the analysis of time series with gaps from pulsating stars located in the classical instability strip such as \dss. However, some studies have pointed out that this technique might introduce biases in the results of the frequency analysis.
  }
   {This work aims at studying the biases that can result from pre-whitening in asteroseismology. The results will depend on the intrinsic range and distribution of frequencies of the stars. The periodic nature of the gaps in CoRoT observations, just in the range of the pulsational frequency content of the \dss, is shown to be crucial to determine their oscillation frequencies, the first step to perform asteroseismology of these objects. Hence, here we focus on the impact of pre-whitening on the asteroseismic characterization of \dss.
   }
   {We select a sample of 15 \dss\ observed by the \corot\ satellite, for which ultra-high quality photometric data have been obtained by its seismic channel. In order to study the impact on the asteroseismic characterization of \dss\ we perform the pre-whitening procedure on three datasets: gapped data, linearly interpolated data, and data with gaps interpolated using Autoregressive and Moving Average models (ARMA).
   }
   {The different results obtained show that at least in some cases pre-whitening is not an efficient procedure for the deconvolution of the spectral window. Therefore, in order to reduce the effect of the spectral window to the minimum it is necessary to interpolate with an algorithm that is aimed to preserve the original frequency content, and not only to perform a pre-whitening of the data.
   }
   {}

  \keywords{Asteroseismology --
            Methods: data analysis --
            Stars: oscillations}
   
	\maketitle
%

\section{Introduction \label{sec:intro}}
Even when ultra-precise space satellites like \corot\ \citep{BAG06} or \kepler\ \citep{GIL10} are used to observe the variation of brightness of the stars, these time series might present gaps due to operational procedures or environmental effects. Gaps introduce correlation between the frequency bins of a periodogram \citep{GAB94} and may have a significant impact on seismic studies which are based on frequency analyses. An unbiased estimation of the power spectrum is necessary for seismic studies in order to obtain realistic models of the stars.

There exists a few studies in the literature about gaps in time series from the Sun, solar-like or red giant stars \citep[e.g.][]{BRO90,FOS99,STA08,GAR14} or, more generally, about unevenly sampled data \citep[e.g.][]{DEE75,SCA82} and how to diminish its contribution to the spectral window \citep{FOS95}.

A new approach to fill the gaps was proposed in \citet[][hereafter PG15]{pg15} using a method based on ARMA models, which is aimed to preserve the original frequency content of the signal. This technique has been used successfully in several studies \citep[e.g][]{GH15,GH17}. However, in asteroseismic studies of \dss\ the usual procedure to deal with the effects of the spectral window is to perform a pre-whitening \citep[e.g][]{GH09, POR09, GH13, BAR15}, which is similar to the CLEAN method introduced by \citet{HOG74}. However, this technique is not without its challenges \citep{BAL14, MAR05}.

Here we want to check the consistency of pre-whitening as a technique for spectral window deconvolution of light curves of \dss. The test is based on a comparison of the frequency content of gapped data and data interpolated with an algorithm that is aimed to preserve the information. In order to perform the interpolation we make use of the gap-filling algorithm \miar. The consistency check, then, consists in verifying the compatibility between both distributions.

For this study we have chosen a sample of light curves of \dss\ from \corot\ Seismofield (with a sampling of 32 secs.). \corot\ has a duty cycle of 90\% approx. with a 10\% loss of data due to the passing through the South Atlantic Anomaly (SAA). Therefore, roughly a 10\% loss in amplitude is usually expected for periods shorter than the duration of the passing through the SAA, which is about 9 min ($\nu > 160~d^{-1}$), and a minor contribution is expected for periods longer than 18 minutes ($\nu < 80~d^{-1}$) \citep[see][]{APP08}. Periods from 9 to 18 minutes are the most contributing to the spectral window but for \dss\ periods no shorter than 18 minutes are expected. Then, the correction originally implemented in \corot\ pipeline was a simple linear interpolation which was expected to be enough to avoid the contribution of the gaps to the spectral window. In PG15, however, it was shown that linear interpolation introduce spurious frequencies that might affect seismic studies of these stars.

We perform the frequency analysis of light curves with gaps and filled with two different techniques: \miar\ algorithm, which is aimed to preserve the frequency content, and linear interpolation, which is not. Linear interpolation is not so often used in frequency analyses of \dss\ as in solar-like studies \citep[e.g.][]{GAR09, ZWI11, BEN09} but by introducing in the consistency check also an interpolation method that is aimed to maximize computational efficiency and not to preserve the original frequency content, we can study the impact of the interpolation on the distribution of frequencies. Our results show that: (i) the pre-whitening process is not efficient enough in the deconvolution of the spectral window, (ii) a linear interpolation can give in some cases a similar distribution of frequencies to gapped data.

The paper has the following structure: we first present in Sect. 2 the sample of \dss\ analyzed and the characteristics of the observations. In Sect. 3 we describe the methodology followed in this study, we outline the gap-filling methods, the frequency analysis procedure, and a cleaning procedure for the frequency combinations and non-linear interactions performed to avoid biases in the analysis. In Sect. 4 we show the results of the gap-filling, the frequency spectra and the histogram of detected frequencies through the methodology described in the previous section. Sect. 5 is devoted to a discussion of the results and finally, in Sect. 6 we present conclusions and plans for a future publication concerning the impact on periodicity studies performed over the periodogram in order to find patterns that help in the modal identification (see Garcia Hernandez et al. 2009 for a detailed explanation).


\section{CoRoT data \label{sec:sample}}
\begin{table*}
\caption[]{Selected sample of \dss\ observed by \corot. From left to right: observing run, HD number, CoRoT number, spectral type, visual magnitude, effective temperature, absolute magnitude, rotational velocity and observation interval.} \label{tab:nodes}
  \centering 
  \begin{tabular}{lccccccccc}
   \hline Run &  HD & ID & SpT     &  $m_v$    &   log $\teff$ (K) & $M_V$  &  $\vsini$ (km/s) & Obs.time (d) \\ 
    \hline
IRa01      &  50844$^1$       	&            123  &  A2     &  9.1   &     3.88   &    1.31   &      64   &     57.713  \Tstrut \\ 
SRc01      &  174936$^2$      	&           7613  &  A2     &  8.58  &     3.9    &    1.88   &     170   &     27.194   \\  
SRc01      &  174966$^3$      	&           7528  &  A3     &  7.72  &     3.88   &    1.95   &     125   &     27.197   \\  
LRc01      &  181555$^{*,4}$        &           8669  &  A5 V   &  7.52  &     3.85   &    -0.72   &     200   &     156.645  \\ 
LRa01      &  49434$^5$         &            100  &  F1 V   &  5.74  &     3.86   &    2.74   &      87   &     136.890  \\ 
LRc02      &  172189$^{6,7}$    &           8170  &  A2     &  8.73  &     3.89   &      1.04    &      78    &     149.013  \\  
SRc02      &  174532$^{*,8}$        &           7655  &  A2     &  6.90  &     3.86   &    1.38   &      32    &     26.239   \\  
SRc02      &  174589$^*$       	&           7663  &  F2 III &  6.09  &     3.85   &    1.45   &     100   &     26.168   \\ 
LRa02      &  51722$^*$         &           1022  &  A5     &  7.53  &     3.86   &    1.13   &     127   &     117.375   \\ 
LRa02      &  51359$^*$       	&           1320  &  A5     &  8.50  &     3.9    &    0.89   &      -    &     117.41   \\  
LRa02      &  50870$^9$       	&            546  &  F0     &  8.88  &     3.88   &    1.67   &      17   &     114.413   \\ 
LRc0506    &  170699$^9$        &           8301  &  A2     &  6.95  &     3.88   &    1.49   &      270    &      89.282  \\  
IRLRa04    &  GSC00144-03031$^{1,10}$&          21960  &  A8     &  10.1  &       3.89    &      -    &     -     &     79.133   \\ 
IRLRa05    &  41641$^*$        	&           5685  &  A5     &  7.9   &     3.882  &    1.92   &    28     &      94.432  \\  
SRa05      &  48784$^*$        	&           3619  &  F0     &  6.66  &     3.84   &    1.87   &     108   &      25.305   \Bstrut \\
\hline
\end{tabular} 
\tablebib{(1) \citet{POR09}; (2) \citet{GH09}; (3) \citet{GH13}; (4) \citet{REE13}; (5) \citet{BRUN15};
		  (6) \citet{MRUIZ05}; (7) \citet{CREEV09}; (8) \citet{FOX10}; (9) \citet{MANT12}; \citet{POR05}; (10) \citet{RAI16};
(*) \corot\ archive (http://idoc-corot.ias.u-psud.fr/sitools/client-user/COROT\_N2\_PUBLIC\_DATA/project-index.html) accessed through the
 Seismic Plus portal (http://voparis-spaceinn.obspm.fr/seismic-plus/).) 
}
\end{table*}

\corot\ had two channels: one for the study of pulsating stars (Seismofield) and the other one for exoplanet detection (Exofield). The observations in the Seismofield are of greater precision since the targets are brighter (i.e. lower noise levels) and the cadence is higher (32 secs.). We have chosen Seismofield light curves since they have a better signal-to-noise (SNR) ratio due to the reduced readout noise because of the large number of pixels involved, and this is interesting in order to compare the distribution of frequencies minimizing the contribution of noise to the results.

The orbital frequency of the satellite is on average 13.972 $d^{-1}$ and the passage through the SAA occurs twice a day \citep{AUV09}. This has a direct impact in the observed stellar power spectra in the form of spurious peaks with their multiples, as well as in the form of combinations with frequencies of the intrinsic stellar oscillations produced by the convolution with the spectral window \citep{DEE75}. In the case of \dss,  pulsations are excited in the range between 10 and 80 $d^{-1}$ \citep[see][for evidence of this]{MOY17}. As a consequence the light curves of \dss\ are affected the most by these spurious frequencies.
    
The main criterion for the selection of stars was the type of variable, so the sample is composed of 15 stars observed in the Seismofield during the Initial Run (IR), Short Run (SR) and the Long Run (LR) both in galactic center and anticenter directions : 14 \dss\ and 1 \gd\ which present excited frequencies in the \dss\ regime \citep{CHA11}. Most of these stars (11) are A-type stars, and 4 are F-type stars (see Table~\ref{tab:nodes} for the specific characteristics of the runs and stars). 

As mentioned in the introduction, the most frequent gap duration in the light curves is 9 min but some gaps can last much longer. One example of this is HD170699, which is observed in the run LRc0506 (i.e. 5th/6th Long Run with galaxy center orientation), where a gap of about 2 hr is found in the light curve. This is the only star observed in LRc0506 in the sample of stars we have selected but in HD174589 there is a gap of 5 hr of duration. Both HD174589 and HD174532 have been observed in SRc02 (i.e. 2nd Short Run with galactic center orientation) so they have the same spectral window. In these stars the contribution of gaps to the spectral window is much greater. See Table~\ref{tab:gaps} for a complete statistical characterization of the gap distribution for the selected \corot\ dataset.


\begin{table*}
\caption[]{Statistical parameters of the gap distribution for the light curves observed by CoRoT used in this paper. In columns 3 to 5 $\tau_g$ refers to the duration of the gaps in hours, where column 3 is the maximum, column 4 is the mean duration, column 5 is the standard deviation of the gap distribution, and the last column is the most frequent gap duration.}
  \centering
  \begin{tabular}{lccccc}
  \hline Star ID  &  duty cycle $(\%)$ &  $\tau_g^{ max }$ (hr) &  $\bar{ \tau_g }$ (hr) & $\sigma( \tau_g )$ (hr)  &  Mode (hr) \\
  \hline 
HD50844    &        88.69        &        1.751        &        0.173        &        0.126        &        0.266          \Tstrut \\
HD174936   &        88.83        &        0.409        &        0.162        &        0.116        &        0.267          \\
HD174966   &        89.55        &        0.409        &        0.175        &        0.099        &        0.267          \\
HD181555   &        88.95        &        2.516        &        0.175        &        0.125        &        0.267          \\
HD49434    &        88.44        &        13.858       &        0.182        &        0.359        &        0.258          \\
HD172189   &        89.07        &        7.031        &        0.079        &        0.162        &        0.009          \\
HD174532   &        87.18        &        5.280        &        0.072        &        0.265        &        0.009          \\
HD174589   &        87.22        &        5.289        &        0.072        &        0.266        &        0.009          \\
HD51722    &        89.30        &        3.476        &        0.106        &        0.150        &        0.009          \\
HD51359    &        88.90        &        3.476        &        0.094        &        0.138        &        0.009	   \\
HD50870    &        89.78        &        3.476        &        0.102        &        0.142        &        0.009	   \\
HD170699   &        88.46        &        4.044        &        0.067        &        0.139        &        0.009	   \\
GSC00144-03031   &        88.46      &    7.636        &        0.066        &        0.227        &        0.009          \\
HD41641    &        78.78        &      144.249        &        0.119        &        2.552        &        0.009          \\
HD48784    &        88.55        &        1.582        &        0.161        &        0.137        &        0.267	   \Bstrut \\
\hline
\end{tabular}
 
 \label{tab:gaps}
 \end{table*}

\section{Methodology \label{sec:methods}}

Since we are interested in studying the impact of gaps in the usual harmonic analysis performed in asteroseismology, we will follow what can be considered a standard workflow. The only deviation from a standard procedure consists in filling the gaps in the light curves with two different methods. We do this with the objective of comparing the results with gapped data. In summary the workflow is:

\begin{enumerate}
\item Correction for the instrumental drift \citep{AUV09} by performing a polynomial fit to the light curves.
\item Gap-filling with ARMA and linear interpolation
\item Pre-whitening analysis of the interpolated and gapped data
\item Detection and removal of frequency combinations due to non-linear interactions between independent modes.
\end{enumerate}

\subsection{Gap-filling}

Linear or polynomial interpolation is still widely used for filling the gaps present in \corot\ light curves \citep[see e.g.][]{APP08, GUT09, KAL10} but also for datasets from other missions like \emph{SOHO} \citep{SEL11}. Re-sampling of irregular data with such analytic methods can be justified in a few cases because they are more robust \citep{WAE00} than more sophisticated interpolation methods that may have divergence issues, but the variance is erroneously estimated and the reconstruction of the original signal can be poor. Indeed, linear interpolation does not preserve the information (see, for example, Figs.~\ref{fig:lc1}, \ref{fig:lc2}, \ref{fig:lc9}, \ref{fig:lc10}, \ref{fig:lc11}, \ref{fig:lc12}, \ref{fig:lc13}). In Sect. 4, we will show that in some cases linear interpolation alters the frequency content in a similar way as no interpolation.

Here we use \miar\ algorithm (see PG15) which is aimed to preserve the original frequency content of the signal, thereby minimizing (and even avoiding) the contribution to the spectral window by the gaps that causes spurious variations in amplitudes and phases. The order of the ARMA model is selected through the Akaike Information Criterion \citep[AIC, ][]{AKA74} and the coefficients of the model are obtained through an optimization algorithm. In contrast to analytic methods, \miar\ guarantees that no bias due to ad-hoc hypothesis about the signal is introduced when filling the gaps (see full details in PG15).

So far some scholar studies assumed that all the spurious frequencies produced by the convolution of the signal with the spectral window \citep{DEE75} are mitigated during the pre-whitening cascade. Therefore, gap-filling was unnecessary. We have checked this assumption by comparing the pre-whitening of data with gaps and filled with two opposed gap-filling techniques: one is aimed at preserving information (\miar) and the other one is not (linear interpolation).

\subsection{Pre-whitening}
The classical pre-whitening technique \citep{PON81} was modified by \citet{REE07} in order to analyze \corot\ data using a rigorous statistical treatment of how to determine the significance of a peak in a DFT.

Light curves pre-processed as described above were subsequently analyzed using \sigspec. This algorithm, has been extensively used by the asteroseismic community \citep[see e.g.][]{PAP16, WEI16, MOL17, ZWI17}. It is based on an iterative sequence of frequency detection, least squares fitting, and a pre-whitening cascade. The iterative sequence stops when a significance threshold is reached (by default $sig=5.0$, that is $\approx SNR = 4$).

The range of frequencies explored by the algorithm is chosen taking into account several considerations. First, the lowest frequencies saturate the power spectrum and, though a deeper study of the selected stars would require including mixed and g modes, to simplify our study we focus here only on the fundamental radial mode and frequencies above (p modes). On the other side, \dss\ have their main frequency content below 80 $d^{-1}$. Therefore, we have used a frequency interval 2-100 $d^{-1}$ for the frequency detection. 

With these parameters \sigspec\ calculations lasted for several months in some cases (e.g. HD\,181555) to release a list of significant frequencies. Such lists were then \emph{cleaned} for spurious frequencies through the process explained below.

\subsection{Non-linear interactions}
In order to detect and remove non-linear interactions and spurious frequencies from the oscillation spectra we followed the heuristic method that we outline below:
\begin{itemize}
 	\item[$\bullet$] Estimate independent frequencies. We calculated the set of independent frequencies, $f_i$, in the range $2 \ge f_i \ge 100$ using the \sigspec\ addon \combi\ \citep{REE07}. The set of independent frequencies is arbitrarily truncated to 12 frequencies. When less than 12 frequencies were found we appended frequencies to the set as sorted by their amplitude. Frequencies below $2~\cd$ were excluded because most of the signal there comes from instrumental effects giving rise to trends rather than pure harmonic components.
 
 	\item[$\bullet$] Searching for harmonics and combinations up to 3rd order within a bin with the size of the Rayleigh dispersion ($f_R=1/T_{obs}$).
\end{itemize}

We differentiate spurious frequencies from combinations that appear due to the non-linear response of the physical system. Thus, spurious frequencies are identified with the orbital frequency, aliases of $1\,\cd$, and the harmonics and combinations of these with independent frequencies.

In the second part our algorithm calculates for each independent frequency found the following combinations:
\begin{enumerate}
 \item Harmonics up to the 5th order.
 \item First-order combinations including absolute value differences.
 \item Second- and third-order combinations, i.e: 
    \begin{align*}  
      2f_1 \pm f_2,...,2f_1 \pm 2f_2,.., \\
      3f_1 \pm f_2,...,3f_1 \pm 2f_2,...,3f_1\pm 3f_2,...    
    \end{align*} 
 \item Combinations with the frequency of the satellite's orbit ($f_s=13.972~d^{-1}$) and their harmonics up to 4th order and $f_s/2$, $f_s/3$, $f_s/4$.
 \item Combinations of the harmonics of $f_s$ up to the 3rd order with the first 4 frequencies (only for the first order).
 \item $1\,\cd$ aliases around $f_s$ and their harmonics up to the 4th order, and the harmonics of the aliases up to the 5th order, i.e:
    \begin{align*}
      f_s\pm 1,\pm 2,\pm 3,\pm 4,\pm 5, \\
      2f_s \pm 1,...,\pm 5, \\
      3f_s\pm 1,...,\pm 5, \\
      4f_s\pm 1,...,\pm 5
    \end{align*}
 \item Aliases up to the 5th order of the combinations calculated in step 5.
 \item The two highest amplitude frequencies ($f_b$) in the range 0-2 $d^{-1}$ are used to calculate interactions with the satellite orbit.
 \item Combinations up to the 5th order between $f_s$ and $f_b$ (1st order).
 \item $1\,\cd$ aliases around combinations calculated in step 8, up to the 5th order.
 \item Harmonics of the main frequency up to the 14th order.
\end{enumerate}

Since the passing of the satellite through the SAA is twice each sidereal day we include $1\,\cd$ aliases at step 5. Notice that, though we excluded frequencies below $2\,\cd$ from the calculation of independent frequencies since we are considering only p modes, we included this range in step 8 since there might be interactions between low frequencies and p modes.

Due to the density of significant frequencies found in \dss, we follow a conservative approach and identify a reduced set of harmonics and combinations with the orbital frequency in order to guarantee the unambiguous and robust identification of spurious frequencies. It is possible that extending the order of harmonics and combinations these number would increase but this is out of the scope of this paper since the aim is to test the impact of the gap-filling on the classical procedure for pre-whitening and cleaning used in asteroseismology.

In summary, the algorithm calculates for each star a set of 703 frequencies of potential non-linear interactions (including spurious frequencies) that are searched for in the list of significant frequencies obtained by \sigspec. When coincidences of the order of the Rayleigh dispersion are found those frequencies are removed from the list.

\section{Results \label{sec:results}}
In this section we present the results of gap-filling, power spectrum and statistical characterization for the selected set of stars. Figures corresponding to HD174966 are shown here as an illustrative case, the rest of the figures for the other 14 stars are shown in the Appendices.

In order to work optimally, radix-2 FFT algorithms need that the number of datapoints is a power of 2, so we have truncated N to the power of 2 closer to the original number of datapoints. Additionally, light curves are normalized statistically so the amplitudes are dimensionless.

Periodograms are plotted in log-scale in amplitudes and $\cd$ in frequencies in the range $2-100\,\cd$. Power spectra of gapped and ARMA interpolated data are compared in the Appendices C together with power spectra of linearly interpolated data.

\begin{figure}[ht]
   \includegraphics[width=8cm]{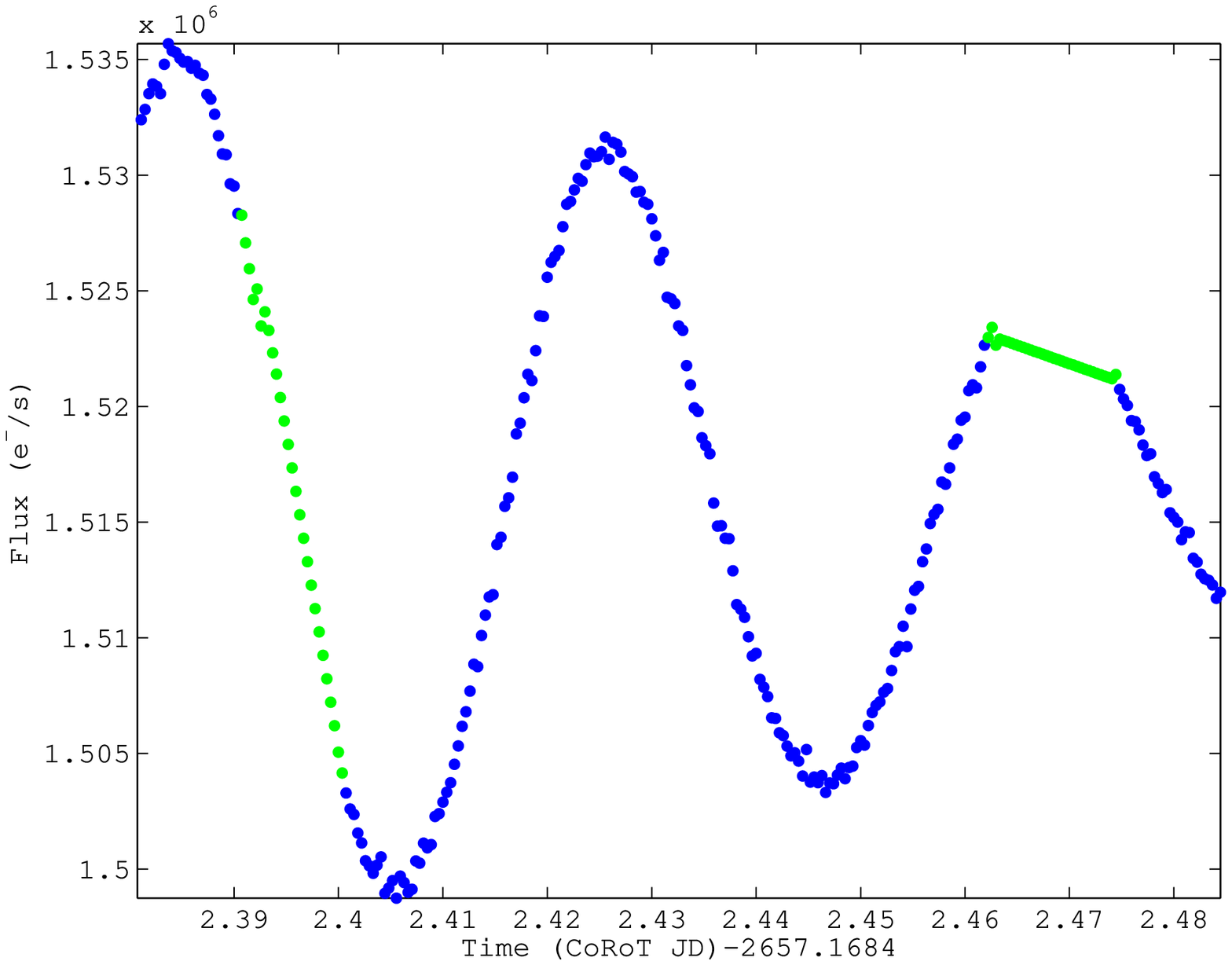}
   \includegraphics[width=8cm]{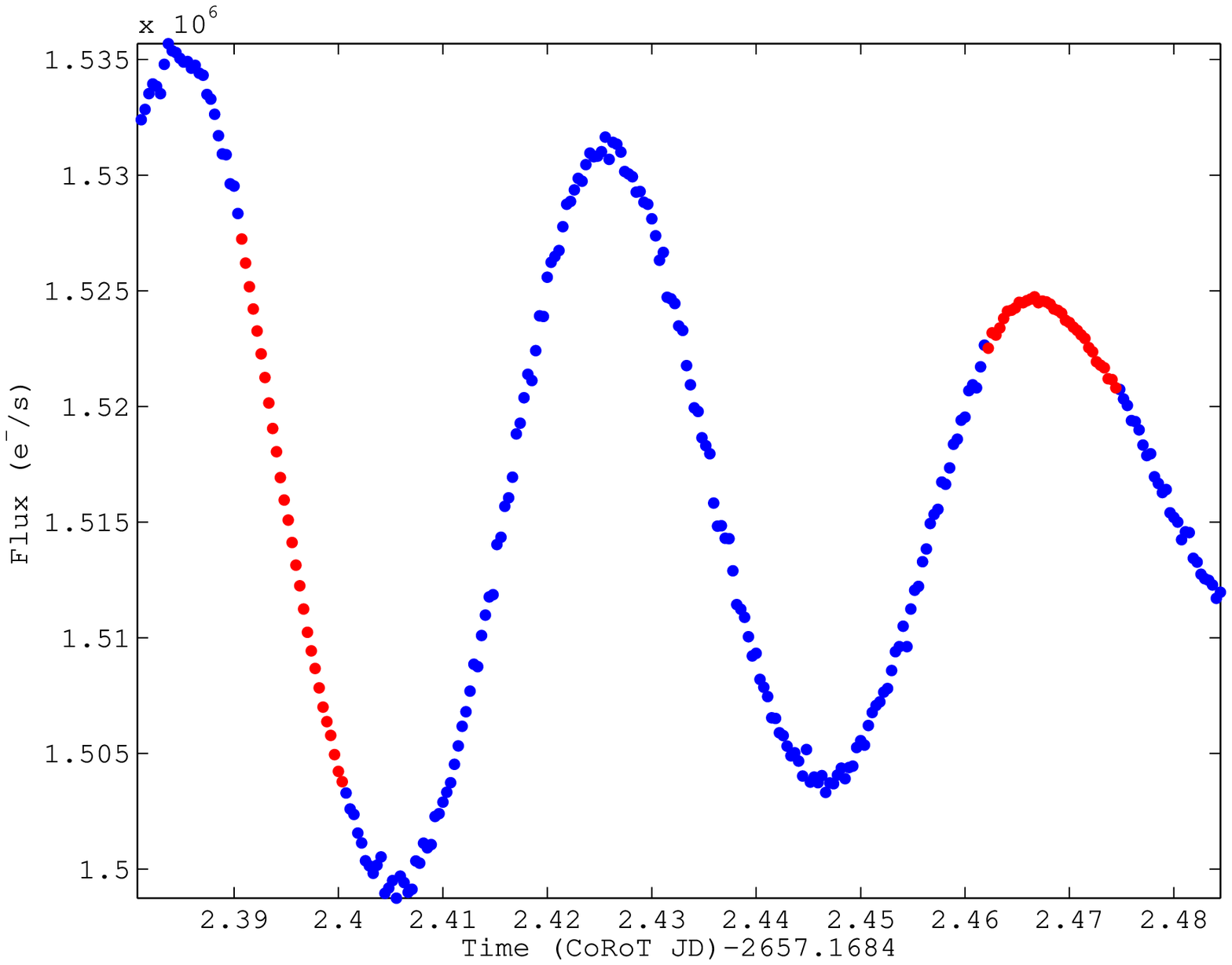}  
   \caption{Illustration of a gap in HD 174966 that has been interpolated linearly (upper panel, in green), and with \miar\ (lower panel, in red).}
   \label{fig:lc}%
\end{figure}

In Fig.\ref{fig:lc} we show a typical gap in the light curve of HD174966 that has been interpolated linearly and with \miar.

In this case it is clear that a linear interpolation does not preserve the frequency content of the signal and as a consequence, spurious peaks will appear in the power spectrum that might introduce a bias in the parameter estimation of the harmonic components and its related physics. Whether a pre-whitening process permits to overcome these difficulties in the frequency analysis or not is something that we will discuss in the following section.

Some stars shown in the Appendices (see, for example, \ref{fig:lc3}, \ref{fig:lc4}, \ref{fig:lc14}) present no apparent differences in the light curves interpolated by ARMA and linearly, and between their corresponding power spectra (see \ref{fig:ps3}, \ref{fig:ps4}, \ref{fig:ps14}). This is in contrast to other stars (see e.g. \ref{fig:lc7}, \ref{fig:lc10}, \ref{fig:lc12}) where ARMA models appear to preserve the signal and linear interpolation not (see also their corresponding power spectra \ref{fig:ps7}, \ref{fig:ps10}, \ref{fig:ps12}). If we compare the first set of cases with the last one, we can see that linear interpolation appears to work better when the signal shows longer scale variations inside the gaps (i.e. derivatives do not change significantly). This can give us an idea of when interpolating linearly is a risk to take seriously into account.

\begin{figure}[ht]
   \includegraphics[width=9.5cm]{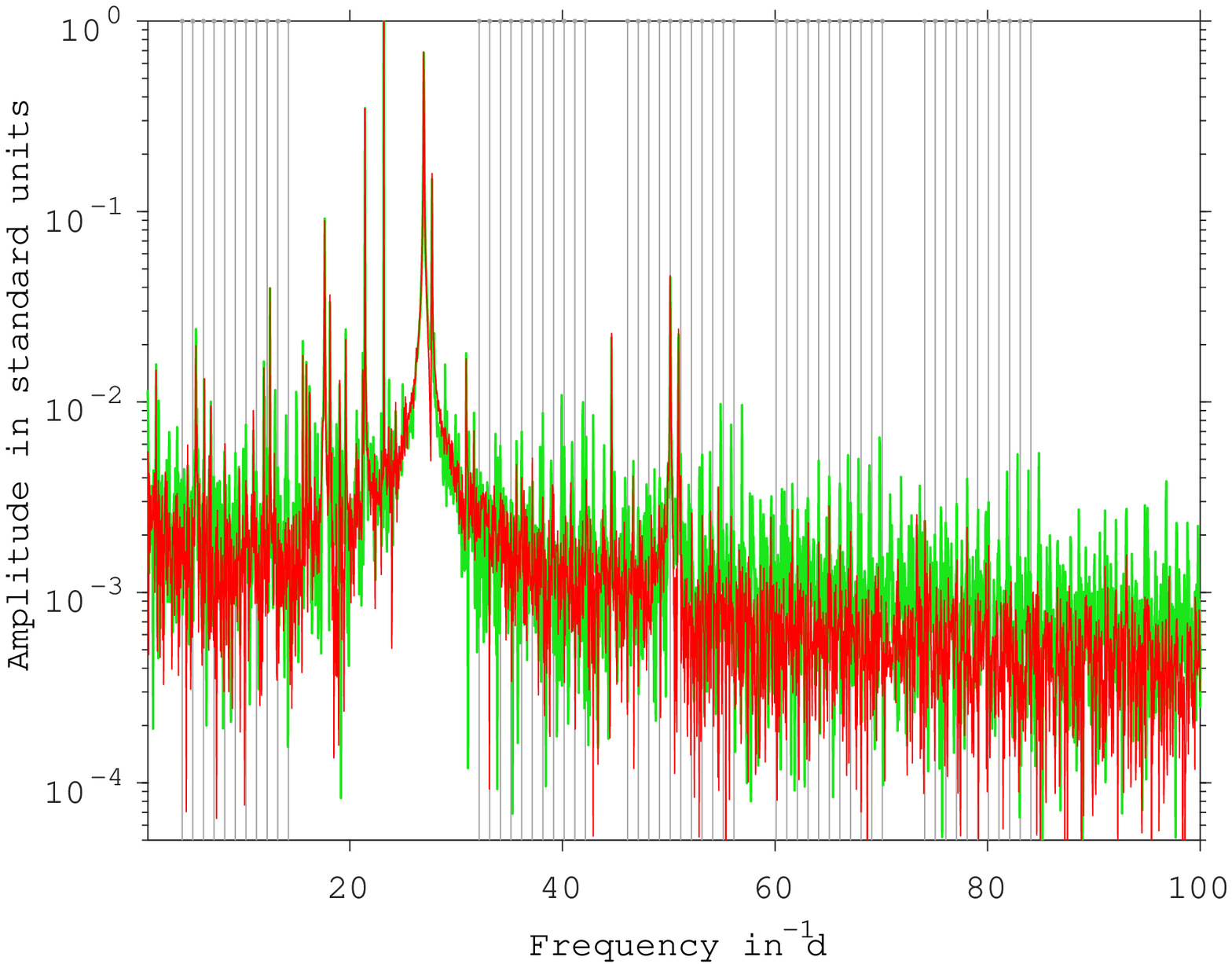}
   \includegraphics[width=9.5cm]{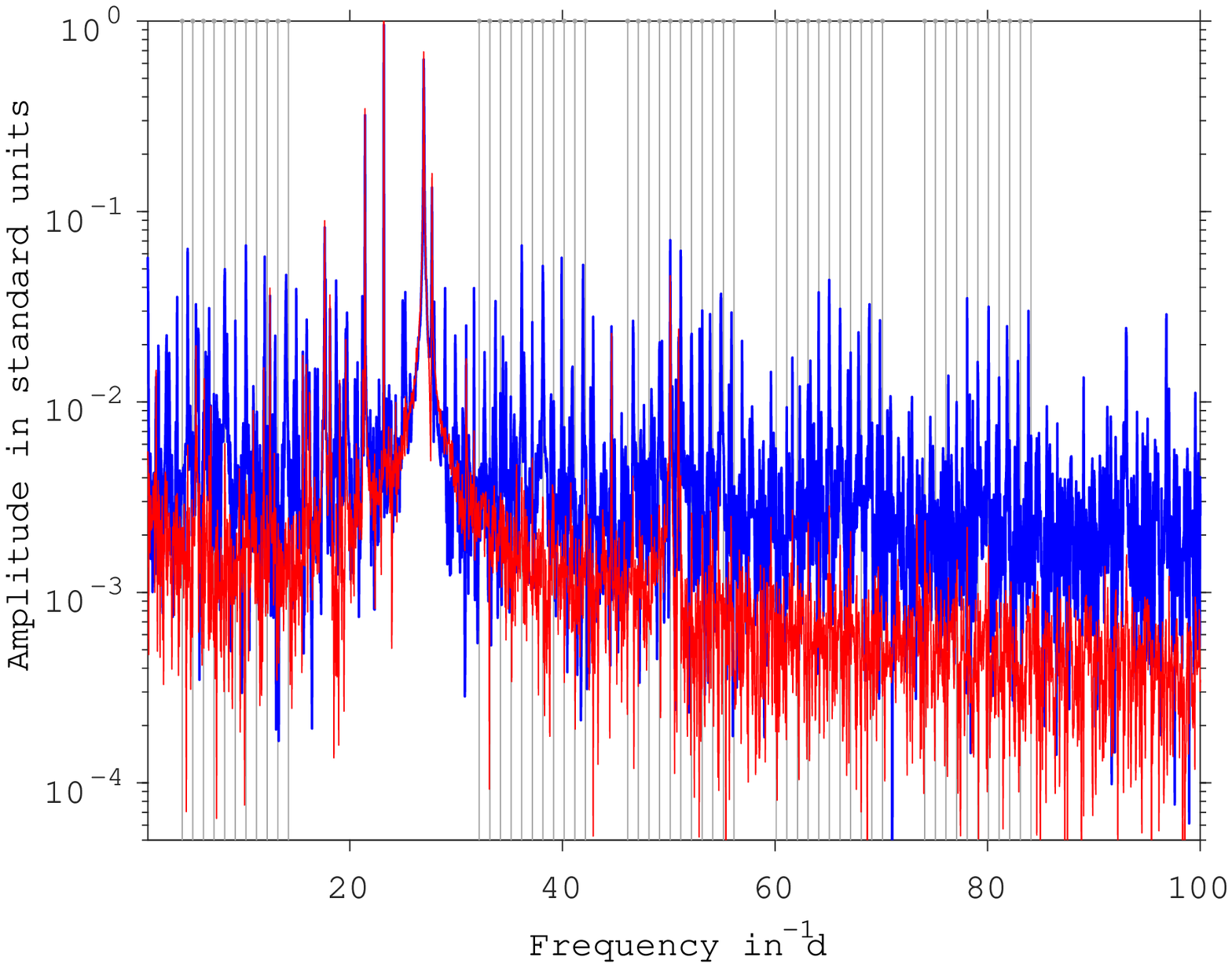}
	\caption{Power spectra of the light curves from HD 174966: lower panel shows gapped data in blue and ARMA interpolated data in red, upper panel shows linearly interpolated data in green and ARMA also in red. Vertical gray lines show the main peaks of the spectral window.}
   \label{fig:ps}
\end{figure}

In Fig.\ref{fig:ps} we show these plots for HD174966. Both power spectra of gapped data and linearly interpolated data, calculated through a Lomb-Scargle periodogram \citep{SCA82}, show clearly the effect of the satellite orbital modulation and SAA, that is, patterns of frequencies around $13.97\,\cd$ and their multiples (vertical gray lines). These patterns do not appear in the power spectrum of ARMA interpolated data.

In order to perform a statistical characterization we apply the cleaning procedure described in the previous section.

\begin{figure}[ht]
   \centering
   \includegraphics[width=9.5cm]{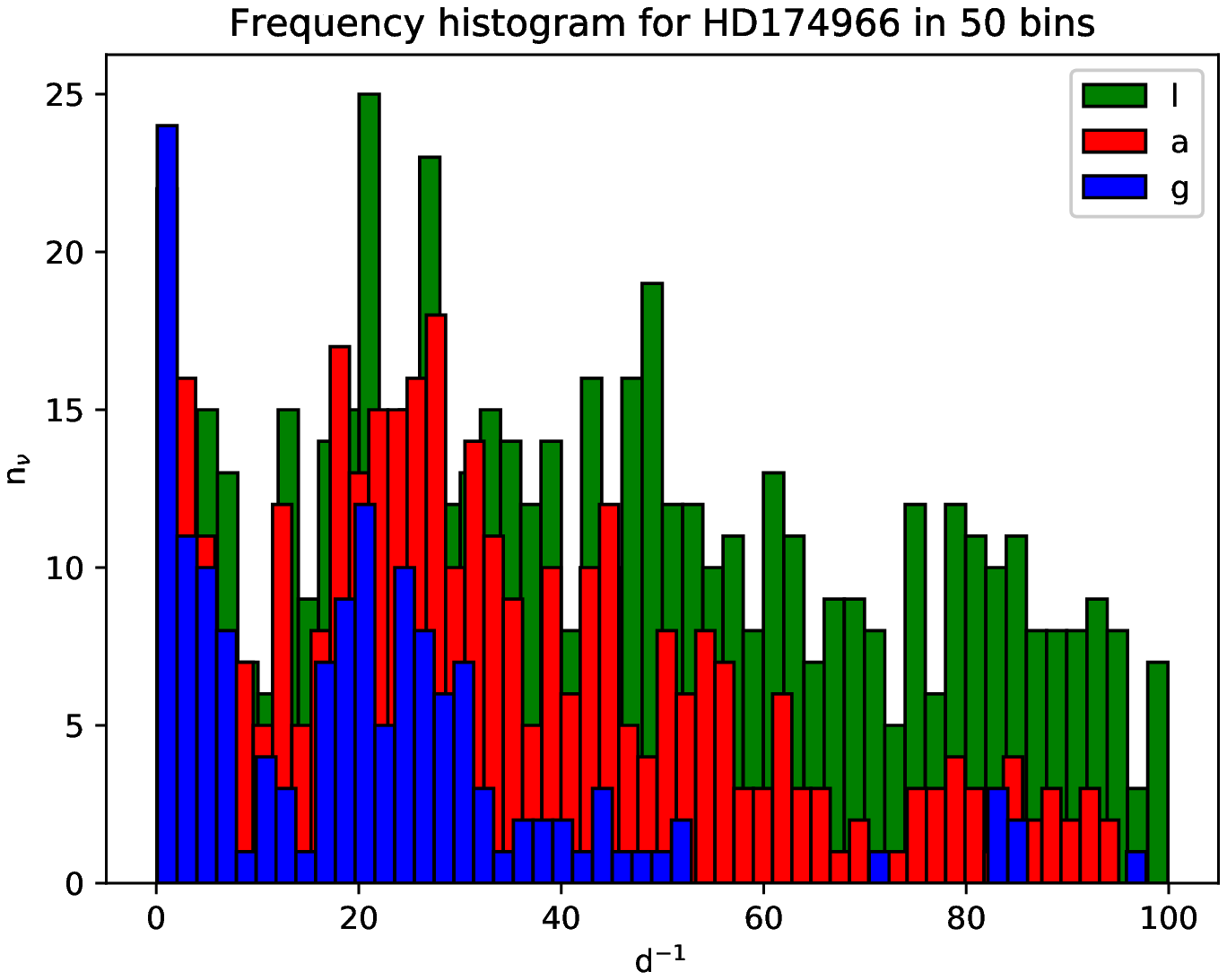}
   \includegraphics[width=9.5cm]{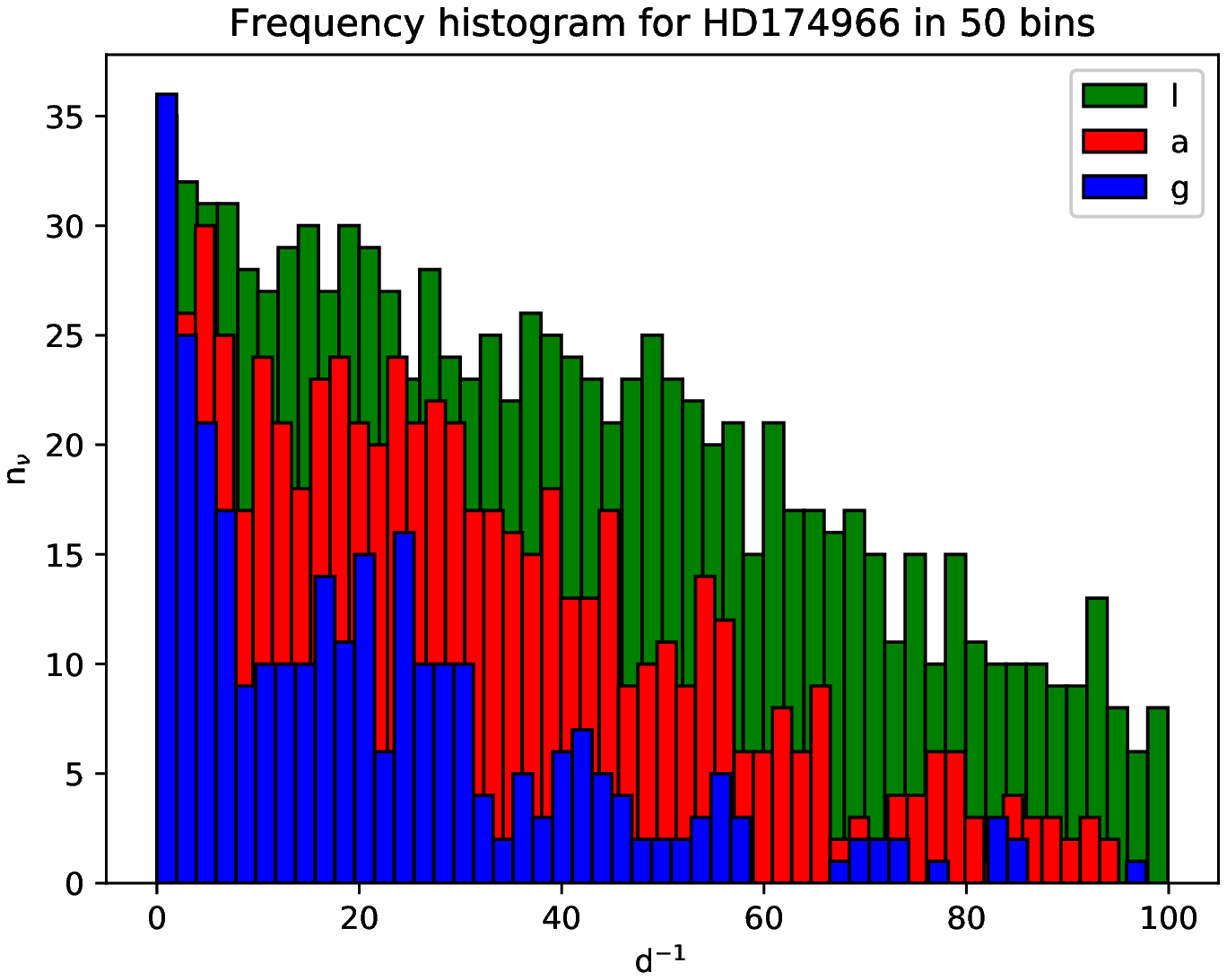}
   \caption{Histograms of frequencies detected in the light curves of HD174966. Upper panel shows cleaned frequencies whereas lower panel shows frequencies before the cleaning is applied. Blue bars correspond to gapped data (g), red bars to ARMA (a) interpolated data, and green bars to linearly (l) interpolated data.
   }
   \label{fig:h}%
\end{figure}

In Fig.\ref{fig:h} we show the histogram of the number of cleaned frequencies for each light curve (ARMA, linear, gapped) of HD174966 resulting after applying the cleaning procedure (upper panel) and the same for the frequencies obtained before the cleaning (lower panel). The cleaning procedure change considerably the distribution and the differences between histogram bins. Therefore, the cleaning is necessary to avoid biases in the analysis. From now on we will refer only to cleaned frequencies.

Notice that the number of cleaned frequencies resulting in gapped data is lower than in ARMA interpolated data, and it is lower in that case than in linearly interpolated data. A possible explanation is that, due to a lower duty-cycle in gapped data, the significance of some of the harmonic components detected in ARMA interpolated data is insufficient.

On the other side, although the cleaning procedure is aimed for removing spurious and combination frequencies, due to the conservative limits used for the harmonic number and possible interactions, it is very probable that some spurious frequencies are not removed as mentioned above. Then, the higher number of frequencies detected in linearly interpolated data might be originated by a higher number of spurious frequencies appearing in the power spectrum. This is compatible with the results shown in Table~\ref{tab:numfreq} that will be discussed thoroughly in the next section for HD174966 and shown for the other stars in the Appendices.

\section{Discussion}
Whether the frequency content is preserved or not when filling gaps with interpolated data is strictly dependent on the algorithm used and the bandwidth of frequencies present in the signal. 

Gaps in \corot\ data cause a reduced amplitude of the peaks when non-interpolated data is analyzed but, as can be seen in the figures included in the Appendices, linear interpolation also change the original frequency content of the signal introducing spurious peaks that leak the power and change the original phases that might be crucial for modal identification due to the correlation between frequency bins \citep[see][and Appendix A]{GAB94}.
	
Note that not all data collected during the passing through the SAA is corrected. The algorithm for data correction implemented in the pipeline of CoRoT N2 level (version 0.2/1.3) is conservative and interpolates only data points with a certain deviation from the mean. This means that in some light curves wrong data points appear that are not interpolated (see e.g. Figs.~\ref{fig:lc5} and \ref{fig:lc6}) and this cause an additional spurious contribution to the power spectrum. However, this is a marginal effect in our sample.

The impact of the gaps depends on the spectral window associated to the gap distribution and sizes. In Table~\ref{tab:gaps} we present the statistical parameters characterizing this distribution. The most frequent gap appearing in the light curves of 9 stars is $9 \times 10^{-3}$~hr (32 secs), that is just one datapoint, meaning that they are dominated by single outliers. The rest have its most frequent gap at $0.267$~hr which is $16.02$ min, but their median is around $0.168$ ($\sim 10$ min) which is more coherent with the duration of the pass through the SAA. However, some stars have much longer gaps (e.g. a $\sim 6$~d length gap in HD41641 or a 13.86~hr length gap in the light curve of HD49434). The contribution of these gaps to the spectral window is much greater.

The contribution of the interpolation to the power spectra shown in Fig.\ref{fig:ps} and Figs.\ref{fig:ps2}, \ref{fig:ps6}, \ref{fig:ps7}, \ref{fig:ps8}, \ref{fig:ps10} are very similar. This is consistent with the values of the statistical parameters shown in Table~\ref{tab:gaps} since these three groups of stars have been observed each during the same run of observations: HD174936 and HD174966 (SRc01), HD174532 and HD174589 (SRc02), HD51722 \& HD50870 (LRa02). Only HD51359 (see Fig.\ref{fig:ps9}), observed during LRa02 too, shows a different behavior since it appears to have only low frequency variability. 

The numbers of frequencies classified as spurious (NS), combinations (NC) and independent (NI) for each star and each light curve (gapped, linearly interpolated and ARMA interpolated) are collected in Table~\ref{tab:numfreq}, where $NI+NS+NC$ amounts the total number of frequencies detected using \sigspec.

Note that the numbers of spurious frequencies detected are a small fraction of the total number of frequencies in most cases. This might be due to the reduced set of harmonics and combinations used.

\begin{table*}
\caption[]{Number of independent frequencies detected after cleaning for non-linear interactions and spurious frequencies (NI); number of frequency combinations found and removed (NC), and number of spurious frequencies found and removed (NS). Columns (A,L,G) correspond, respectively, to the frequencies detected in ARMA interpolated, linearly interpolated, and gapped light curves.
  The last three columns represent the Anderson-Darling (AD) test applied to the histogram densities of each (A,L,G) case compared by
  pairs: A with G, A with L, and L with G, respectively. These values  correspond to the p-value of the AD test. The null hypothesis is that compared \textbf{samples arose from a common distribution}.}
\centering
  \begin{tabular}{l|ccc|ccc|ccc|ccc}
    \hline
              &         &    NI       &           & 
   			  	        &    NC       &           &
                        &    NS       &           &
                        &    A-D Test                \\
    Star ID   &    A    &    L        &    G      & 
   				   A    &    L        &    G      &
                   A    &    L        &    G      &
                   AG   &    AL       &    LG        \\
   \hline
HD50844         &    1189   &     1124   &     1001  &  237 & 214 & 168 & 320 & 319 & 320 & $1.9\times 10^{-5}$ & $7.4\times 10^{-6}$  & $4.6\times 10^{-10}$  \Tstrut \\

HD174936        &    520    &     550    &     348   &   82 &  81 &  68 & 268 & 284 & 164 & $2.3\times 10^{-3}$ & $4.4\times 10^{-2}$  & $2.1\times 10^{-4}$\\

HD174966        &    364    &     583   &     152   &   42 &  21 &   8 & 239 & 412 & 136 & $3.4\times 10^{-15}$ & $7.1\times 10^{-12}$  & $1.4\times 10^{-22}$  \\

HD181555        &    1782   &     1786   &     2054  &  171 & 211 & 196 & 151 & 157 & 250 & $3.5\times 10^{-30}$ & $5.5\times 10^{-11}$  & $3.1\times 10^{-47}$  \\

HD49434         &    1232   &     1284   &     2243  &  287 & 161 & 308 &  93 &  96 & 199 & $2.9\times 10^{-48}$ & $4.7\times 10^{-10}$  & $7.8\times 10^{-41}$  \\

HD172189        &    1360   &     1275   &     1790  &  114 & 284 & 158 & 150 & 177 & 151 & $3.3\times 10^{-40}$ & $9.6\times 10^{-29}$  & $1.1\times 10^{-26}$  \\

HD174532        &    516    &     532   &     405   &  107 & 132 &  89 & 328 & 377 & 284 & $1.5\times 10^{-9}$ & $2.9\times 10^{-8}$  & $3.3\times 10^{-16}$  \\

HD174589        &    237    &     246    &     116   &   58 &  80 &  35 & 208 & 260 & 100 & $5.9\times 10^{-25}$ & $4.3\times 10^{-5}$  & $4.1\times 10^{-34}$  \\

HD51722         &    1461   &     2913   &     1379  &  210 &  53 & 178 & 179 & 284 & 193 & $5.9\times 10^{-23}$ & $5.3\times 10^{-34}$  & $3.1\times 10^{-35}$  \\

HD51359         &    1692   &     1681   &     2634  &  144 & 244 & 175 & 174 & 229 & 306 & $5.7\times 10^{-28}$ & $9.1\times 10^{-10}$  & $6.3\times 10^{-29}$  \\

HD50870         &    2041   &     3294   &     2150  &  204 & 219 & 207 & 239 & 290 & 260 & $3.3\times 10^{-21}$ & $6.3\times 10^{-28}$  & $2.0\times 10^{-18}$  \\

HD170699        &    2878   &     2993   &     2042  &  238 & 150 & 198 & 326 & 392 & 310 & $1.4\times 10^{-12}$ & $1.1\times 10^{-1}$  & $1.5\times 10^{-10}$ \\

GSC00144-03031  &    2971   &     2174   &     1735  &   73 & 304 & 157 & 444 & 429 & 285 & $6.2\times 10^{-12}$ & $8.8\times 10^{-11}$ & $1.5\times 10^{-5}$ \\

HD41641         &    2017   &     2833   &     2301  &  182 & 249 & 238 & 351 & 419 & 389 & $1.6\times 10^{-17}$ & $7.7\times 10^{-14}$ & $6.1\times 10^{-4}$  \\

HD48784         &    183    &     171    &     141   &   28 &  32 &  43 & 253 & 222 & 139 & $2.1\times 10^{-3}$ & $3.0\times 10^{-1}$  & $8.3\times 10^{-3}$ \Bstrut \\
\hline
\end{tabular}
  \label{tab:numfreq}
  \end{table*}

After cleaning spurious frequencies and linear combinations (produced by non-linear interactions) of independent frequencies, the resulting number of frequency components (NI) is still of the order of $\sim1000$ in many cases. Note that there might be yet spurious and combination frequencies hidden in the final list that cannot be resolved unambiguously.

No frequency component has been detected beyond the photometric precision of CoRoT Seismofield cameras which is $0.6-4$ ppm \citep{AUV09}. Therefore the high number of frequencies detected cannot be explained by a lack of precision. Further analysis is required to clarify the origin of the frequency components.

In order to have a quantitative validation of the statistical results we have performed a k-samples Anderson-Darling test \citep{SCH87} to the histograms of each case comparing them by pairs. This test evaluates the hypothesis that 2 time series with n independent samples arise from a common unspecified distribution. The choice of this test is motivated by the fact that it does not require any specific probability distribution.

In the last column of table~\ref{tab:numfreq} we show the corresponding p-values (false alarm probability) for each pair. The null hypothesis is that compared frequency values are originated from a common distribution so when a small p-value is found (here below 0.01) it can be interpreted that the frequencies are not compatible with the same distribution. Only the pair AL for HD170699, HD48784 and HD174936 show non-negligible p-values avoiding the rejection of the null hypothesis. This reinforces the conclusion that gap-filling methods have an impact on the estimation of the frequency content of our sample (i.e. the pre-whitening is not unbiased).

\begin{figure}
    \includegraphics[width=7.5cm]{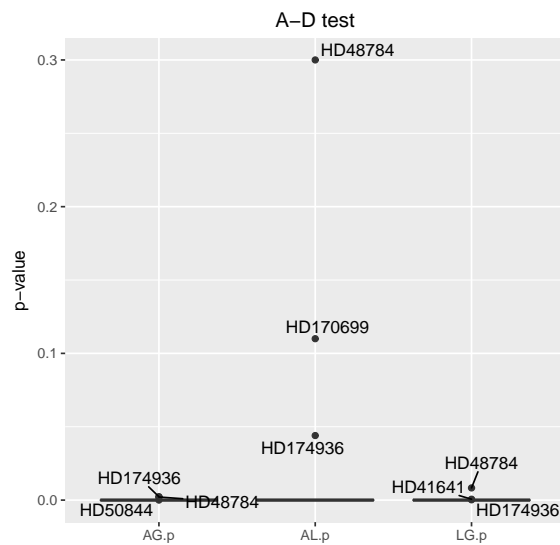}
    \caption{Box plot of the p-values calculated in Table 3. Notice that most of the stars present a p-value of $\sim 10^{-5}$. The outliers of the sample are labelled by their HD number.}
    \label{fig:boxout}
 \end{figure}

We have performed also a box-plot of the p-values to evaluate differences in the distributions of frequencies (Fig.~\ref{fig:boxout}). 
The Anderson-Darling test gives p-values of the order of $10^{-5}$ in most cases. On the other side, three stars (HD48784, HD170699, HD174936) denoted by points, have much higher p-values in the comparison between ARMA and linearly interpolated data. This means that ARMA and linearly interpolated data present similar frequency distributions in these cases. This could be, in fact, an effect produced by the sample selection related to the analyticity of the intervals that are interpolated (see Figs.~\ref{fig:lc13} and \ref{fig:lc2}, for example). Also, other factors that could affect to our sample such as the rotational velocity and the visibility of the modes cannot be assessed here. In any case, these 3 stars can be considered outliers in our sample that are not affecting the statistical characterization.

Finally, in order to confirm that the bias that we have found is originated by the pre-whitening process we have performed an additional AD test applied to the histogram densities of the (A,L,G) power spectrum before any pre-whitening process is applied (see Table 4).

We check the consistency of the analysis (unbiasedness) comparing the AD test of Table 3 and 4 and verifying if the differences in the frequency distribution of A, L, G are always less after the pre-whitening process than before. We call this the consistency condition.

These are the possibilities:
\begin{enumerate}[label=\Alph*]
 \item the frequency distributions are initially similar, and similar at the end
 \item the frequency distributions are initially similar, different at the end
 \item the frequency distributions are initially different, similar at the end
 \item the frequency distributions are initially different, different at the end
\end{enumerate}
 
Cases A and C imply that the pre-whitening process fulfill the consistency condition. B shows an inconsistency in the analysis, i.e. the pre-whitening process bias the results. Finally, though case D fails the consistency condition, its origin is indeterminate. Now, assuming as significance criterion $p > 10^{-2}$ and comparing Table 3 and 4, we can affirm the following statements relative to each comparison:

\begin{itemize}
 \item AG - HD174936 is case D), and the rest are B) cases. No star fulfill the consistency condition.
 \item AL - Only HD170699, HD48784 fulfill the condition by A) and HD174936 by C). HD50870 and HD41641 are D) cases and the rest are B) cases.
 \item LG - HD50870 and HD41641 are D) cases, the rest are B) cases. No star fulfill the consistency condition.
\end{itemize}
 
Since the intersection of the sets AL, AG, LG fulfilling the consistency condition is null we conclude that unequivocally, there is an inconsistency in the pre-whitening process.

Additionally, most of the tests are B) cases showing that during the pre-whitening process spurious residual frequencies appear as a consequence of a bad fitting. This is consistent with the results that \citet{BAL14} obtained for HD50844.

\begin{table*}[ht]
	\caption{Anderson-Darling (AD) test applied to the histogram densities of the (A,L,G) 
			oscillation spectrum before any pre-whitening process is applied. The pairs AG, AL, 
            and LG, represent the compared distributions for the three gap-filling method considered
            in this work. The numbers in brackets are the value of the AD test and its corresponding
            p-value. The null hypothesis is that compared \textbf{samples arose from a common distribution}.} 
	\centering
	\begin{tabular}{l|ccc}
    \hline
             		&           &    A-D Test  &              \\
           Star 	&   AG      &    AL        &    LG        \\
   \hline
  HD50844 & (0, 1) & (0, 1) & (0, 1) \Tstrut \\ 
  HD174936 & (82.7, $1.04\times10^{-45}$) & (82,7, $1.04\times10^{-45}$) & (0,1) \\ 
  HD174966 & (0, 1) & (0, 1) & (0, 1) \\ 
  HD181555 & (0, 1) & (0, 1) & (0, 1) \\ 
  HD49434 & (0, 1) & (0, 1) & (0, 1) \\
  HD172189 & (0, 1) & (0, 1) & (0, 1) \\ 
  HD174532 & (0, 1) & (0, 1) & (0, 1) \\ 
  HD174589 & (0, 1) & (0, 1) & (0, 1) \\ 
  HD51722 & (0, 1) & (0, 1) & (0, 1) \\
  HD51359 & (0, 1) & (0, 1) & (0, 1) \\ 
  HD50870 & (0, 1) & (114,$3.45\times10^{-63}$) & (114, $3.45\times10^{-63}$) \\ 
  HD170699 & (0, 1) & (0, 1) & (0, 1) \\ 
  GSC00144-03031 & (0, 1) & (0, 1) & (0, 1) \\ 
  HD41641 & (0, 1) & (112, $1.29\times10^{-61}$) & (112, $1.29\times10^{-61}$) \\  
  HD48784 & (0, 1) & (0, 1) & (0, 1) \Bstrut \\
   \hline
	\end{tabular}
      \label{tab:s0adtest}
\end{table*}

\section{Conclusions}
In PG15 it was shown that it is essential for filling gaps in time series to use interpolation techniques like \miar, which are aimed to preserve the original frequency content of the time series. Otherwise, the periodogram cannot be an unbiased estimator of the pulsational content of the stars. 

Here we have used that result to investigate whether the classical pre-whitening procedure could give biased results when analyzing \dss. To do that we have applied the widely used algorithm \sigspec\ to a set of 15 \dss\ from the Seismofield of \corot\ to perform a frequency analysis of linearly interpolated, ARMA interpolated and gapped light curves. This allow us to test the efficiency of the pre-whitening cascade that this program applies to suppress the contribution of the spectral window to the power spectrum. 

The differences found between the results of the analyses show that, at least, for \dss\ the pre-whitening cascade is not sufficiently efficient to remove the spurious frequencies caused by the presence of gaps. This is a novel result that enhance the importance of using a gap-filling method that is aimed at preserve the information. 

These results might have a significant impact on asteroseismic studies. In particular for \dss, the study of quasi-periodicities is being used to constraint the internal structure of the star \citep[see e.g.][]{GH09, SUA14, GH15}. These periodicities are highly sensitive to how frequencies are distributed in the periodogram.  Any variation due to an incorrect pre-whitening of the light curves might introduce a bias in these periodicities. We will evaluate this in a future paper (Suárez et al., in prep.) using the same sample of \dss\ observed by CoRoT satellite.

\begin{acknowledgements}
JPG acknowledge support from the "Junta de Andaluc\'{i}a" local government under project 2012-P12-TIC-2469.
JCS and AGH acknowledge funding support from Spanish public funds for research under project ESP2015-65712-C5-5-R (MINECO/FEDER), and from project RYC-2012-09913 under the ‘Ramón y Cajal' programme of the Spanish MINECO.
JCS also acknowledge support by the European FP7 project "SPACEINN" (FP7-SPACE-2011-1).
AM acknowledges support by the Spanish MINECO National Space Program through project ESP2015-65712-C5-1-R and 'Ramon y Cajal' program RYC-2012-09913. This project has received funding from the European Union’s Horizon 2020 research and innovation programme under the Marie Sklodowska-Curie grant agreement No 749962. \\ The CoRoT space mission, launched on 2006 December 27, was developed and is operated by the CNES, with participation of the Science Programs of ESA, ESA's RSSD, Austria, Belgium, Brazil, Germany and Spain.
\end{acknowledgements}
%
\bibliographystyle{aa} 
\bibliography{gapsdscuti} 
\clearpage
\appendix

\section{Spectral response function of linear interpolation}
According to the definition given by \citet{DEE75} the expected value of the classical periodogram is obtained as the convolution of the true power spectrum with the spectral window associated to the sampling used. The aim of gap-filling with any interpolation method is to recover the original regular sampling so, in this sense, the spectral window is just a sinc function. More important is to determine the spectral response function of the model used for the interpolation. Here we discuss the properties of the spectral response function of linear interpolation and its relation to a rectangular window (i.e. no interpolation).

In order to interpolate linearly inside gaps the algorithm may use just two datapoints, one at the beginning and one at the end of the gap. Then, a linear interpolation can be considered just an oversampling of the gap interval.

Let say without loss of generality that the gap interval is the unity and $\xi$ is a number between 0 and 1, then a linearly interpolated value at an inner point is:
\begin{equation}
\hat{X}(n + \xi) = (1-\xi) \cdot X(n) + \xi \cdot X(n+1)
\end{equation}
where $X(n)$ and $X(n+1)$ are the datapoints used for the interpolation. It is easy to see that for $\xi=0$ and $\xi=1$ the linearly interpolated value coincides with the datapoints $X(n)$ and $X(n+1)$ and for $\xi=1/2$ it is just the mean between both values.

The interpolation error of the $\xi$ datapoint,
\begin{equation}
\epsilon_{\xi} = |\hat{X}(n + \xi) - X(n + \xi)|
\end{equation}
is nonzero between the datapoints but it can be $\approx 0$ if X is a linear function between n and n+1 (not necessarily in other intervals).

Now, let say to simplify that $n=0$, then
\begin{equation*}
\hat{X}(\xi) = (1 - \xi) \cdot X(0) + \xi \cdot X(1)
\end{equation*}
If we interpolate L points inside the gap interval we can say the interval $(0,1)$ which has initially unit sampling will be oversampled to $1/L$. Then $\xi = k/L$ with $ 0 \le k < L$. Now the equation above can be expressed as:
\begin{equation*}
\hat{X}(k/L) = (1 - k/L) \cdot X(0) + (k/L) \cdot X(1)
\end{equation*}
and defining $h(k) = 1-k/L$ and $h(k-L) = k/L$ we have finally
\begin{equation}
\hat{X}(k/L) = h(k) \cdot X(0) + h(k-L) \cdot X(1)
\end{equation}
In this way we have defined linear interpolation through the convolution of a triangular filter h of length $2L-1$ with impulse response:
\begin{eqnarray}
h(k) & = & 1-|k|/L, \qquad |k|<L \\
     & = & 0 \qquad \text{otherwise} \nonumber
\end{eqnarray}

If we want to interpolate, for example, 4 points, $L=5$ and the impulse response h has a duration of $N = 2\cdot L -1 = 2 \cdot 5 - 1 = 9$ but only two nonzero samples are coincident with h.

A triangular filter can be defined as the convolution of two rectangular functions:
\begin{equation*}
tri(k) = rect(k) \star rect(k)
\end{equation*}
so the frequency response of the linear interpolation is a product of sinc functions:
\begin{equation}
P(\nu) = N^2 sinc^2 (\nu/N)
\end{equation}

\begin{figure}[ht]
   \centering
   \includegraphics[width=8.5cm]{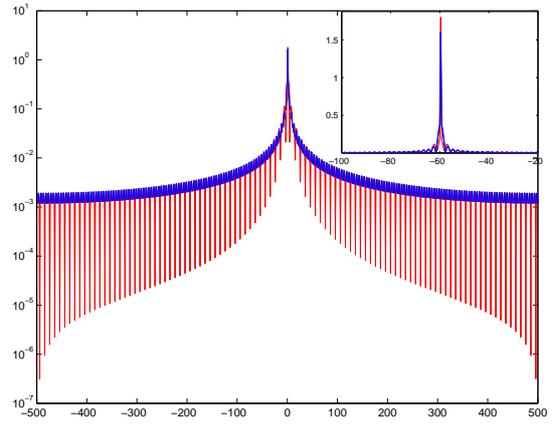}
   \caption{Spectral response function in log scale associated to gapped data (in blue) and linearly interpolated data (in red). See the inset for a zoom of the central peak in linear scale.}
   \label{fig:sw}%
\end{figure}

In Fig.\ref{fig:sw} we compare the spectral response function caused by a gap and the one corresponding to linearly interpolated data. Notice that, though linear interpolation concentrate the power at the central peak and reduce the power at the sidelobes, the shape of the function is very similar.

Notice now that expression (A.1) looks like an autoregressive model. Indeed, this could explain that in some cases linear and ARMA interpolation give similar results since, by serendipity, the signal might have low variability near the gap and the models fitted by \miar\ to local data would have low order. However, in spite of the similarity of expression (A.1) to an AR model the properties of the spectral response function cannot be extrapolated and some serious difficulties appear when trying to identify both methods showing that their spectral response functions are in general very different. 

First of all, \miar\ use only a causal representation of the data, either in forward or backward extrapolation (see PG15 for more details). However, in (A.1) the interpolated value $\hat{X}+\xi$ is obtained using $X(n+1)$, so linear interpolation can be considered a mixed (causal + acausal) representation. Using mixed representations is not an issue as demonstrated in \citet{SCA81}, but we should compare the spectral response function with an AR(1,1) model with coefficients $(1-\xi,1,\xi)$ and this is formally different to what \miar\ does.

Now, for the sake of clarity we could evaluate the spectral response function of the most similar AR representation to (A.1) that is used by \miar\ which is, in fact, an AR(1) model
\begin{equation}
X(n+1) = a \cdot X(n) + E(n)
\end{equation}

where E(n) is an uncorrelated random time series which represents the input of the model, and $a$ is a constant coefficient with $|a|<1$. E(n) can be understood as random pulses feeding the model, this makes the AR model intrinsically different to (A.1) where there is no random component. 

Furthermore, the mathematical form of (A.1) rise some additional restrictions: the same model (A.6) should be fitted for both data segments around the gap, and the weighing between forward and backward extrapolation should be a triangular function. The first one is quite reasonable since it assumes stationarity which is guaranteed in most cases for a short time interval. The weighing function can be freely chosen in \miar\ but the default is a triangular function too, so there is no problem with any of these assumptions. Now, if we follow the same reasoning as before (A.1) is now:
\begin{equation}
\hat{X}(n+\xi) = (1 - \xi) \cdot \hat{X}^f(n+\xi) + \xi \cdot \hat{X}^b(n+\xi)
\end{equation}

where $\hat{X}^f(n+\xi)$ and $\hat{X}^b(n+\xi)$ are the forward and backward extrapolations at $n+\xi$. These can be obtained from (A.6) and with some simple calculations it is easy to see that
\begin{eqnarray}
\hat{X}(k/L) & = & a h(k) X(k-1/L) + (1/a) h(k-L) X(k+1/L) + \nonumber \\  
             & + & a h(k) E(k-1/L) - (1/a) h(k-L) E(k+1/L)
\end{eqnarray}

The differences between (A.3) and (A.8) are clear. In addition to the random input $E(1/L)$ and the coefficients $a$ and $1/a$, note that for each partition $k/L$, the function X is evaluated in $k-1/L$ and $k+1/L$ and not in 1 and 0. That is, \miar\ fits causal models to the data bracketing the gap but each interpolated value depends on the previous and consecutive datapoints and not on fixed values $X(1), X(0)$.

\begin{figure}[ht]
   \centering
   \includegraphics[width=8.5cm]{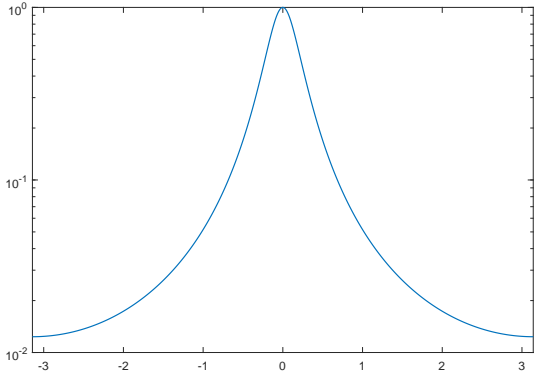}
   \caption{Spectral response function in log scale associated to an AR(1) model with a = 0.8.}
   \label{fig:sw_ar}%
\end{figure}

It is not so simple to calculate the spectral response function of AR(1) as it is for linear interpolation but with a little more effort it can be shown \citep[see, e.g., p. 261 of ][]{SAR87} that it is:
\begin{equation}
 P(\nu) = \frac{\sigma^2}{1 - 2a \cdot \cos (\nu) + a^2}
\end{equation}

where $\sigma^2$ is the variance. In Fig.~\ref{fig:sw_ar} we plot the spectral response function of a AR(1) model with $a=0.8$ to illustrate the differences between the frequency responses of (A.3) and (A.8).

We have studied in this appendix the spectral response function of an interpolation with a linear fitting and with an AR(1) model. Only a single gap have been considered for this study and the full spectral response depends, of course, on the gap distribution. Then, the full spectral response will be in general more complex but we can have a notion of the effect of the interpolation on spectral analysis with the calculations presented here.

%
\section[]{Gap-filling features}
\begin{figure}[ht]
    \centering
   \includegraphics[width=8cm]{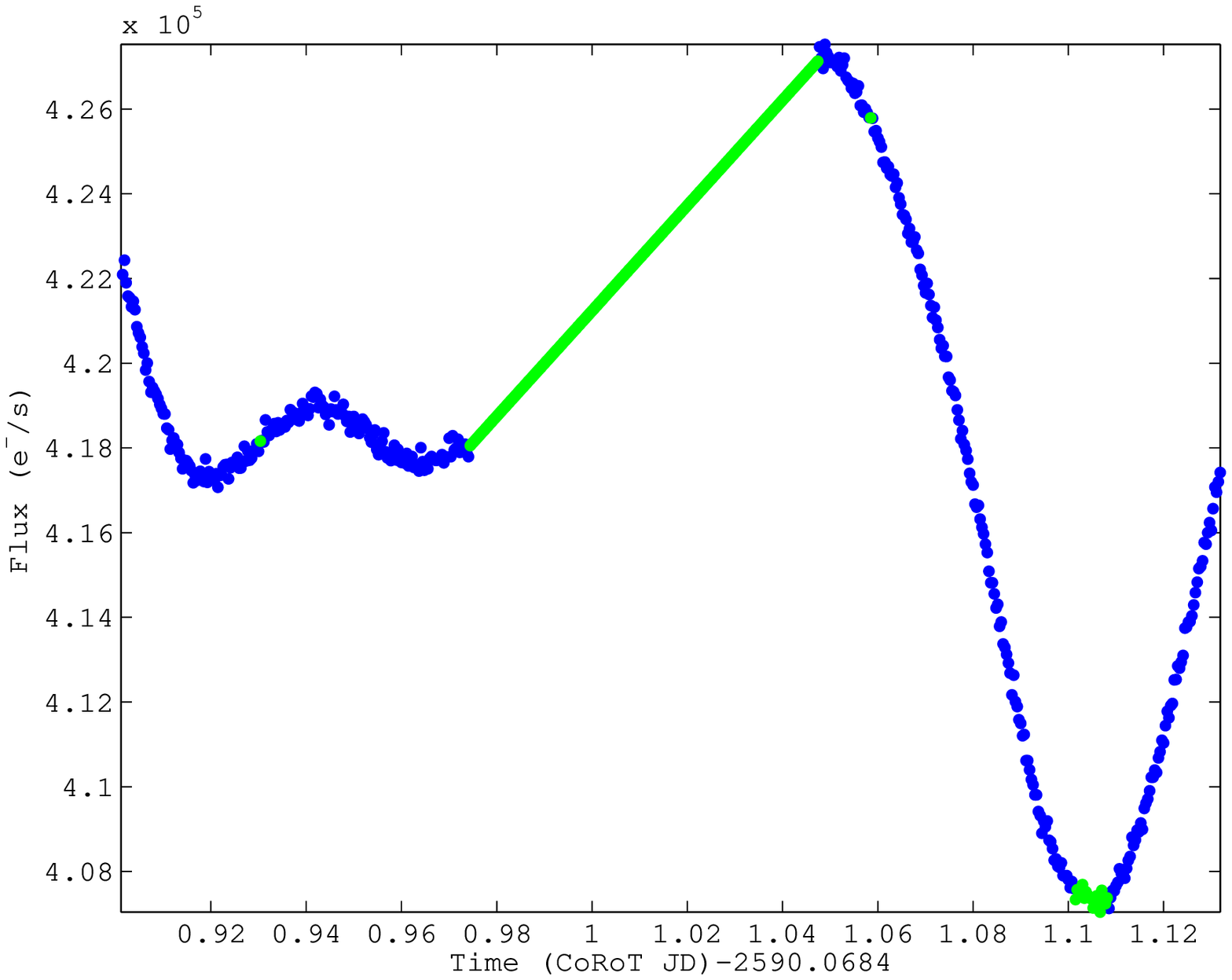}
   \includegraphics[width=8cm]{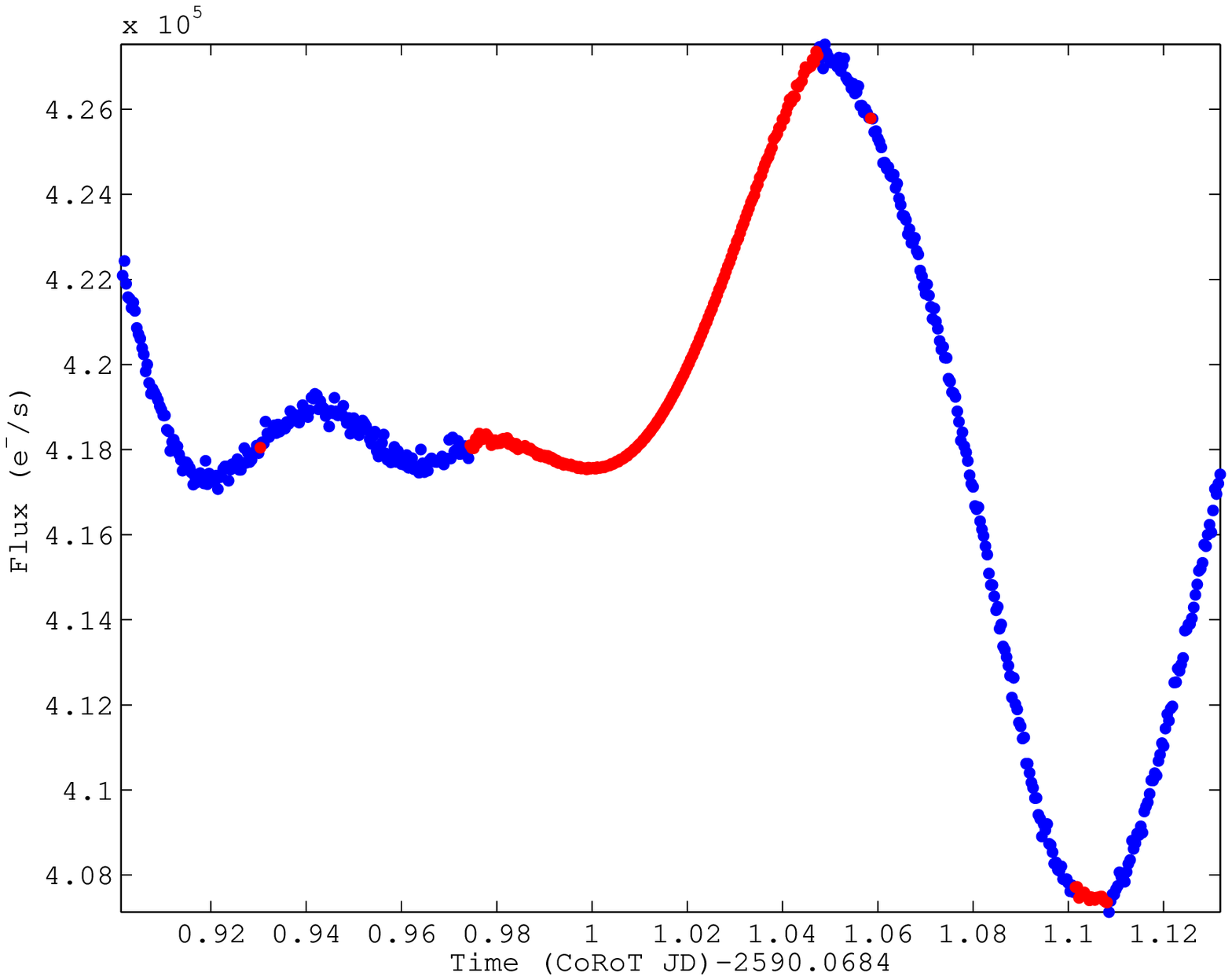}  
   \caption{Illustration of a gap in HD 50844 that has been interpolated linearly (upper panel, in green), and with \miar\ (lower panel, in red).}
   \label{fig:lc1}%
\end{figure}
\begin{figure}
    \centering
   \includegraphics[width=8cm]{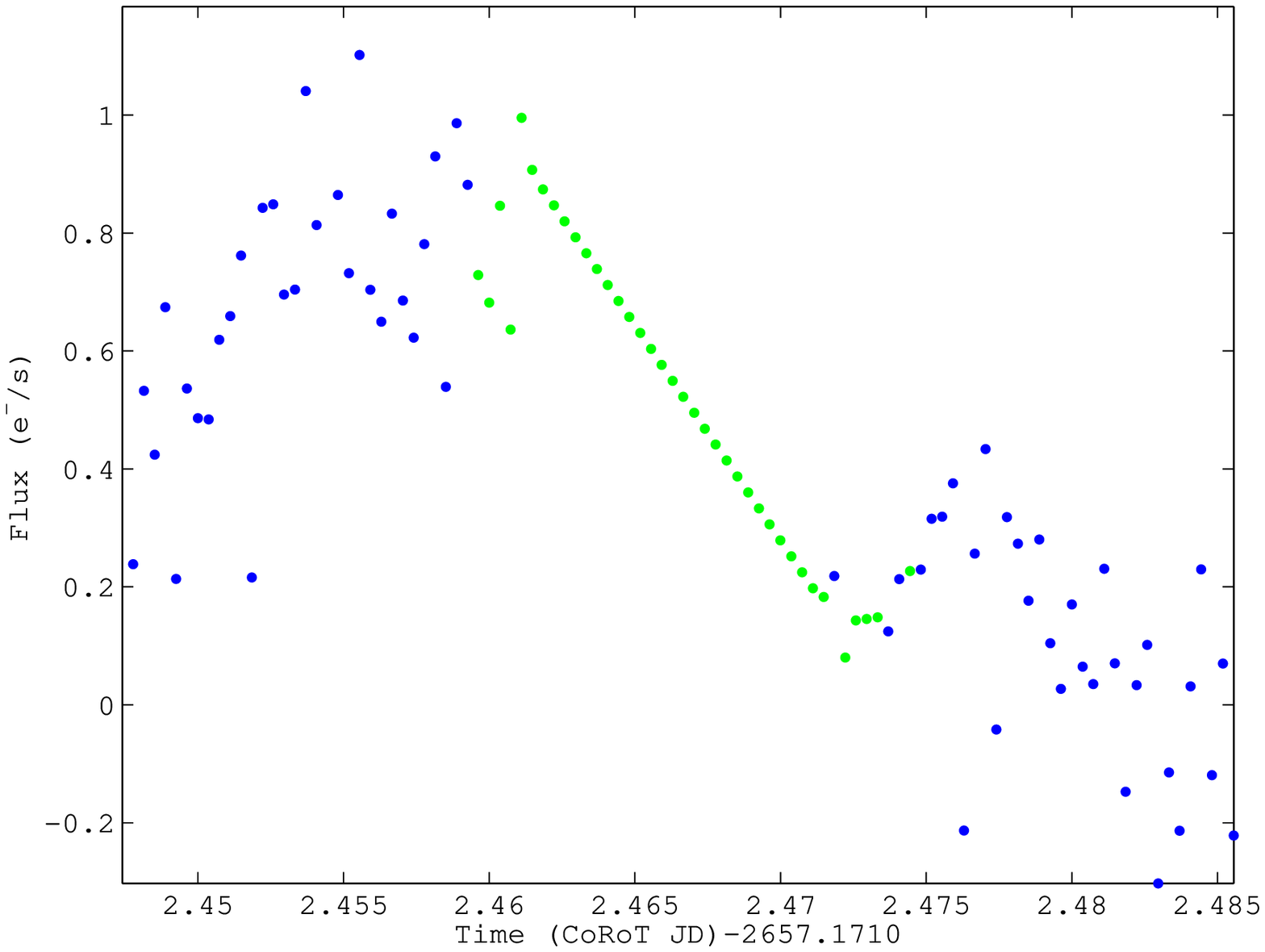}
   \includegraphics[width=8cm]{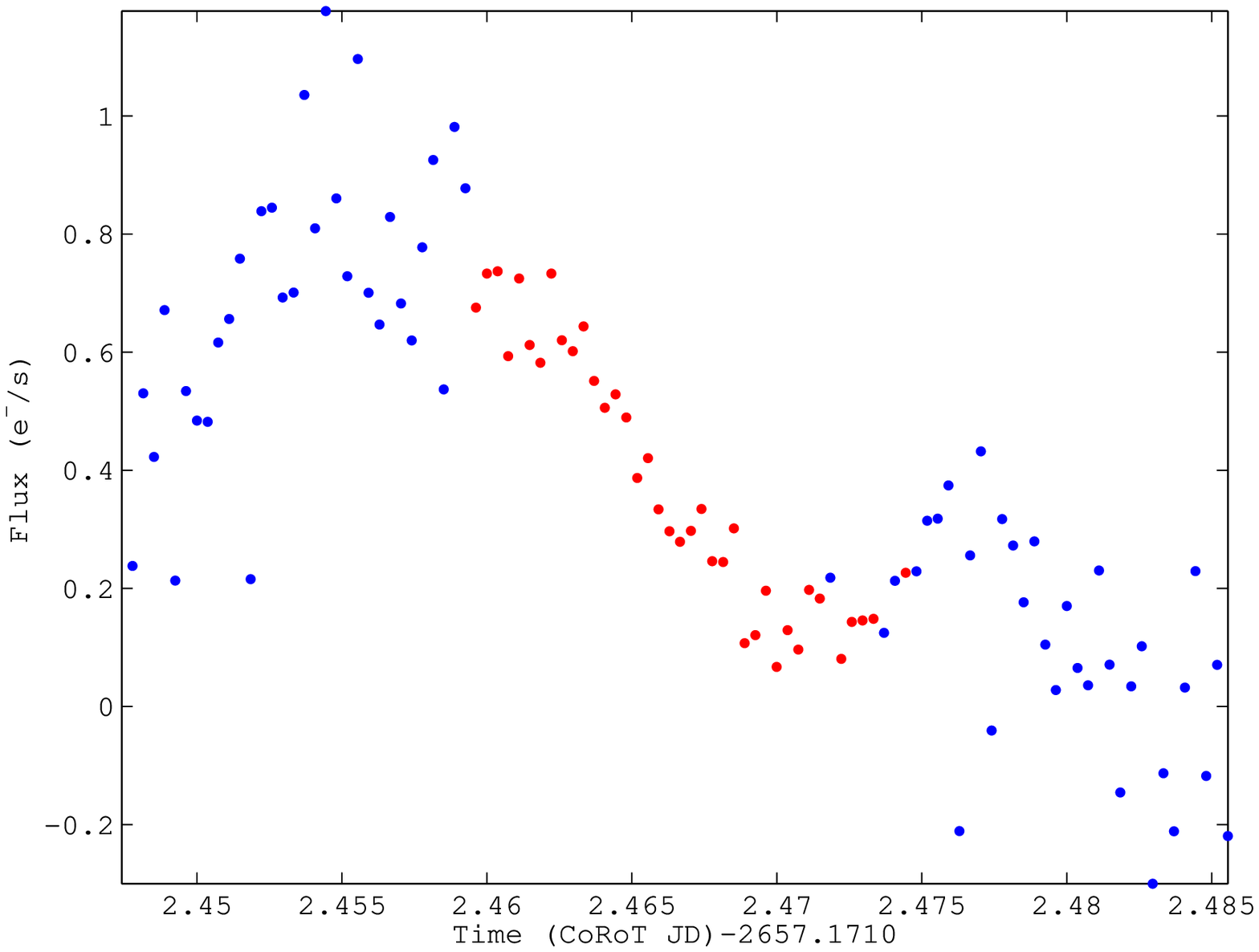}  
   \caption{Illustration of a gap in HD 174936 that has been interpolated linearly (upper panel, in green), and with \miar\ (lower panel, in red).}
   \label{fig:lc2}%
\end{figure}
\begin{figure}
   \centering
   \includegraphics[width=8cm]{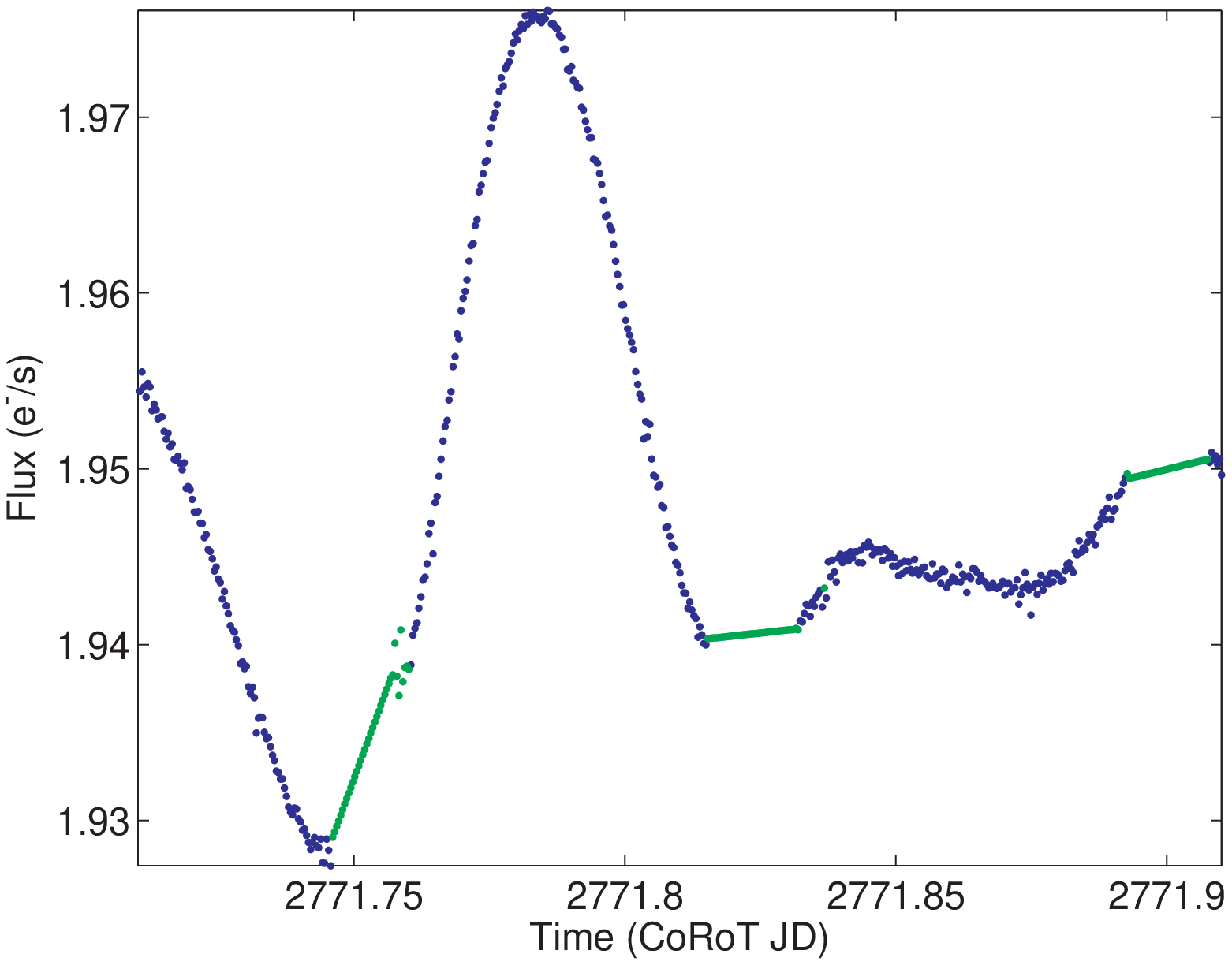}
   \includegraphics[width=8cm]{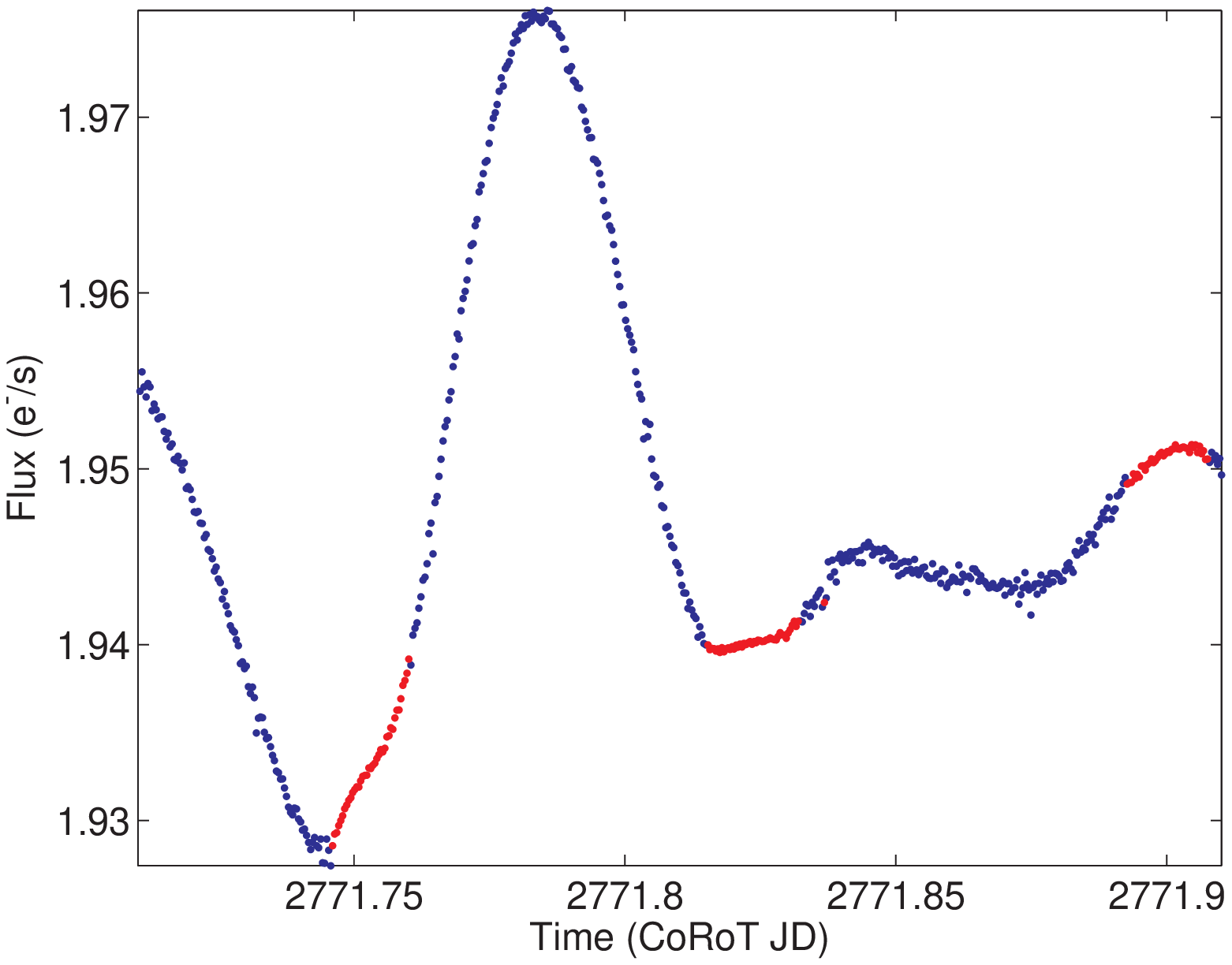}  
   \caption{Illustration of a gap in HD 181555 that has been interpolated linearly (upper panel, in green), and with \miar\ (lower panel, in red).}
   \label{fig:lc3}
\end{figure}
\begin{figure}
    \centering
   \includegraphics[width=8cm]{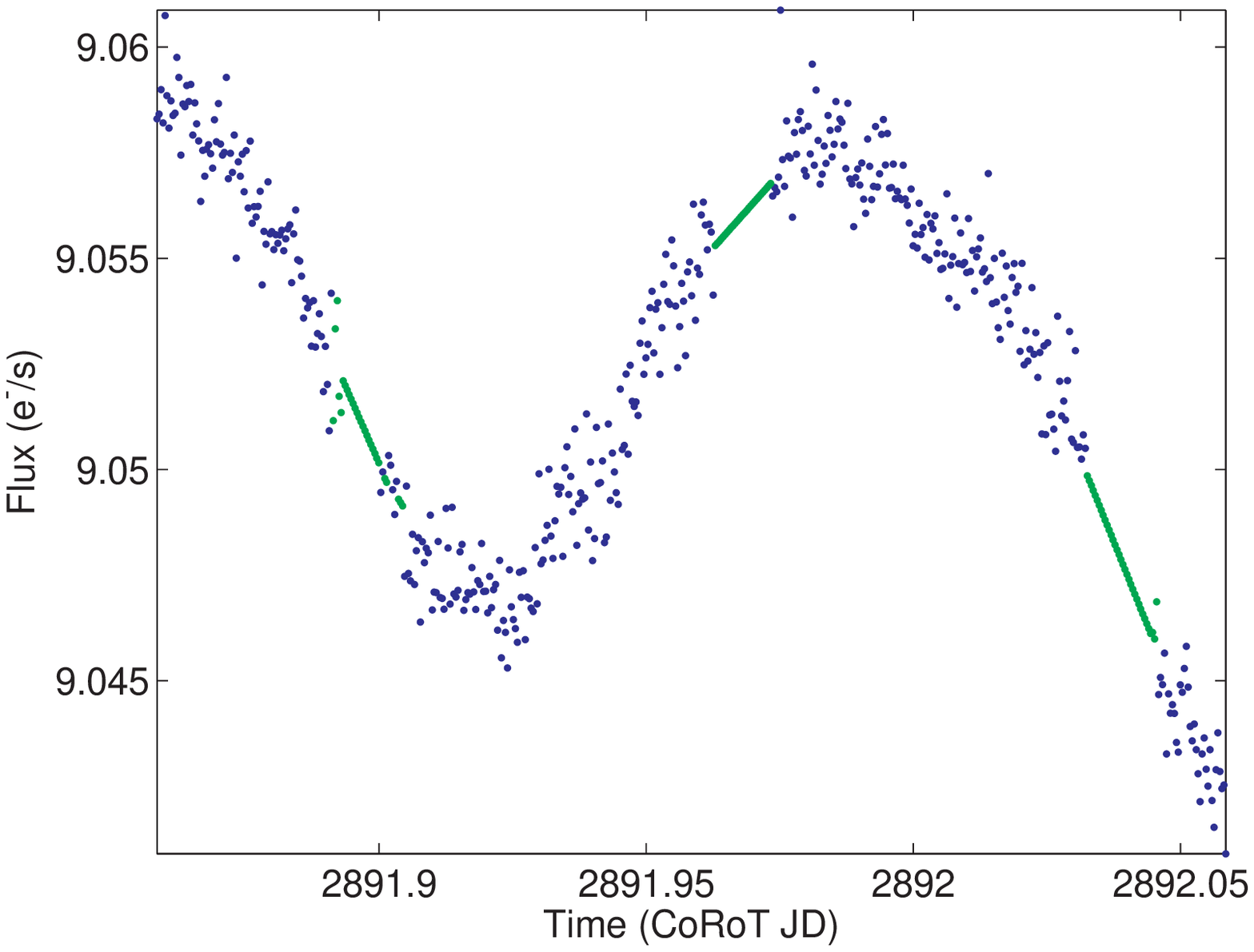}
   \includegraphics[width=8cm]{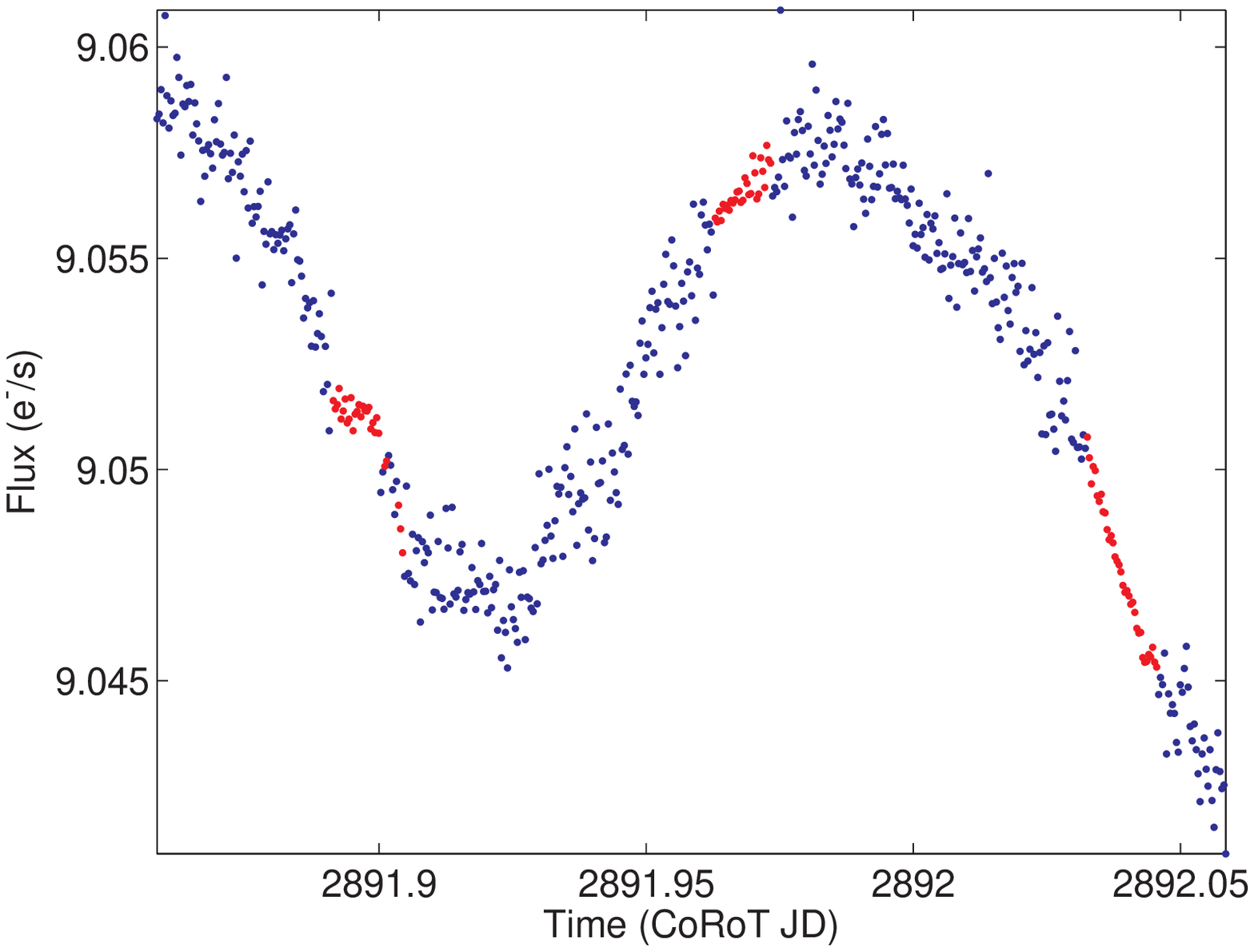}  
   \caption{Illustration of a gap in HD 49434 that has been interpolated linearly (upper panel, in green), and with \miar\ (lower panel, in red).}
   \label{fig:lc4}%
\end{figure}
\begin{figure}
    \centering
   \includegraphics[width=8cm]{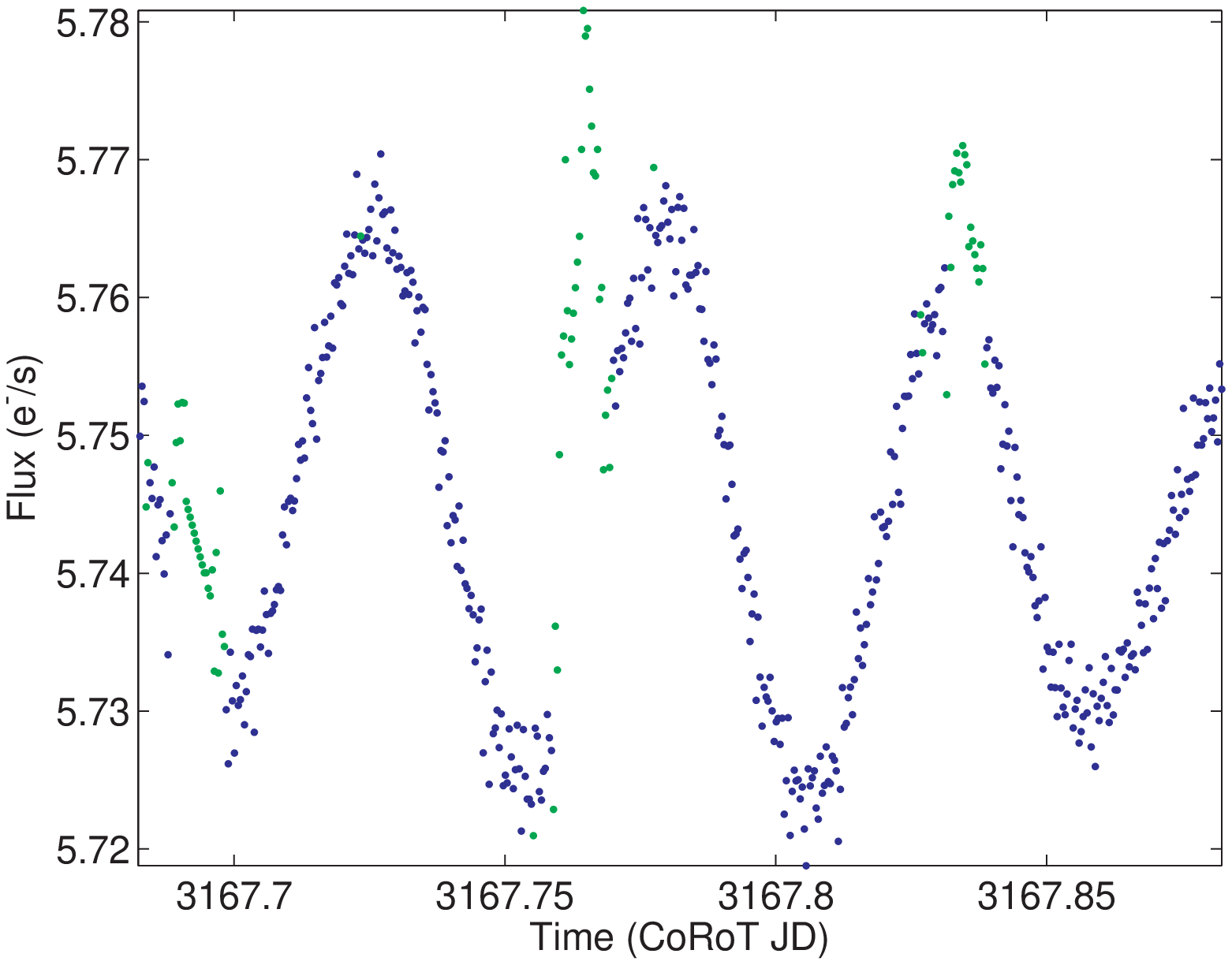}
   \includegraphics[width=8cm]{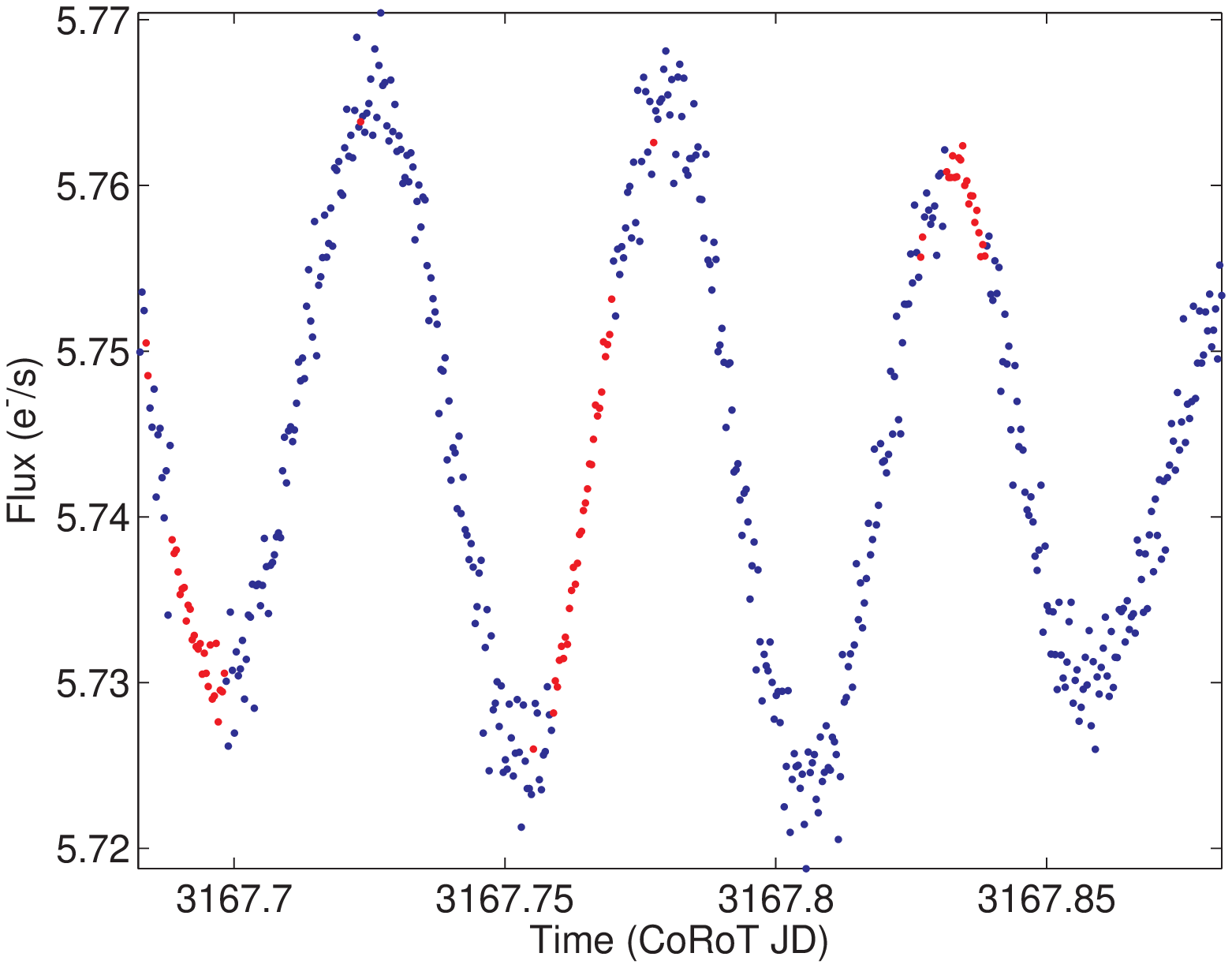}  
   \caption{Illustration of a gap in HD 172189 that has been interpolated linearly (upper panel, in green), and with \miar\ (lower panel, in red).}
   \label{fig:lc5}%
\end{figure}
\begin{figure}
   \centering
   \includegraphics[width=8cm]{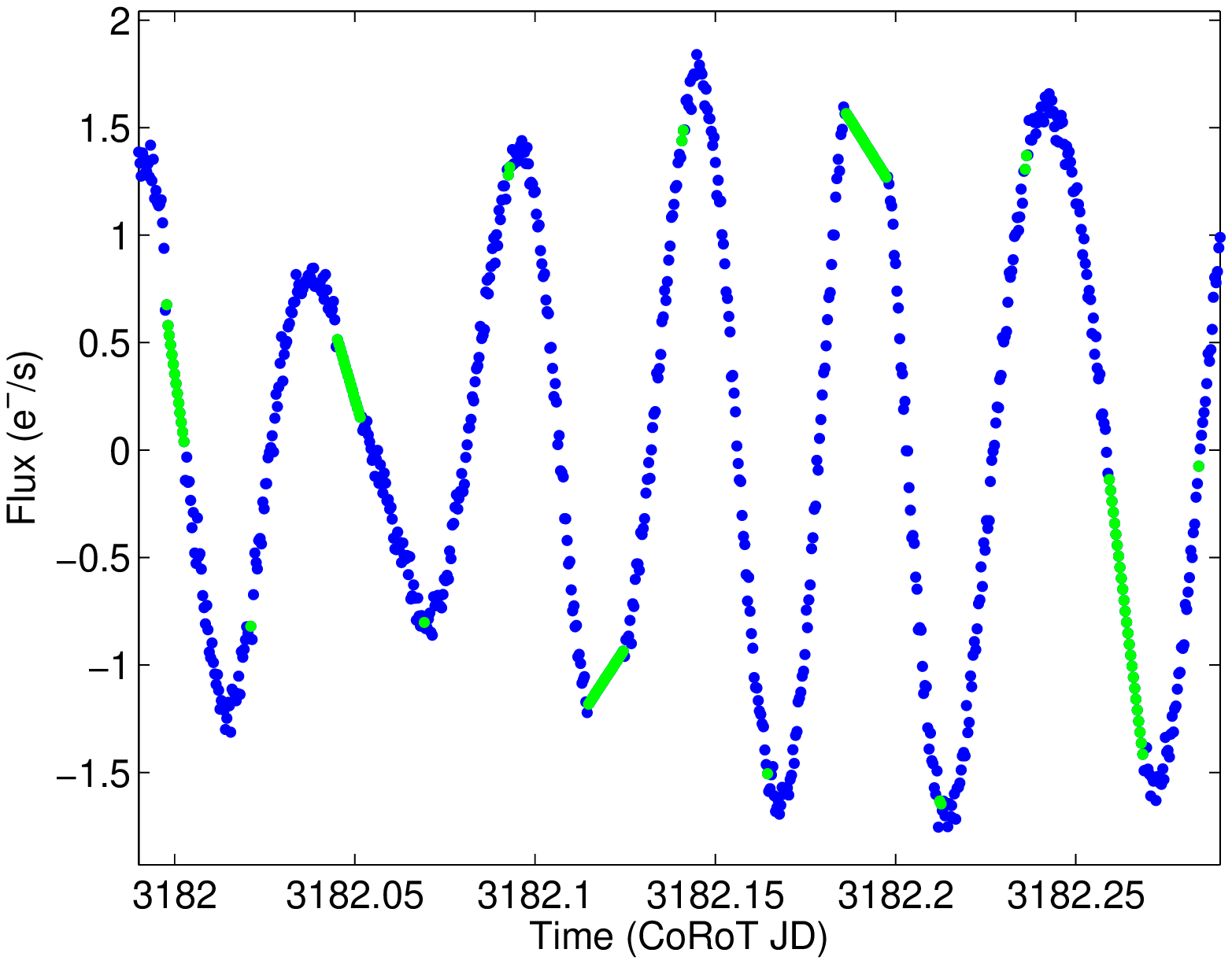}
   \includegraphics[width=8cm]{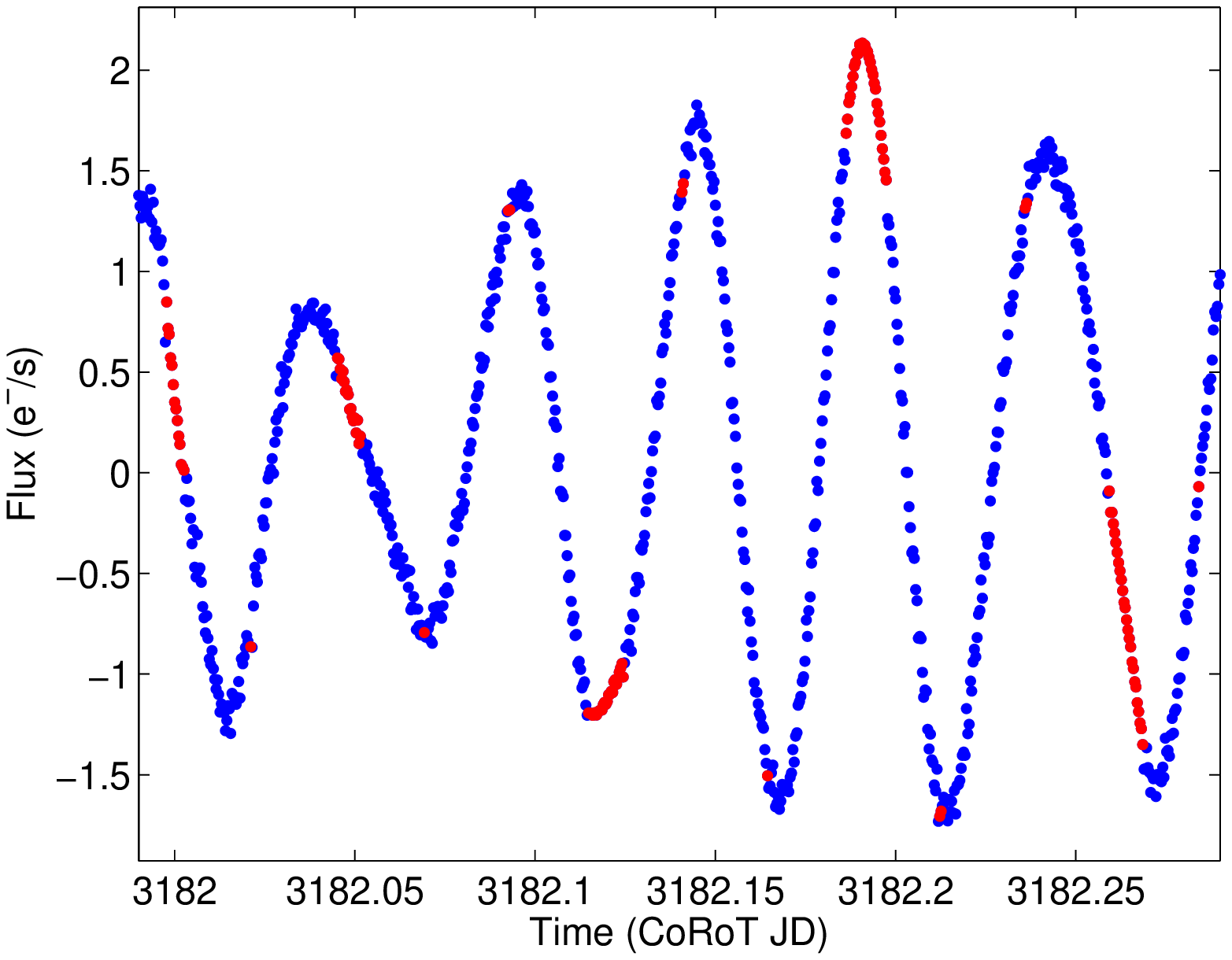}  
   \caption{Illustration of gaps in HD 174532 that has been interpolated linearly (upper panel, in green, and with \miar\ (lower panel, in red).}
   \label{fig:lc6}%
\end{figure}
\begin{figure}
   \centering
   \includegraphics[width=8cm]{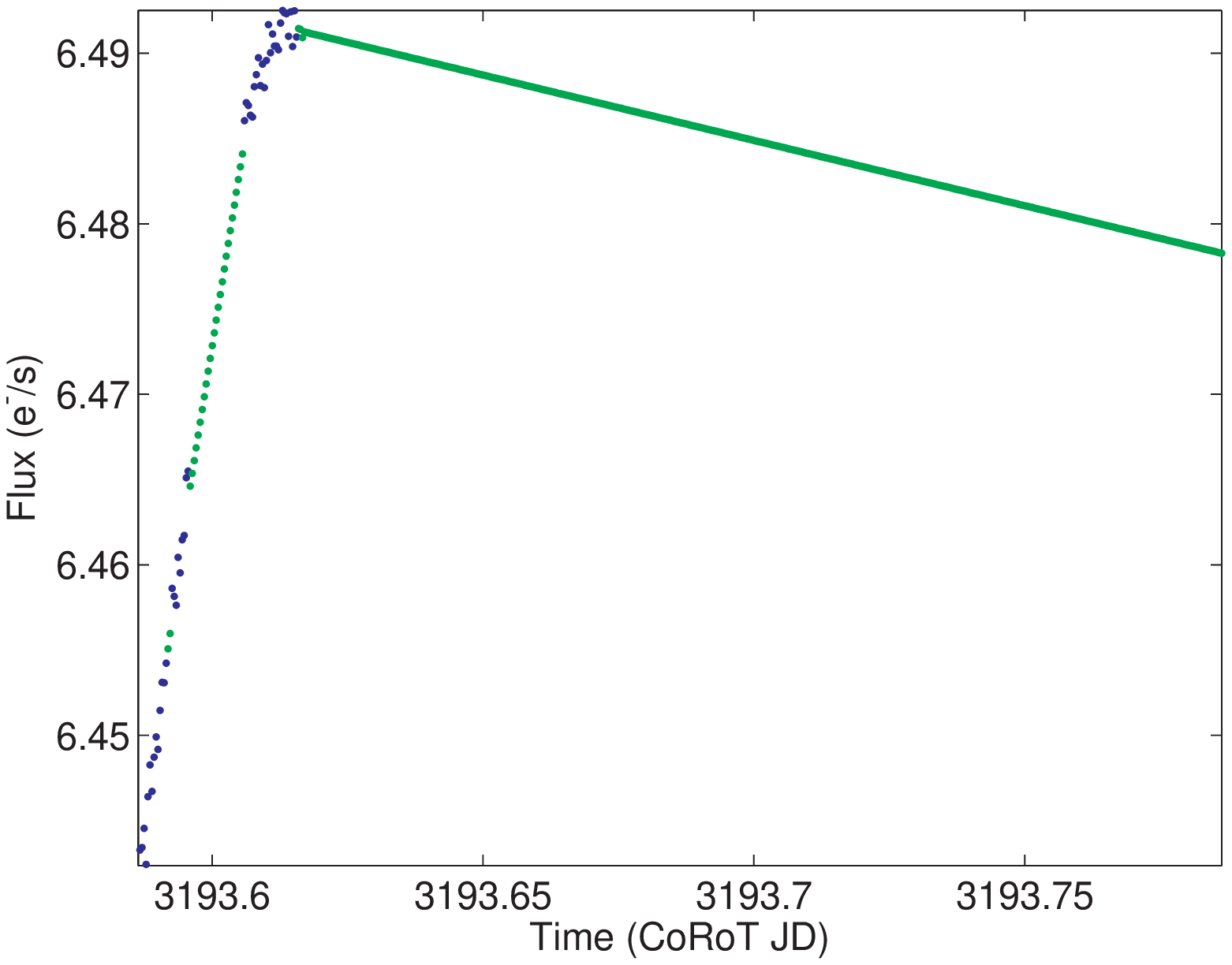}
   \includegraphics[width=8cm]{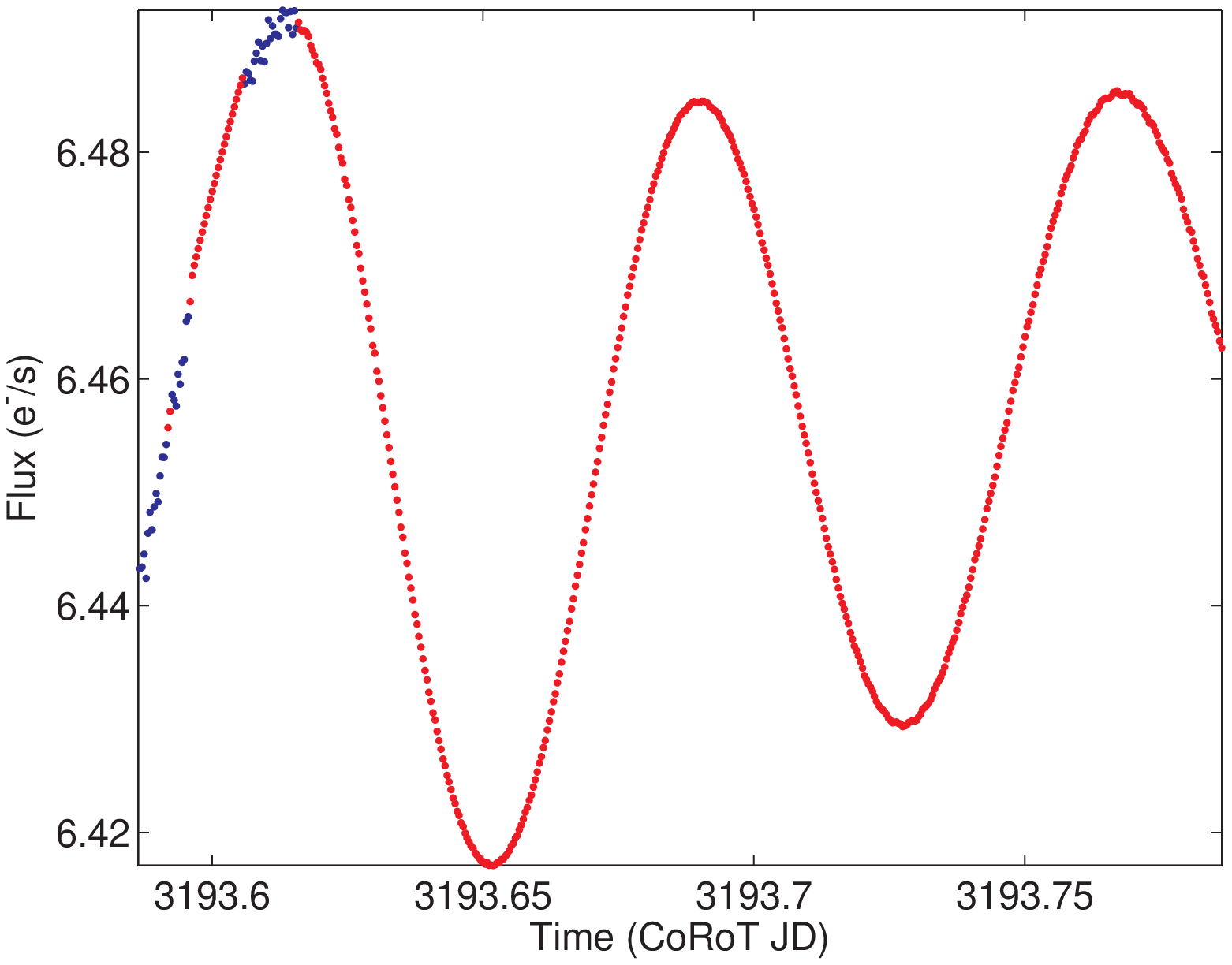}  
   \caption{Illustration of a gap in HD 174589 that has been interpolated linearly (upper panel, in green), and with \miar\ (lower panel, in red).}
   \label{fig:lc7}%
\end{figure}
\begin{figure}
   \centering
   \includegraphics[width=8cm]{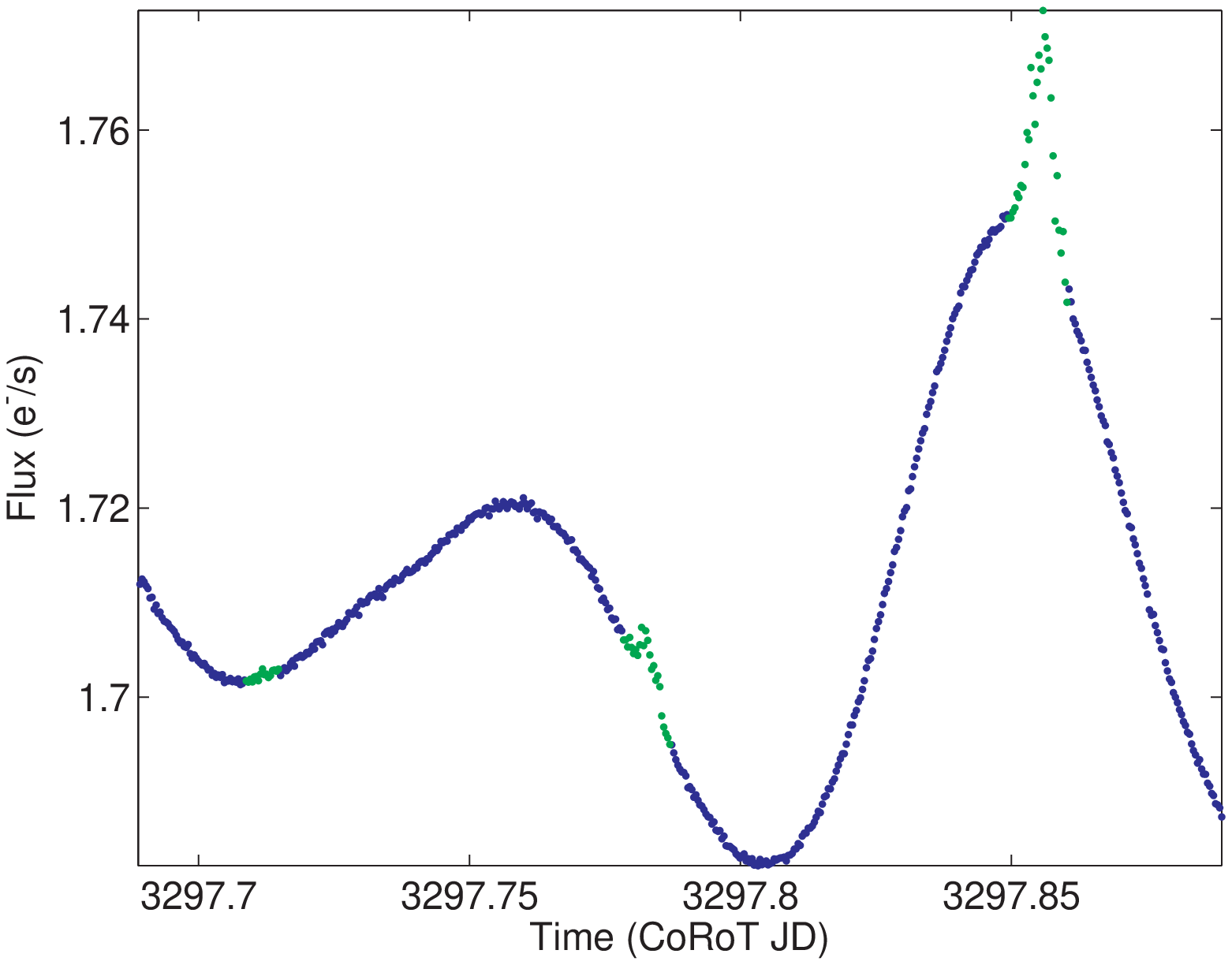}
   \includegraphics[width=8cm]{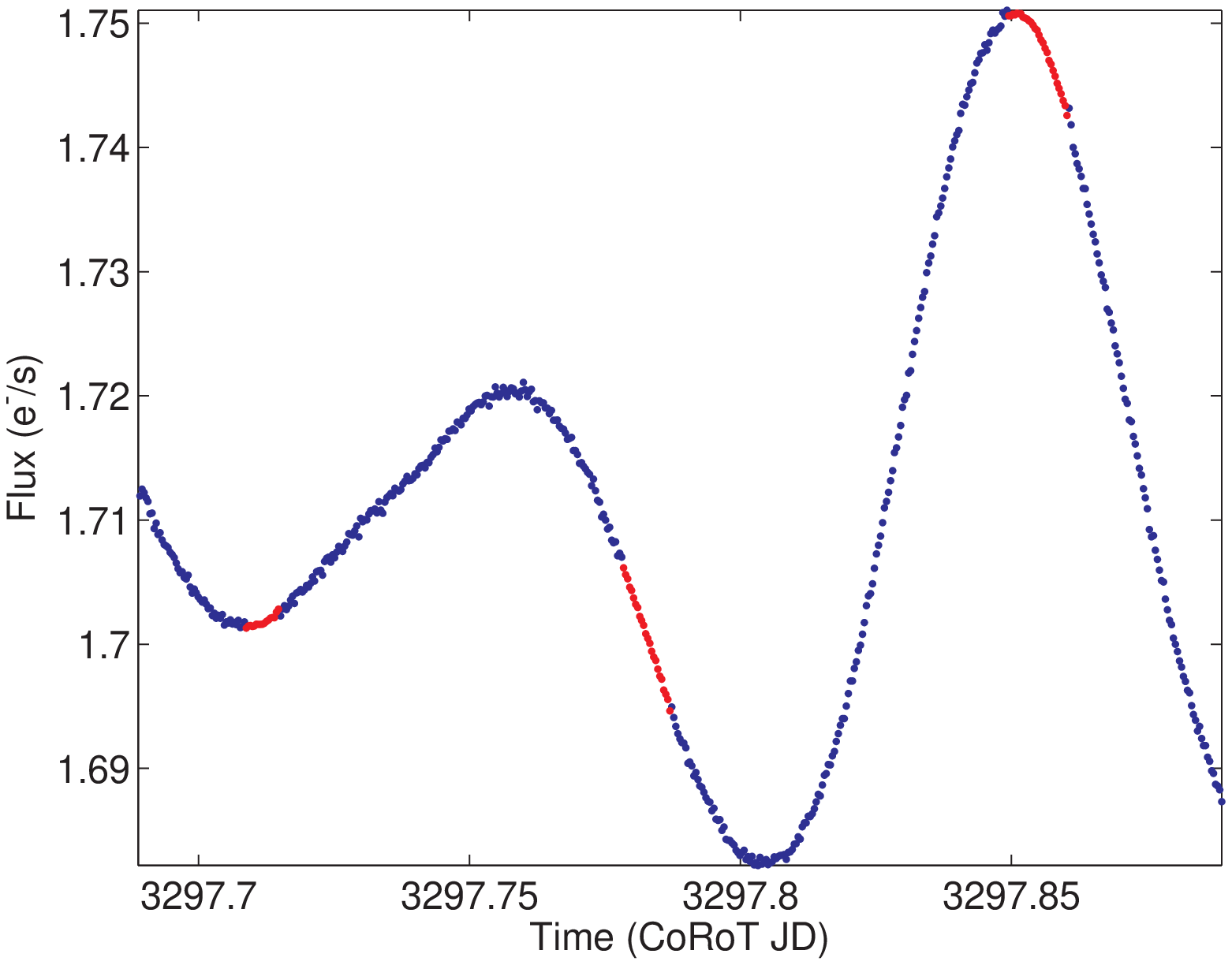}  
   \caption{Illustration of a gap in HD 51722 that has been interpolated linearly (upper panel, in green), and with \miar\ (lower panel, in red).}
   \label{fig:lc8}%
\end{figure}
\begin{figure}
   \centering
   \includegraphics[width=8cm]{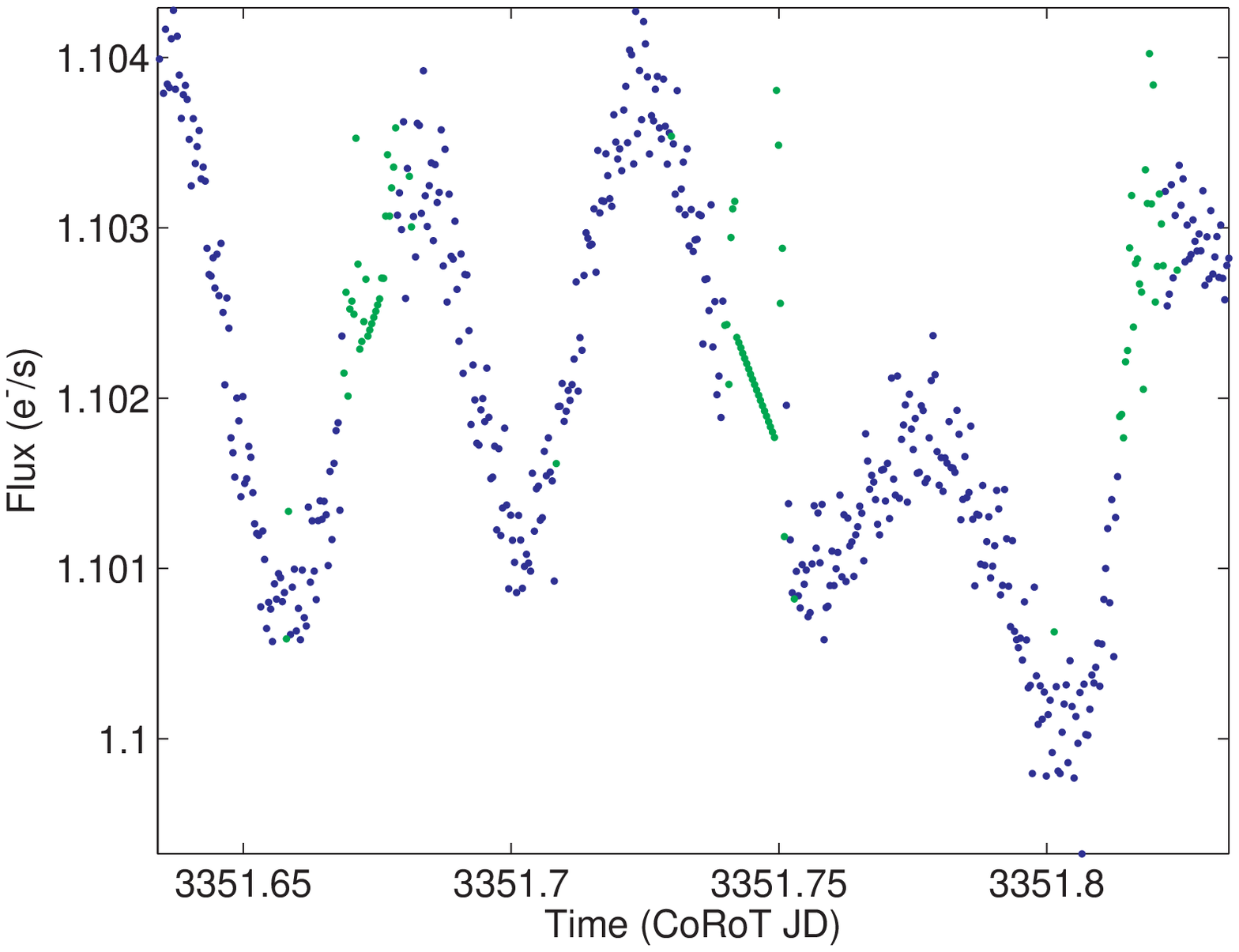}
   \includegraphics[width=8cm]{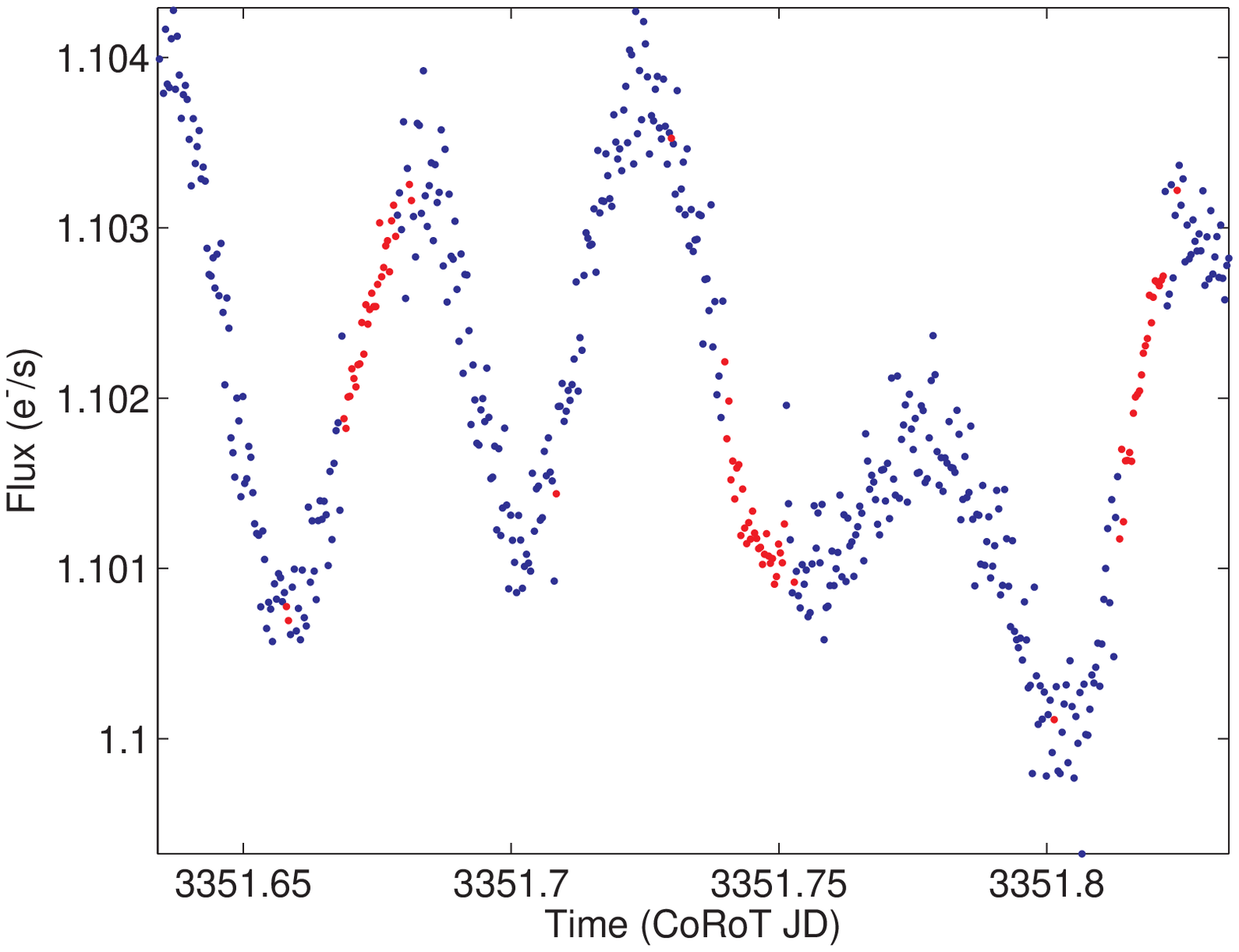}  
   \caption{Illustration of a gap in HD 51359 that has been interpolated linearly (upper panel, in green), and with \miar\ (lower panel, in red).}
   \label{fig:lc9}%
\end{figure}
\begin{figure}
   \centering
   \includegraphics[width=8cm]{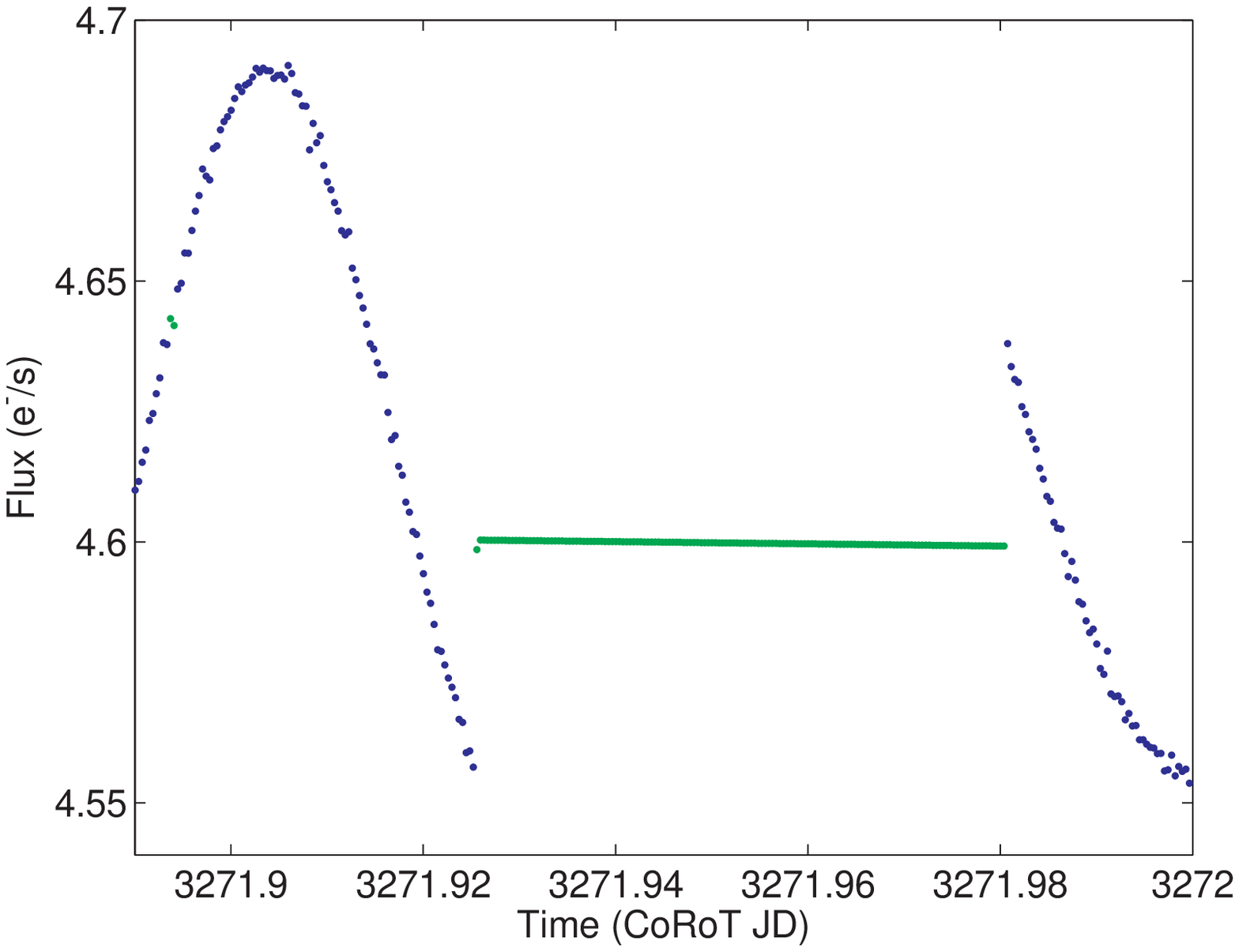}
   \includegraphics[width=8cm]{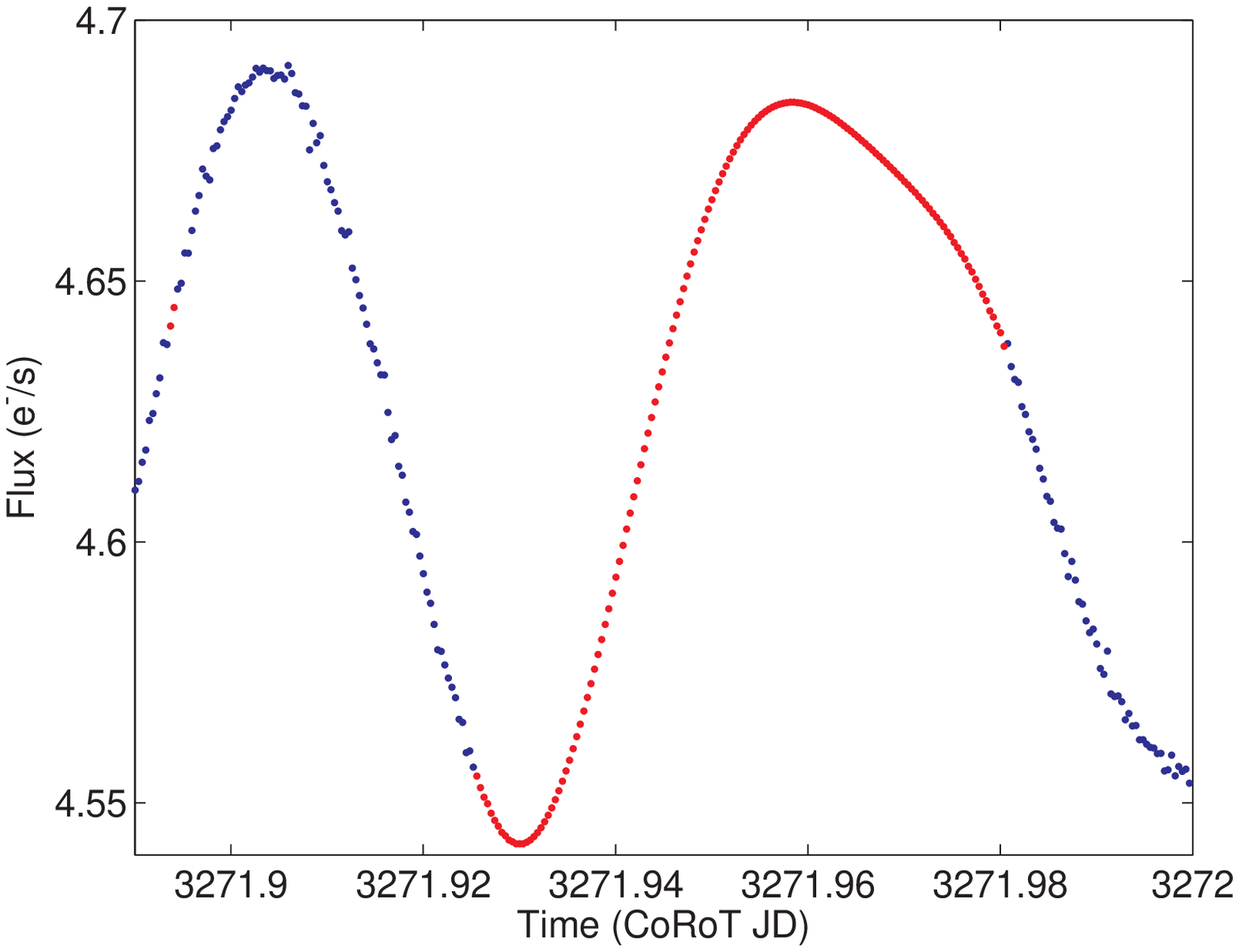}  
   \caption{Illustration of a gap in HD 50870 that has been interpolated linearly (upper panel, in green), and with \miar\ (lower panel, in red).}
   \label{fig:lc10}%
\end{figure}
\begin{figure}
    \centering
   \includegraphics[width=8cm]{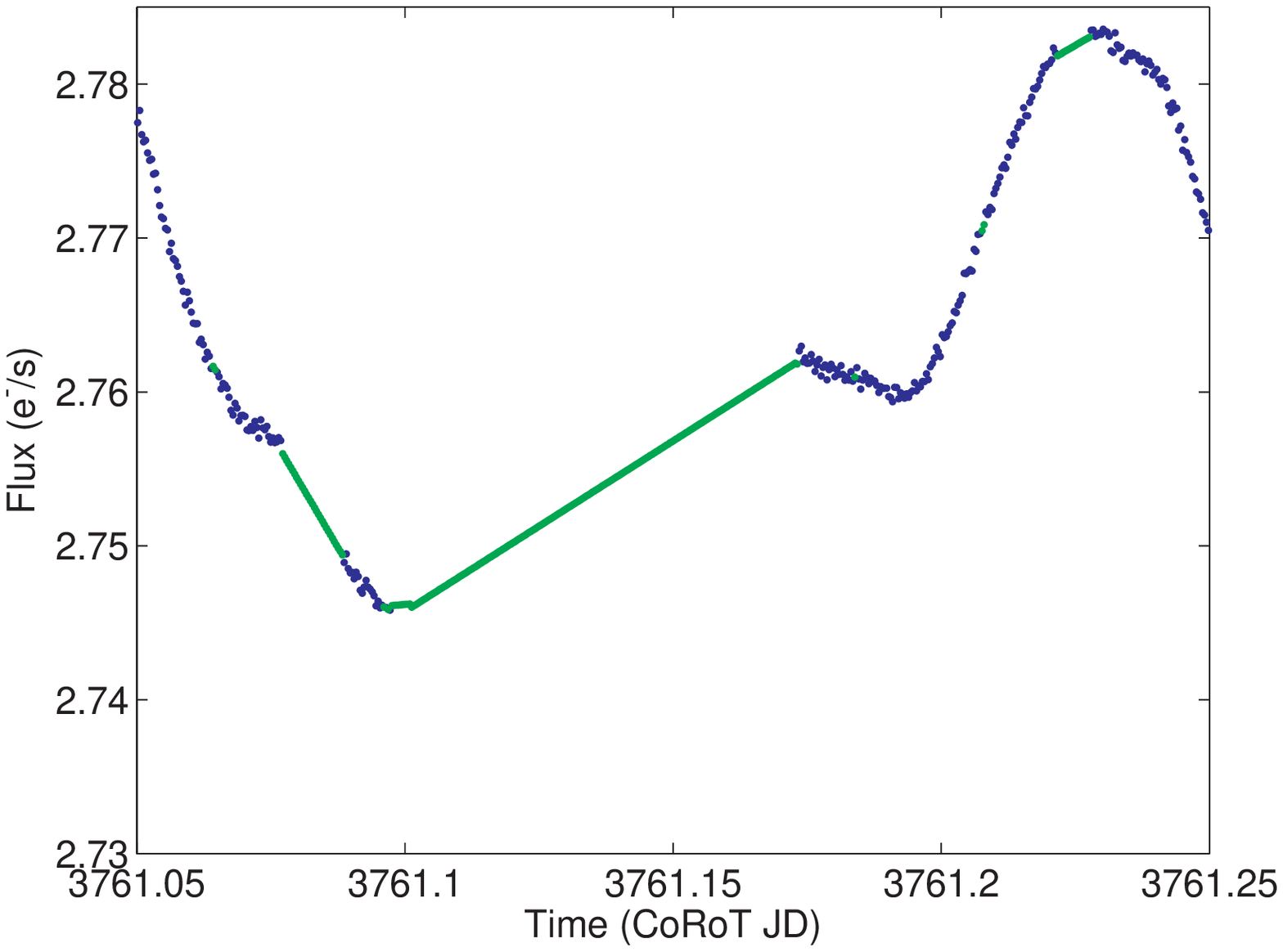}
   \includegraphics[width=8cm]{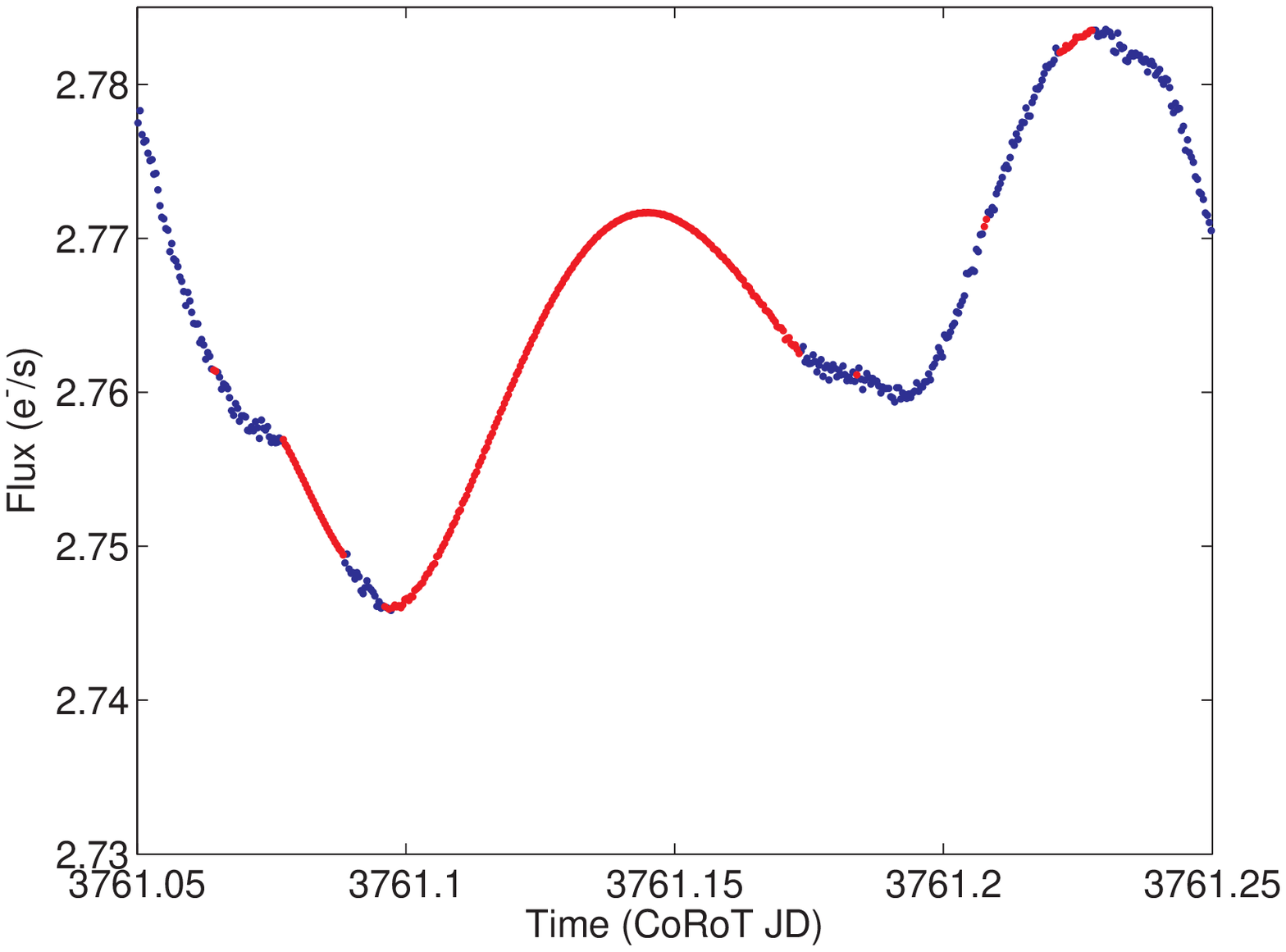}  
   \caption{Illustration of a gap in HD 170699 that has been interpolated linearly (upper panel, in green), and with \miar\ (lower panel, in red).}
   \label{fig:lc11}%
\end{figure}
\begin{figure}
   \centering
   \includegraphics[width=8cm]{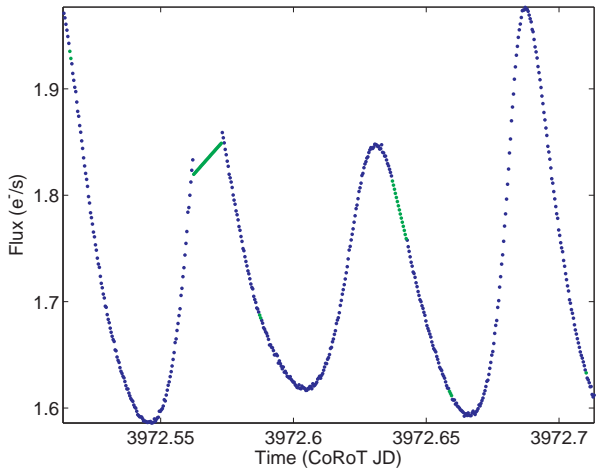}
   \includegraphics[width=8cm]{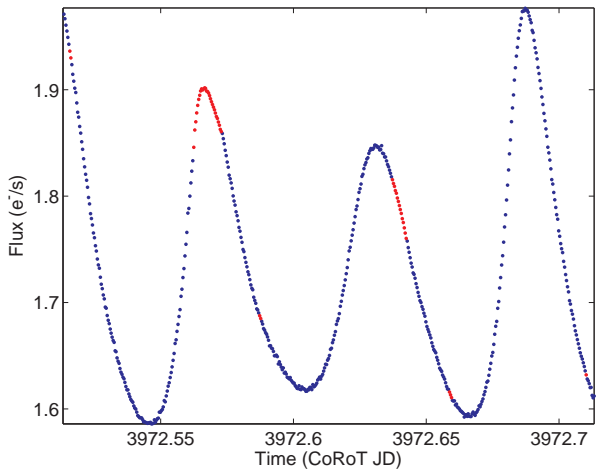}  
   \caption{Illustration of a gap in GSC00244-03031 that has been interpolated linearly (upper panel, in green), and with \miar\ (lower panel, in red).}
   \label{fig:lc12}%
\end{figure}
\begin{figure}
    \centering
   \includegraphics[width=8cm]{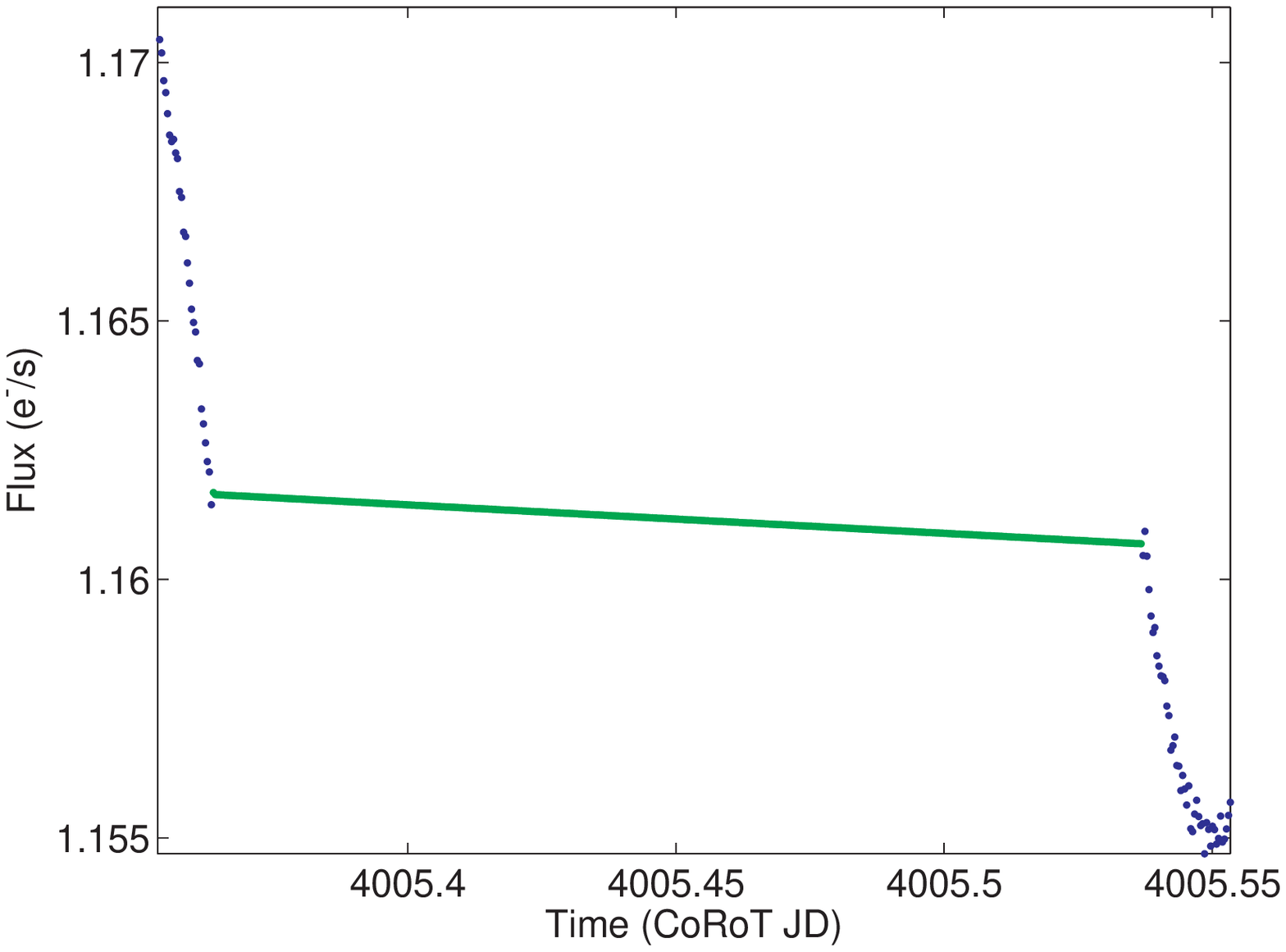}
   \includegraphics[width=8cm]{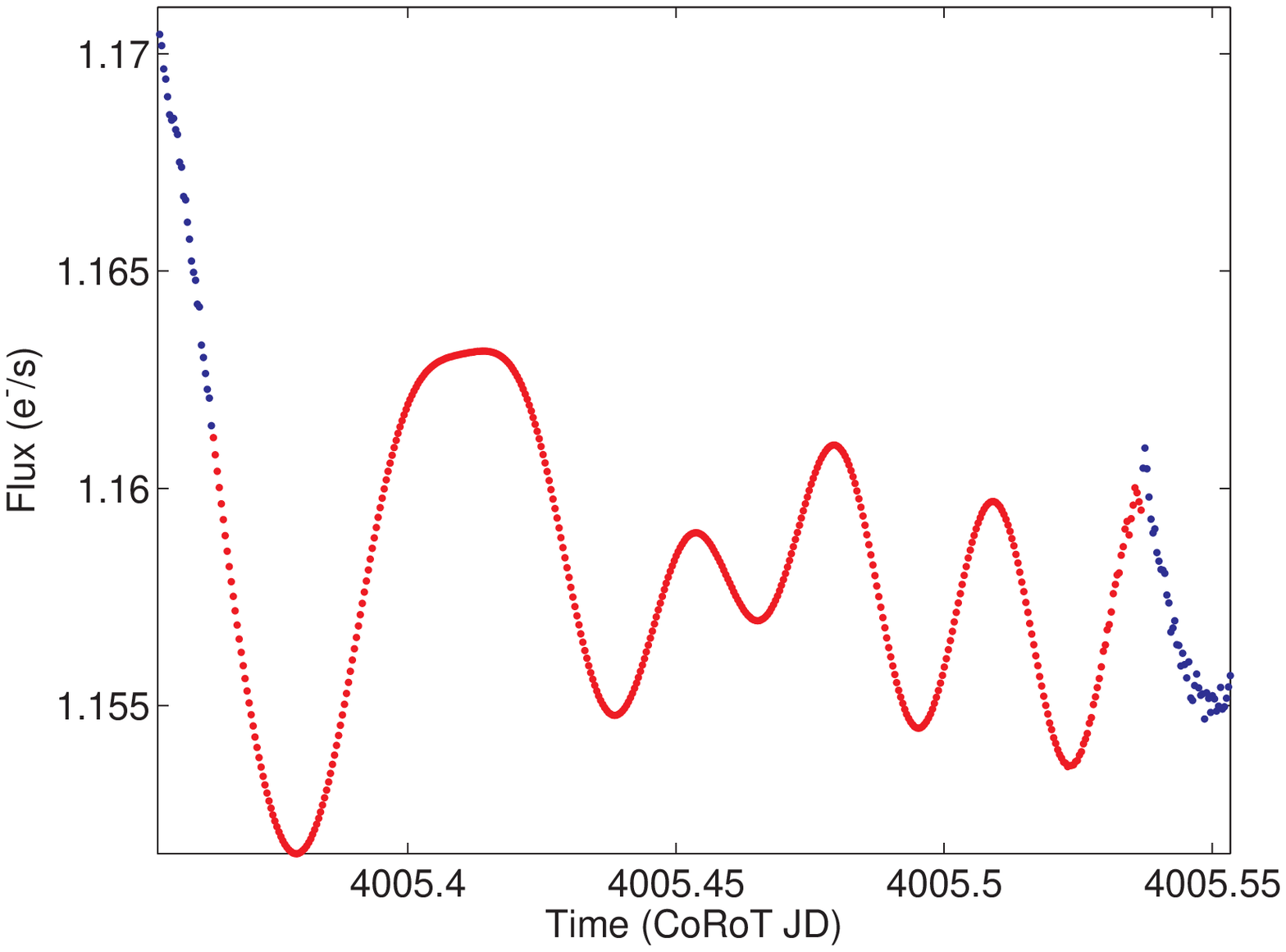}  
   \caption{Illustration of gaps in HD 41641 that has been interpolated linearly (upper panel, in green, and with \miar\ (lower panel, in red).}
   \label{fig:lc13}%
\end{figure}
\begin{figure}
   \centering
   \includegraphics[width=8cm]{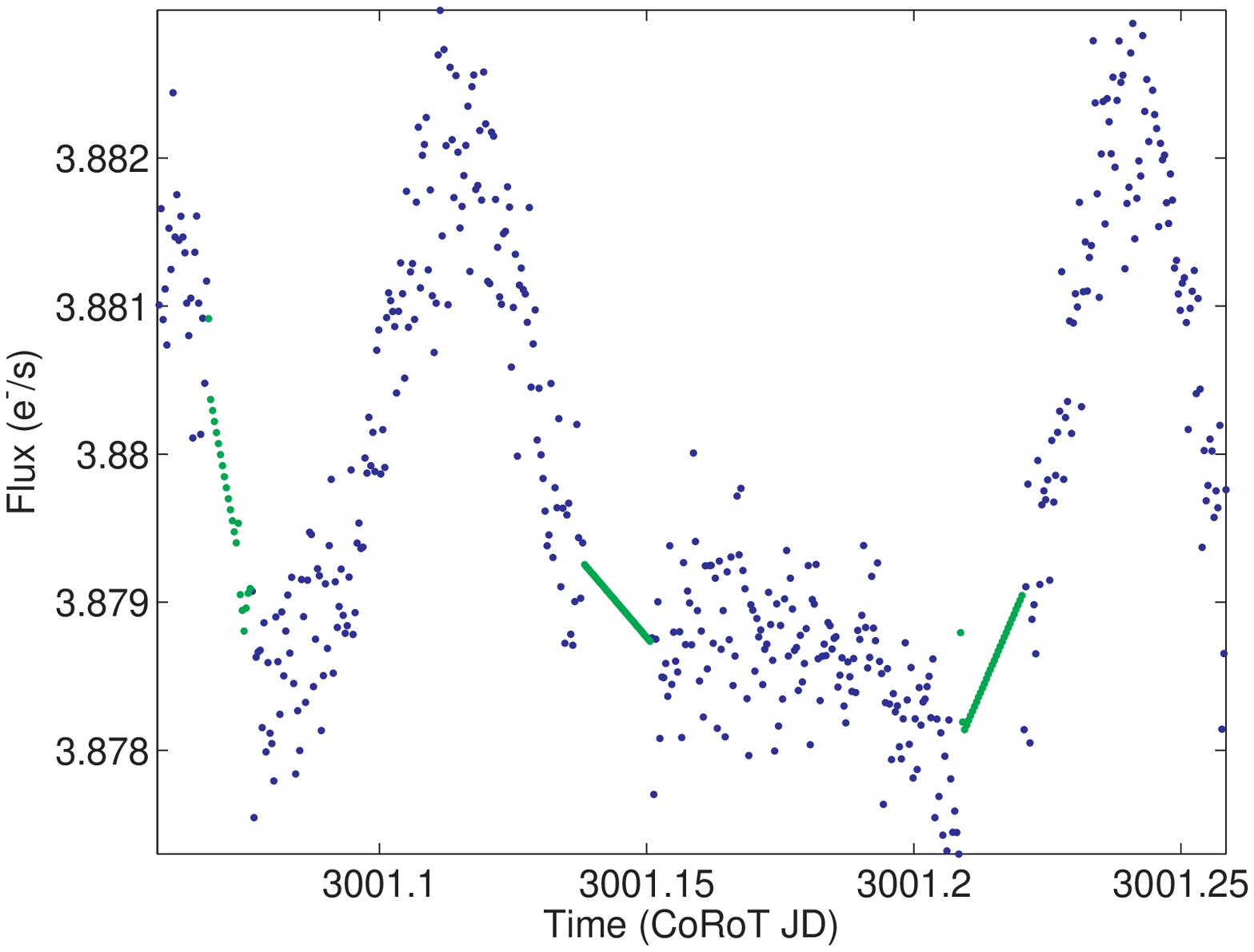}
   \includegraphics[width=8cm]{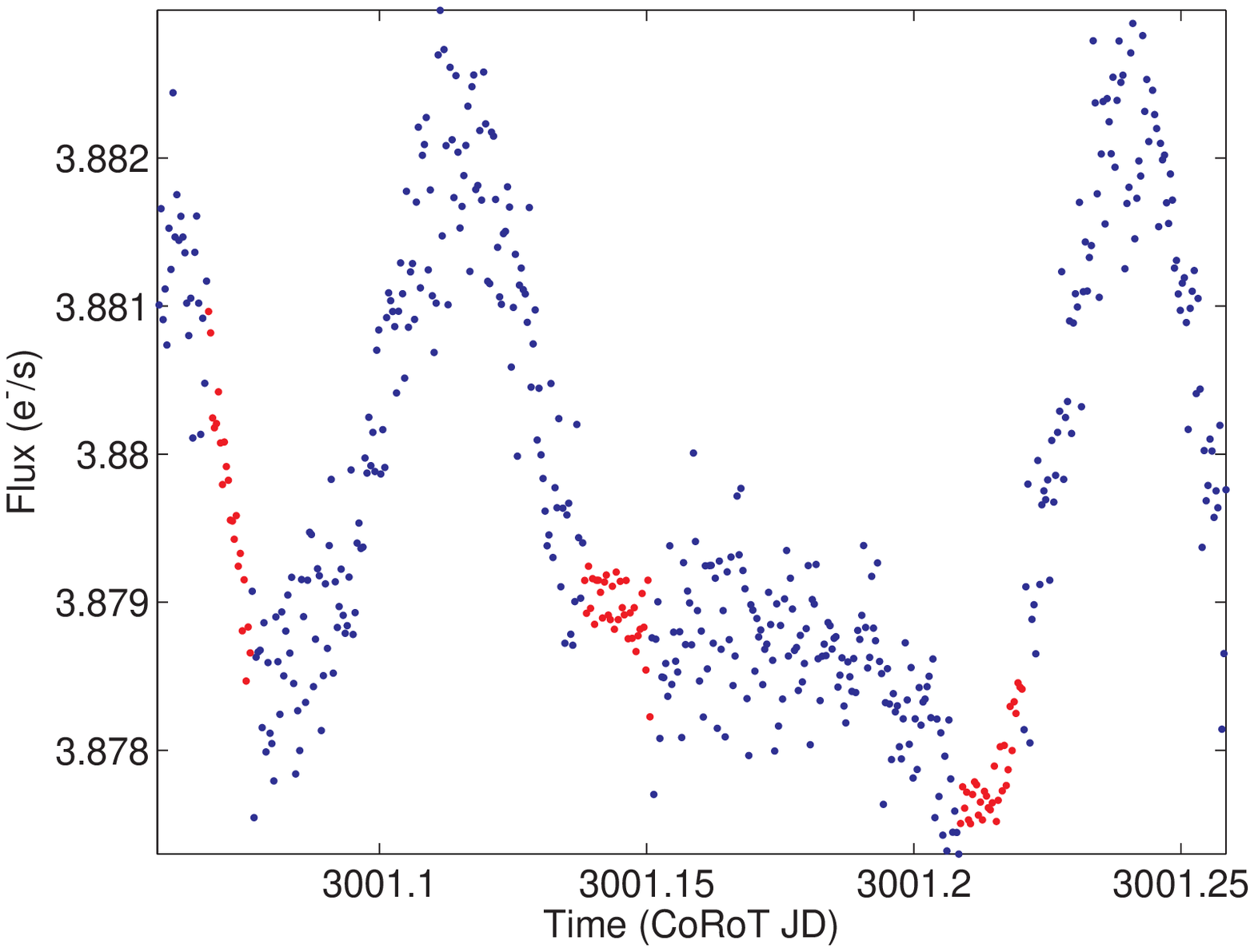}  
   \caption{Illustration of gaps in HD 48784 that has been interpolated linearly (upper panel, in green, and with \miar\ (lower panel, in red).}
   \label{fig:lc14}%
\end{figure}

\clearpage

\section[]{Power spectra}
\begin{figure}[ht]
   \centering
   \includegraphics[width=9cm]{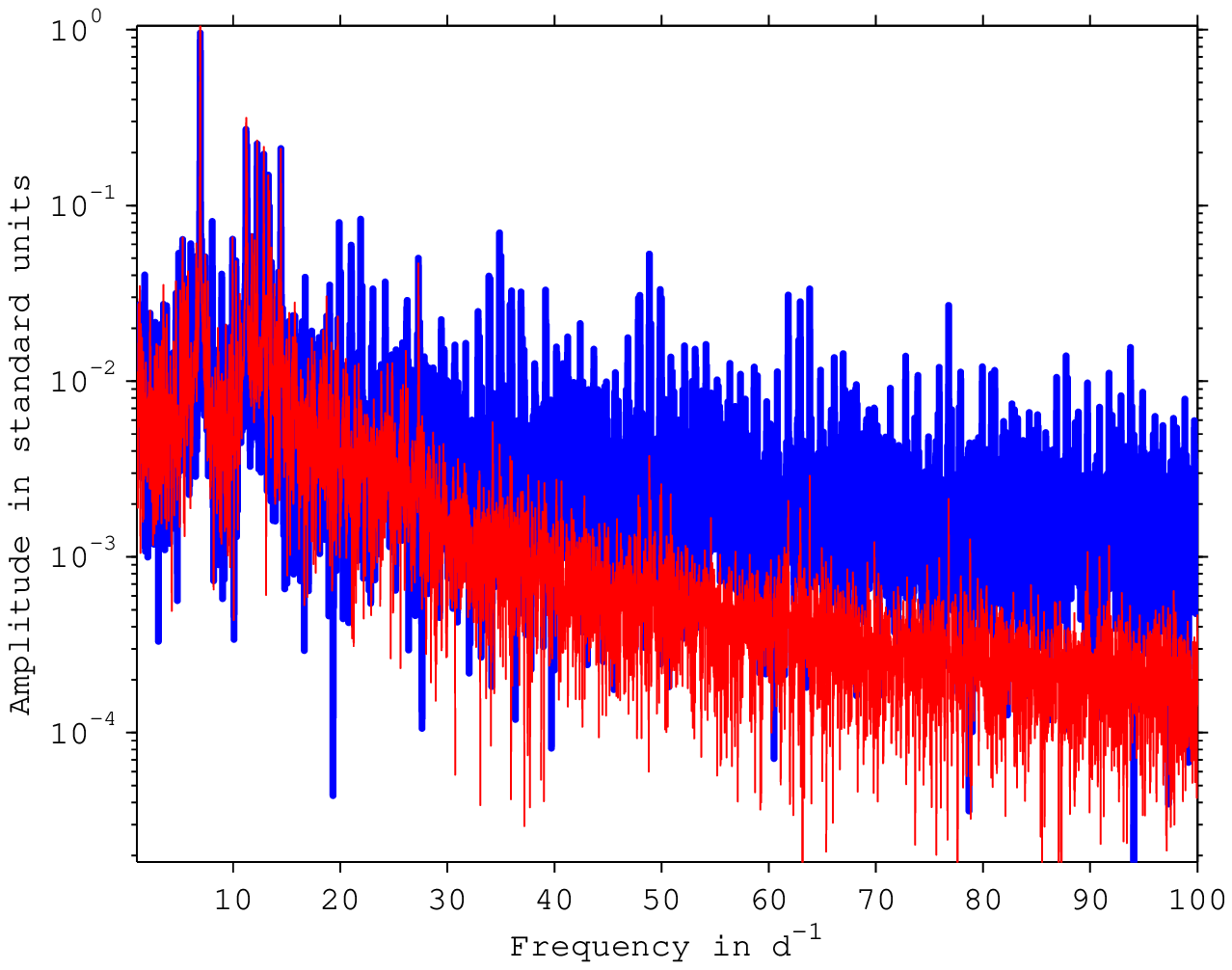}
   \includegraphics[width=8cm]{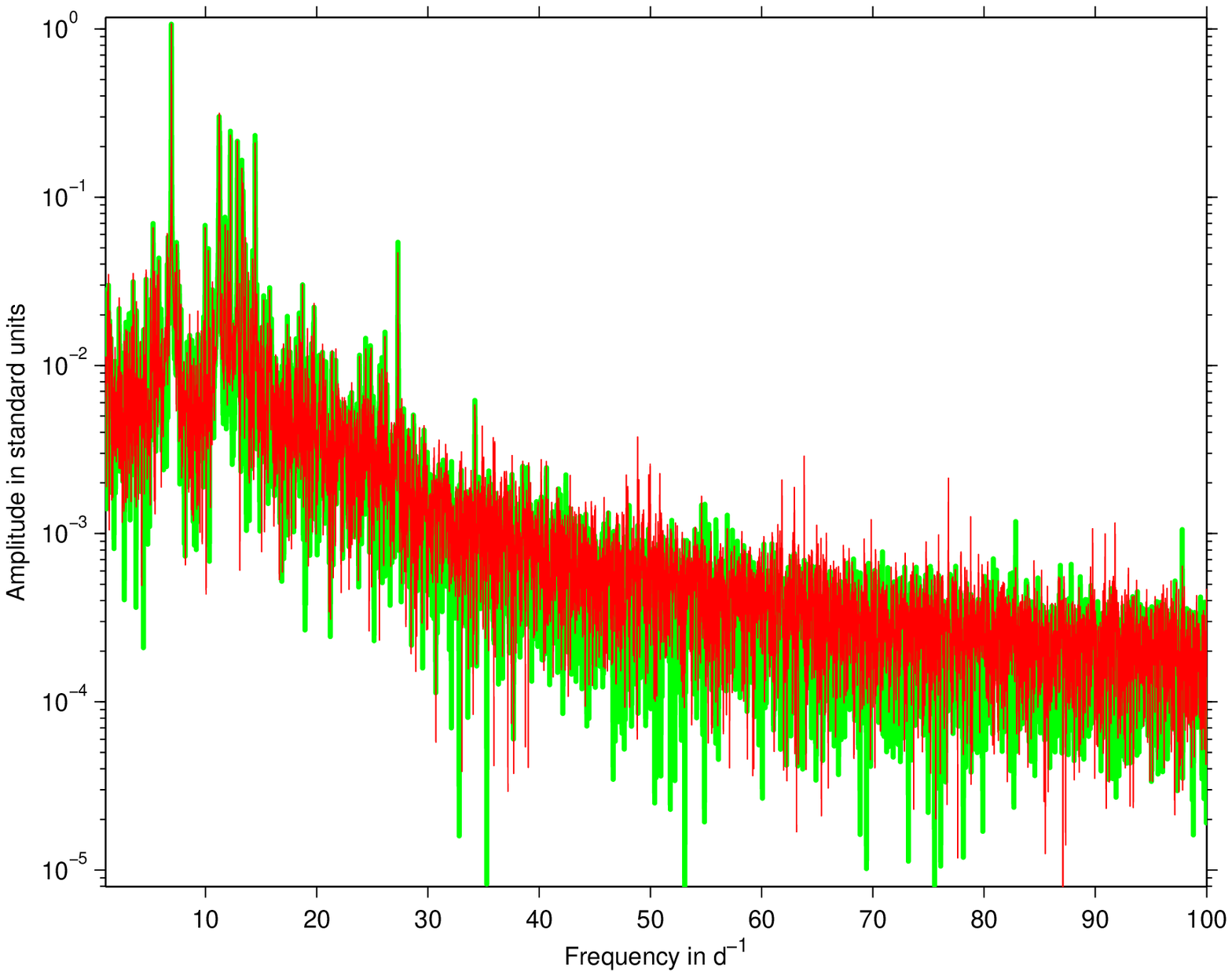}
   \caption{Power spectra of the light curves from HD 50844: upper panel shows gapped data in blue and ARMA interpolated data in red, lower panel shows linearly interpolated data in green and ARMA also in red.}
   \label{fig:ps1}
\end{figure}
\begin{figure}
   \centering
   \includegraphics[width=8cm]{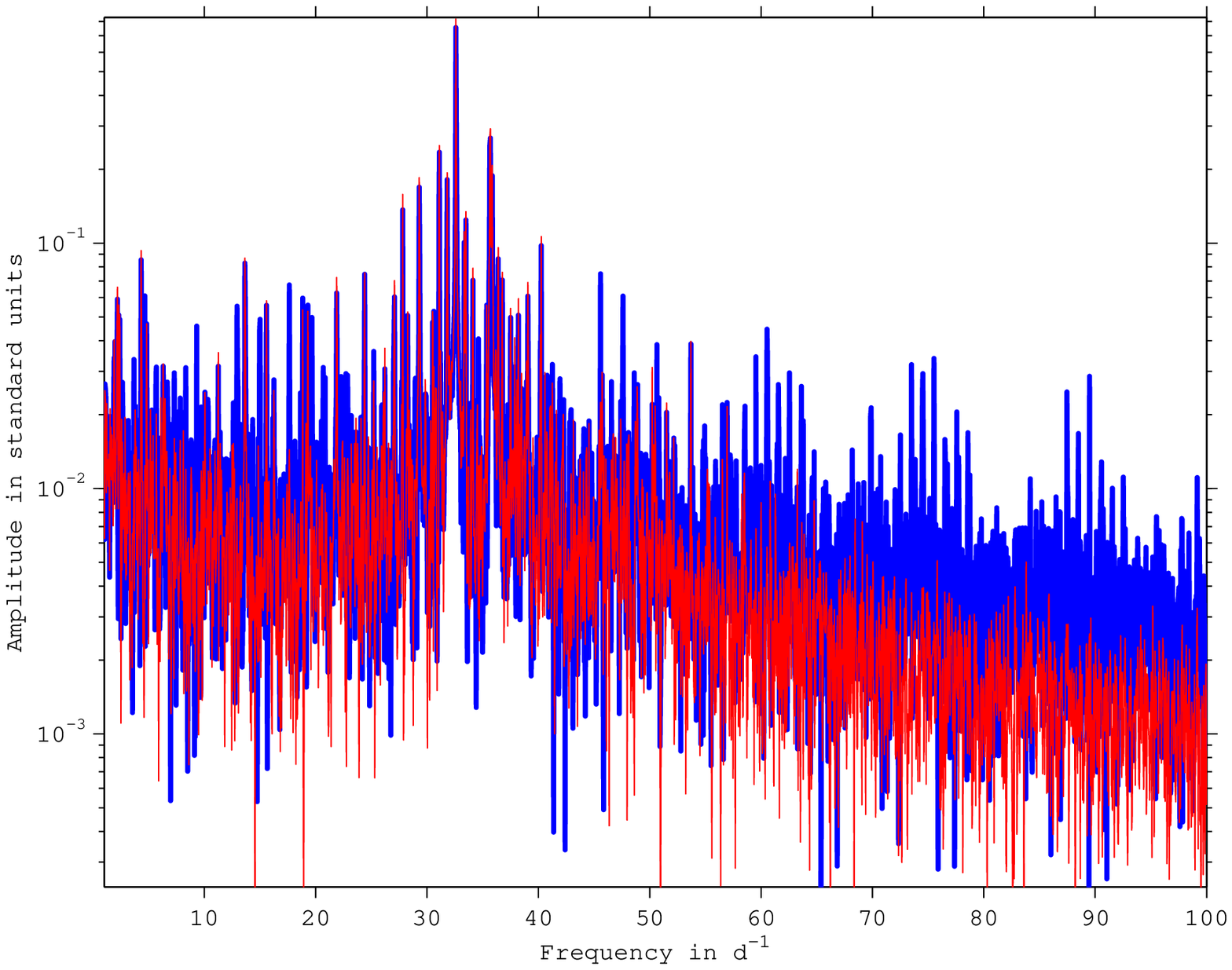}
   \includegraphics[width=8cm]{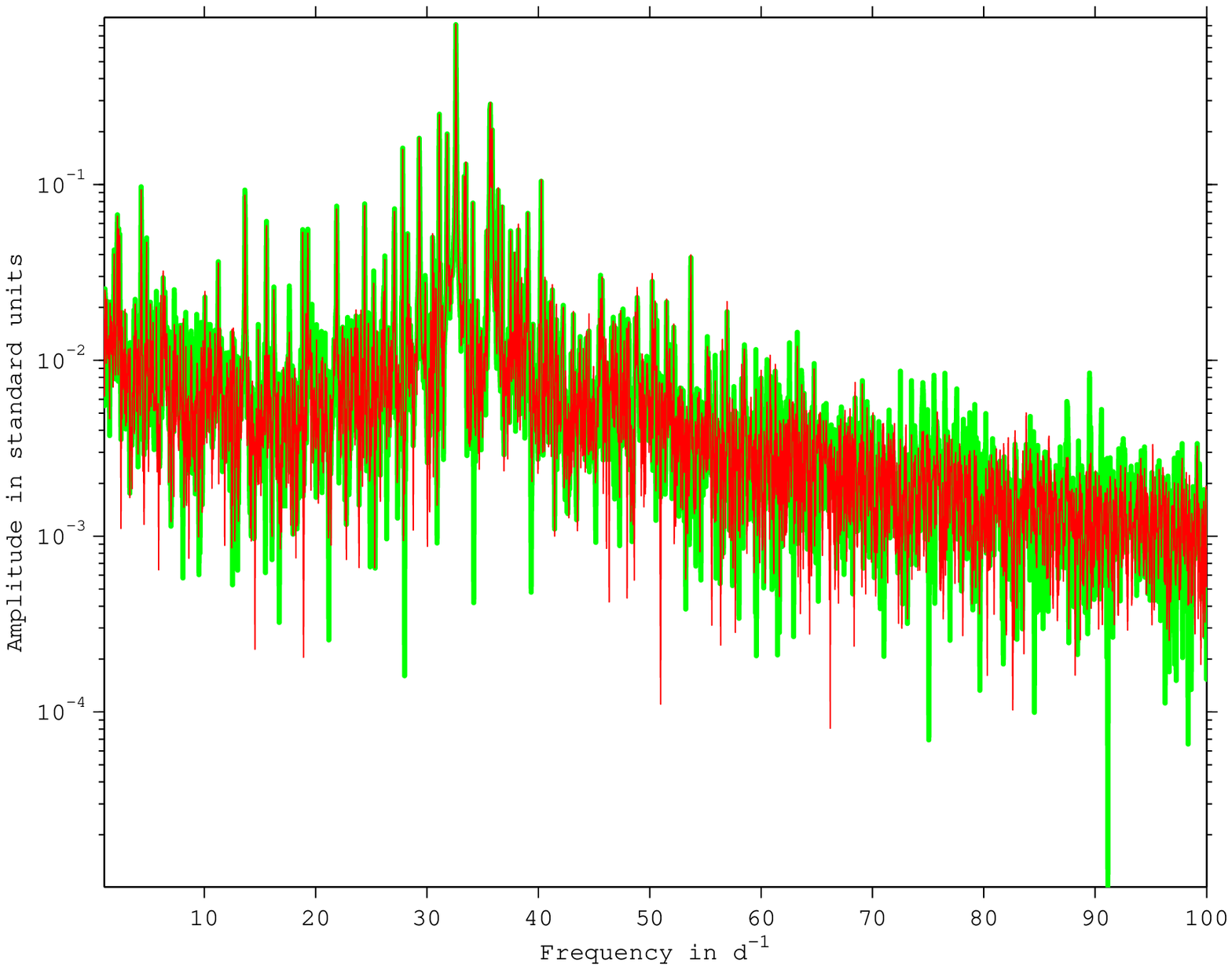}
   \caption{Power spectra of the light curves from HD 174936: upper panel shows gapped data in blue and ARMA interpolated data in red, lower panel shows linearly interpolated data in green and ARMA also in red.}
   \label{fig:ps2}
\end{figure}
\begin{figure}
    \centering
   \includegraphics[width=8cm]{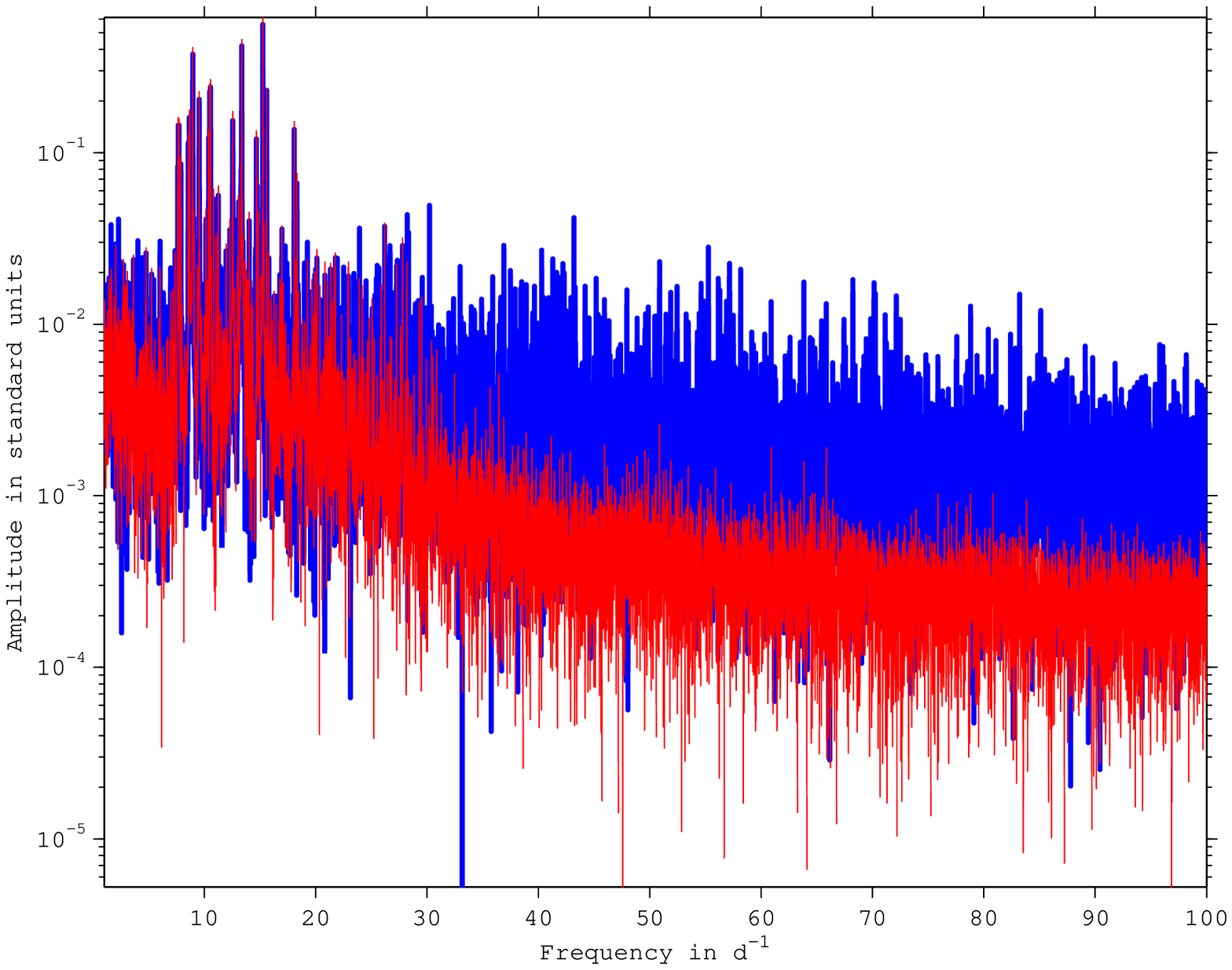}
   \includegraphics[width=8cm]{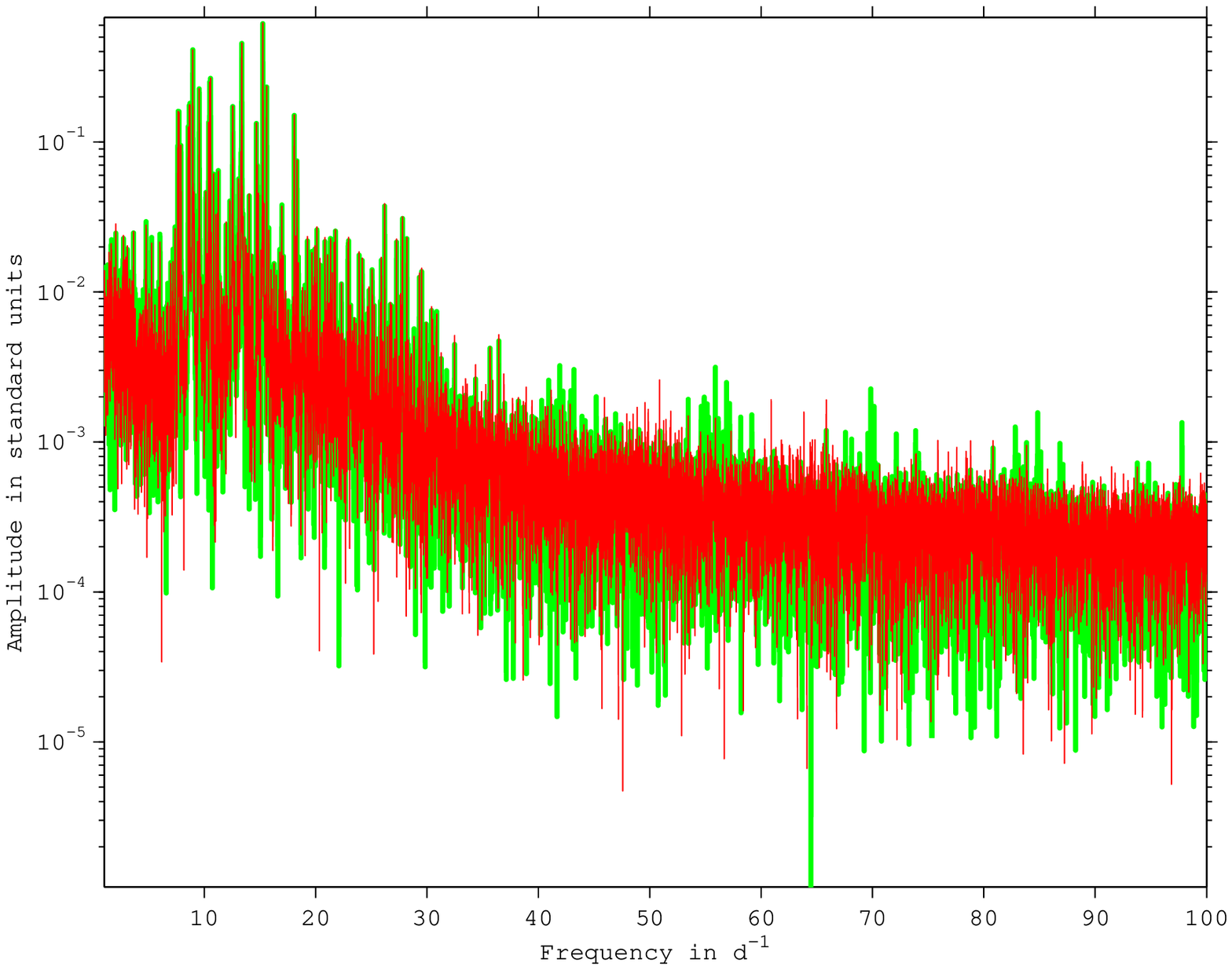}
   \caption{Power spectra of the light curves from HD 181555: upper panel shows gapped data in blue and ARMA interpolated data in red, lower panel shows linearly interpolated data in green and ARMA also in red.}
   \label{fig:ps3}
\end{figure}
\begin{figure}
  \centering
  \includegraphics[width=8cm]{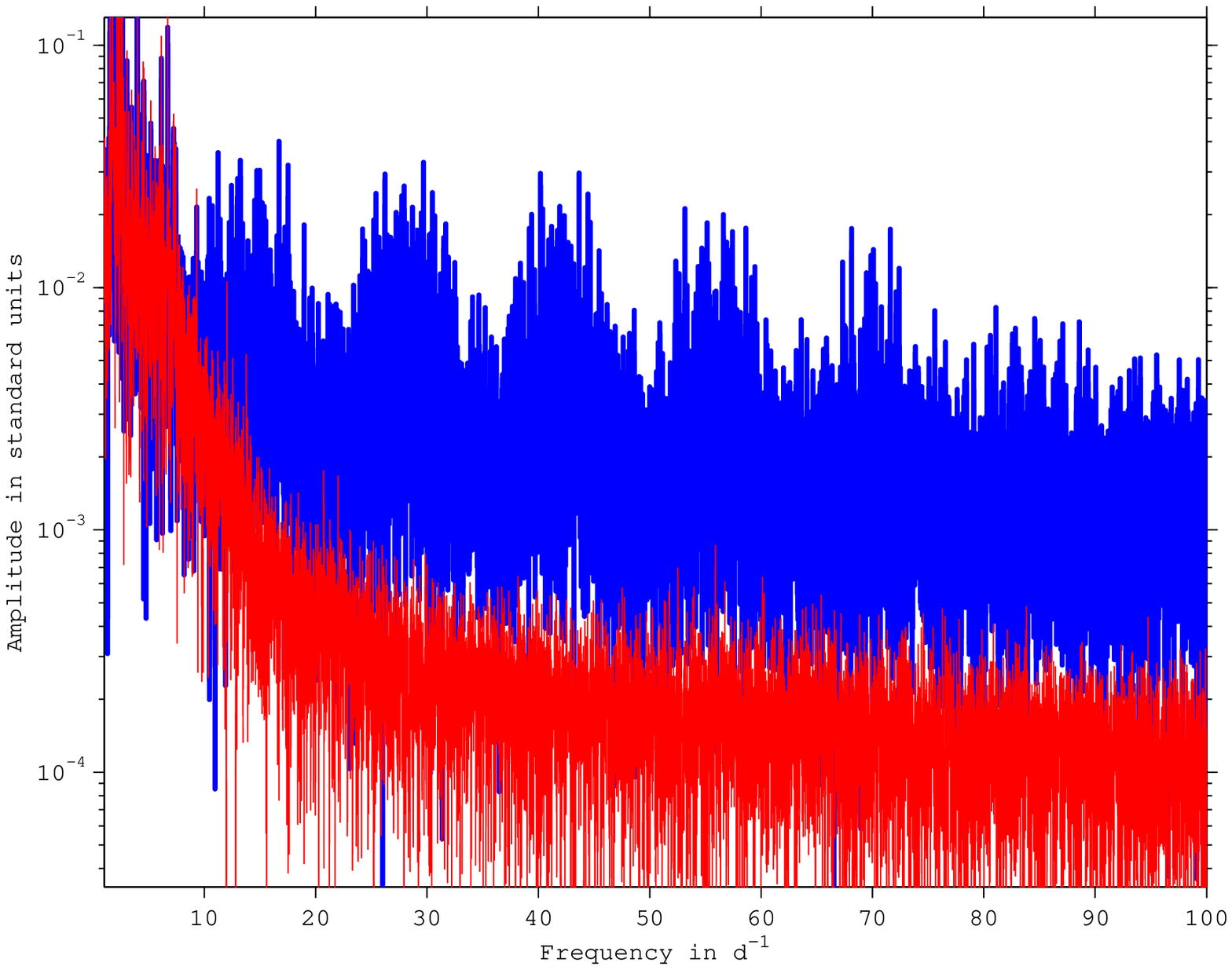}
  \includegraphics[width=8cm]{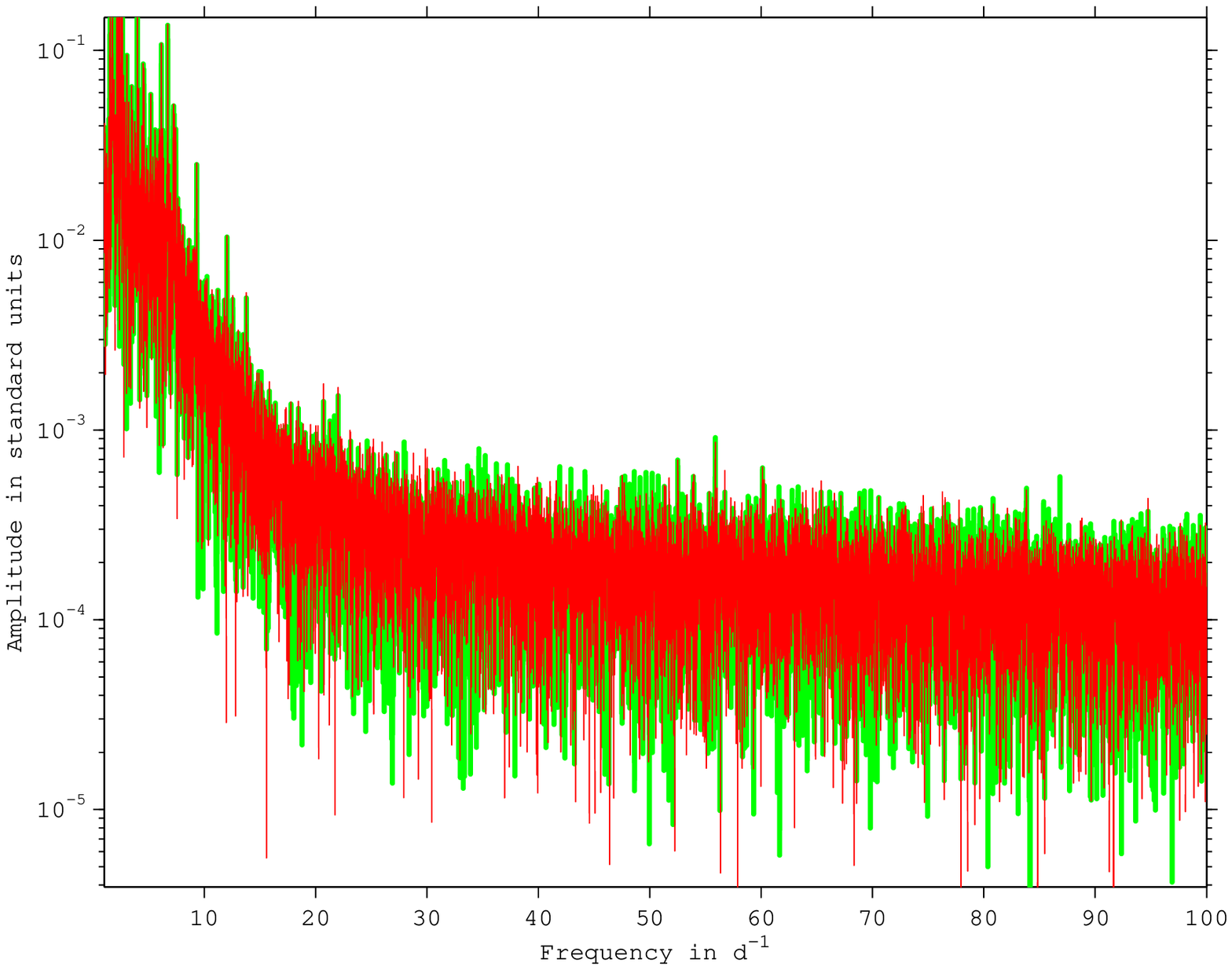}
  \caption{Power spectra of the light curves from HD 49434: upper panel shows gapped data in blue and ARMA interpolated data in red, lower panel shows linearly interpolated data in green and ARMA also in red.}
  \label{fig:ps4}
\end{figure}
\begin{figure}
  \centering
  \includegraphics[width=8cm]{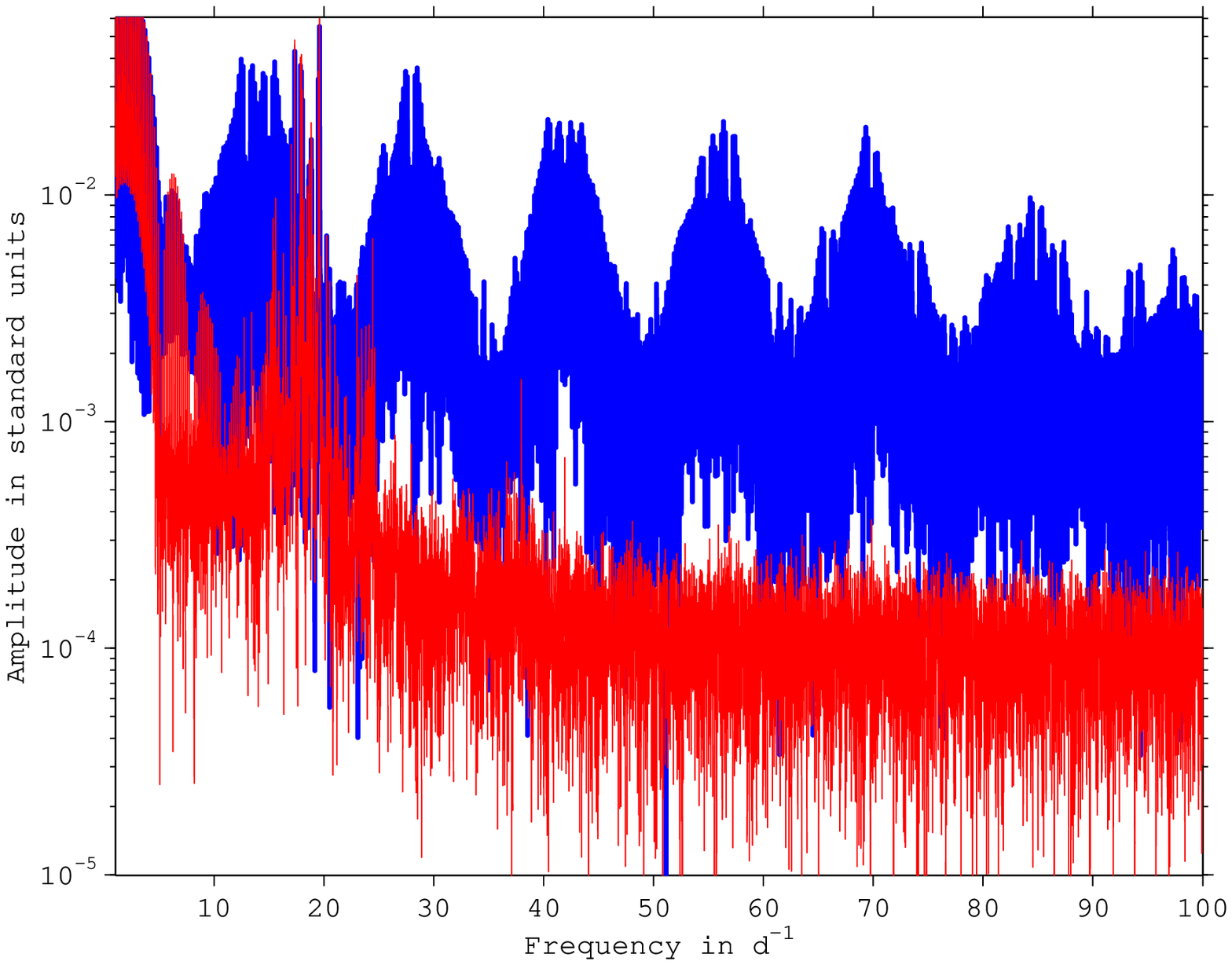}
  \includegraphics[width=8cm]{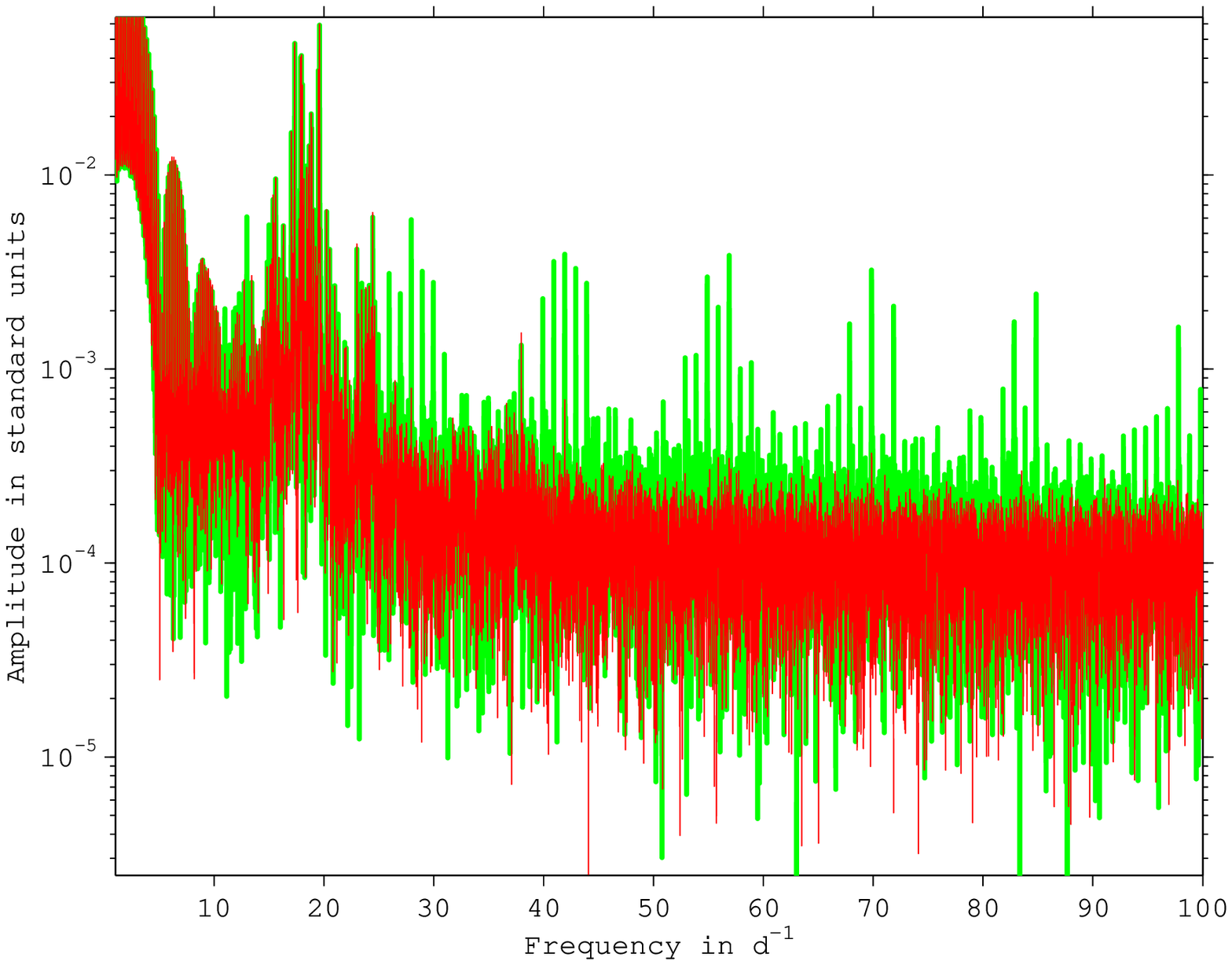}
  \caption{Power spectra of the light curves from HD 172189: upper panel shows gapped data in blue and ARMA interpolated data in red, lower panel shows linearly interpolated data in green and ARMA also in red.}
  \label{fig:ps5}
\end{figure}
\begin{figure}
  \centering
  \includegraphics[width=8cm]{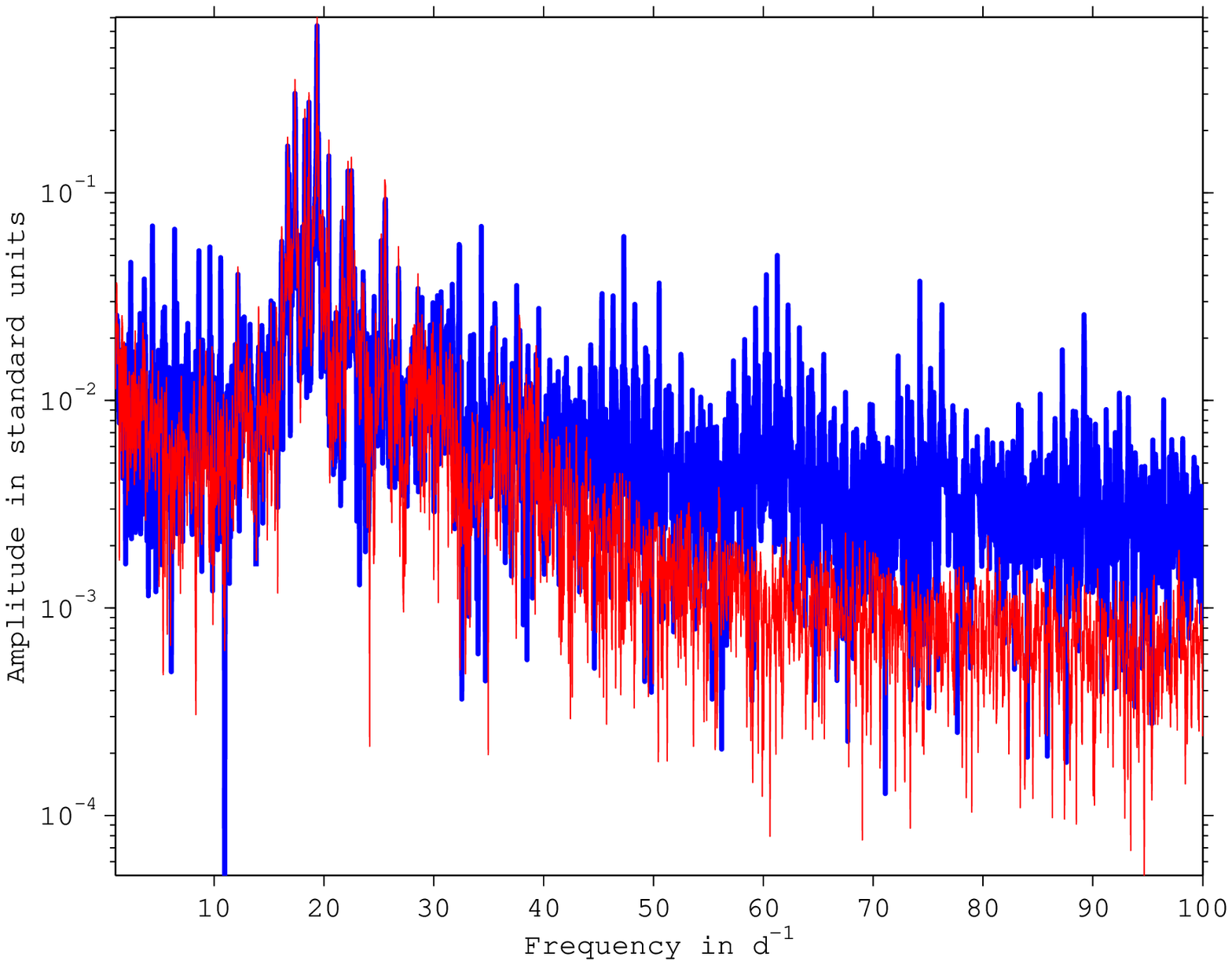}
  \includegraphics[width=8cm]{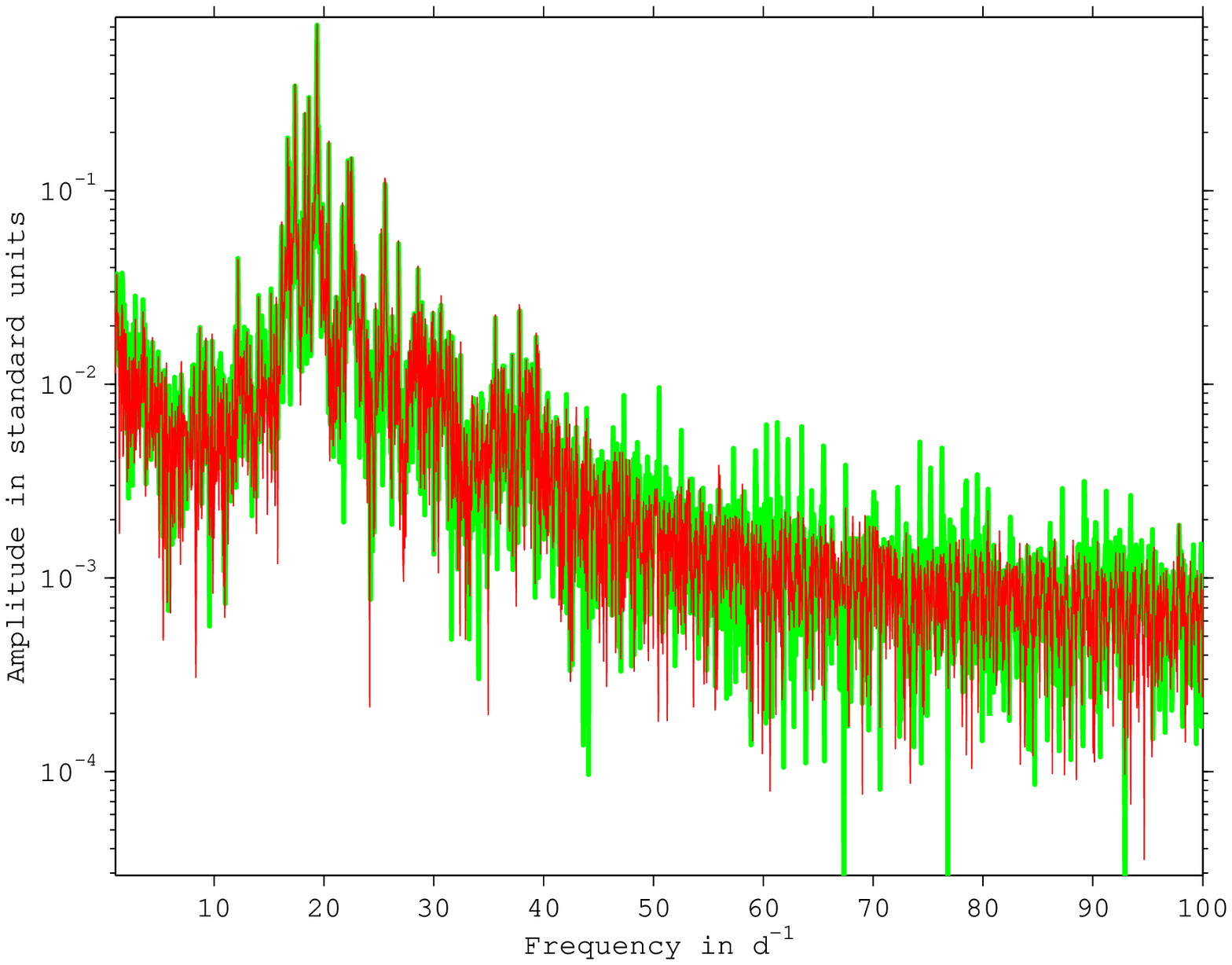}
  \caption{Power spectra of the light curves from HD 174532: upper panel shows gapped data in blue and ARMA interpolated data in red, lower panel shows linearly interpolated data in green and ARMA also in red.}
  \label{fig:ps6}
\end{figure}
\begin{figure}
  \centering
  \includegraphics[width=8cm]{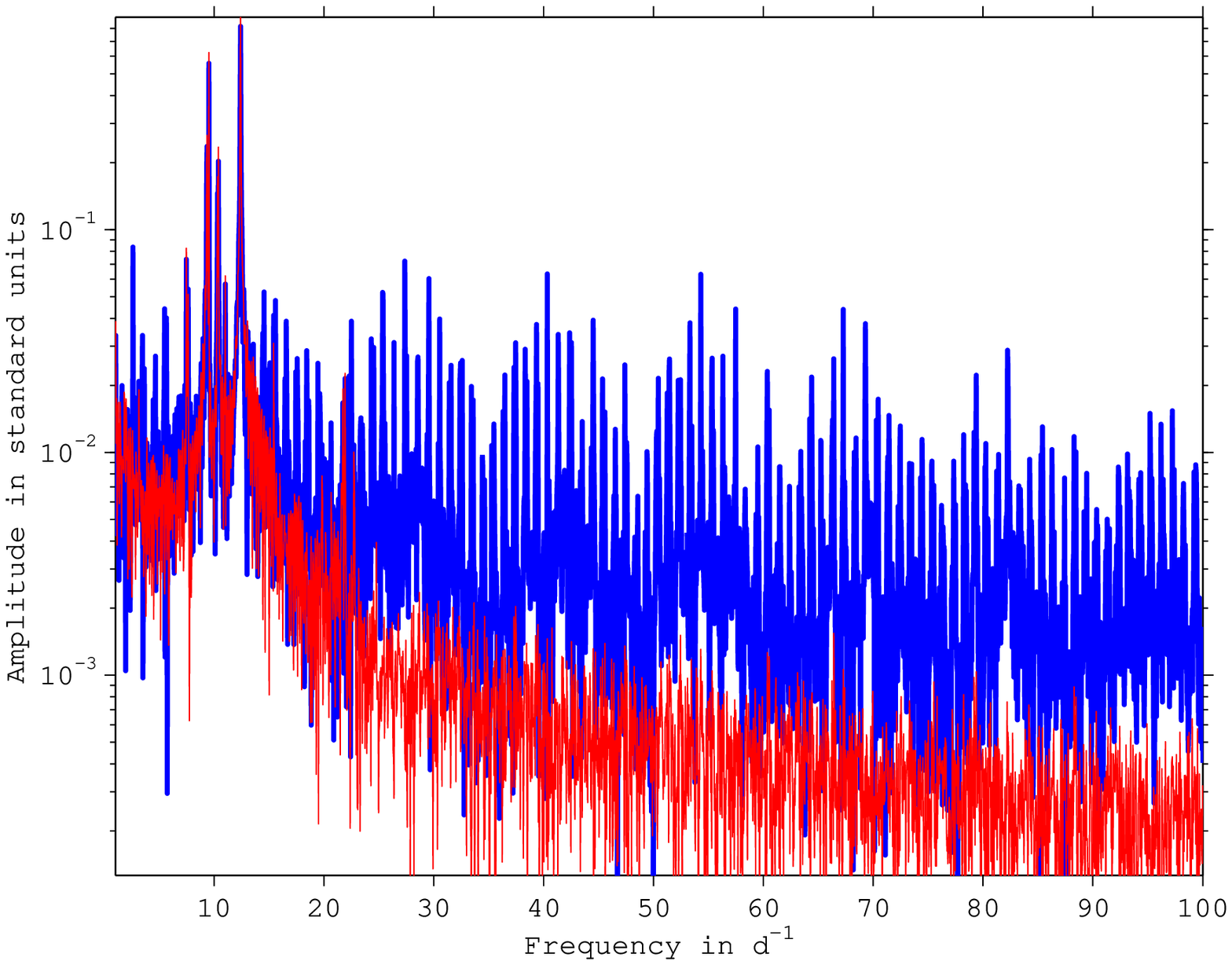}
  \includegraphics[width=8cm]{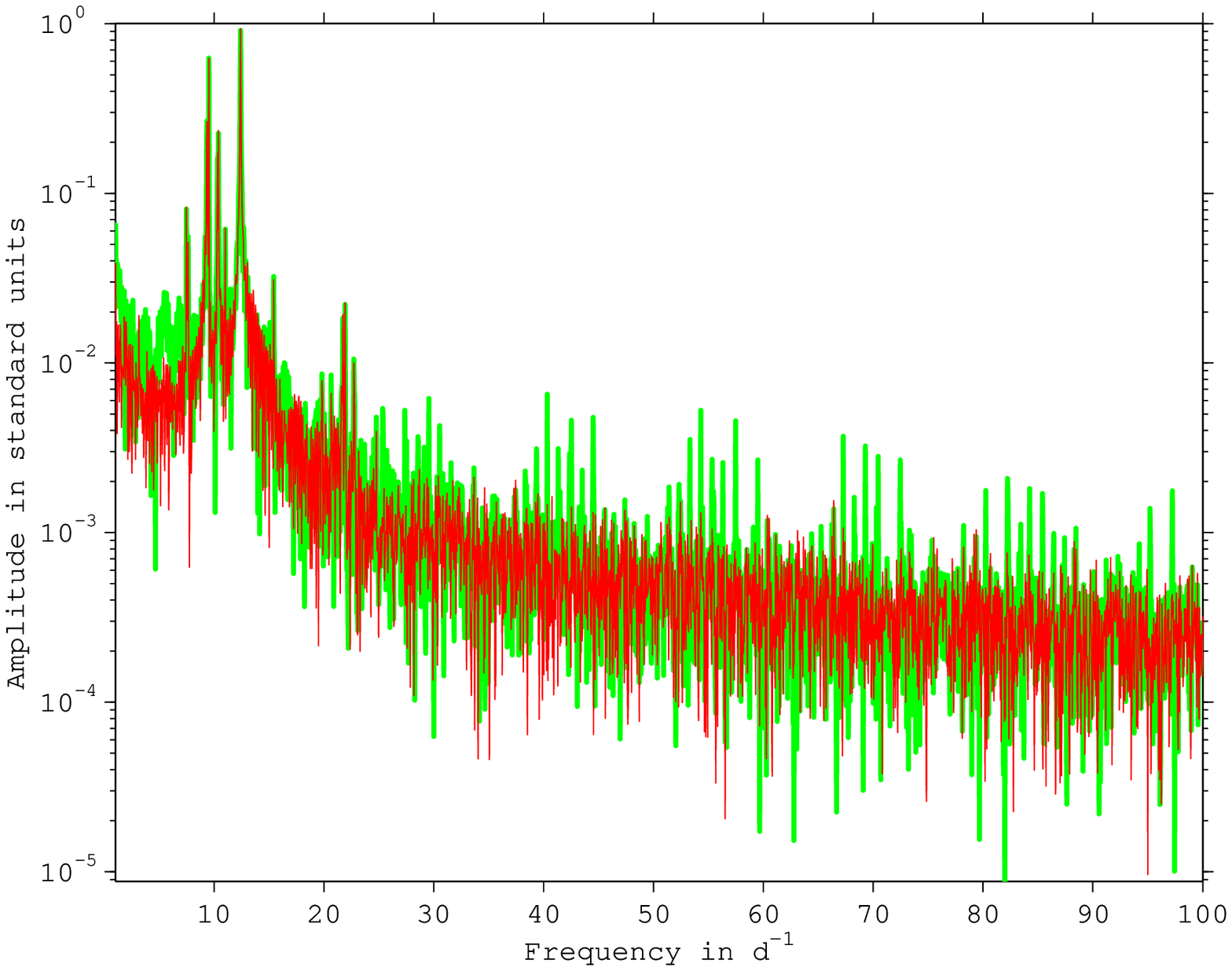}
  \caption{Power spectra of the light curves from HD 174589: upper panel shows gapped data in blue and ARMA interpolated data in red, lower panel shows linearly interpolated data in green and ARMA also in red.}
  \label{fig:ps7}
\end{figure}
\begin{figure}
  \centering
  \includegraphics[width=8cm]{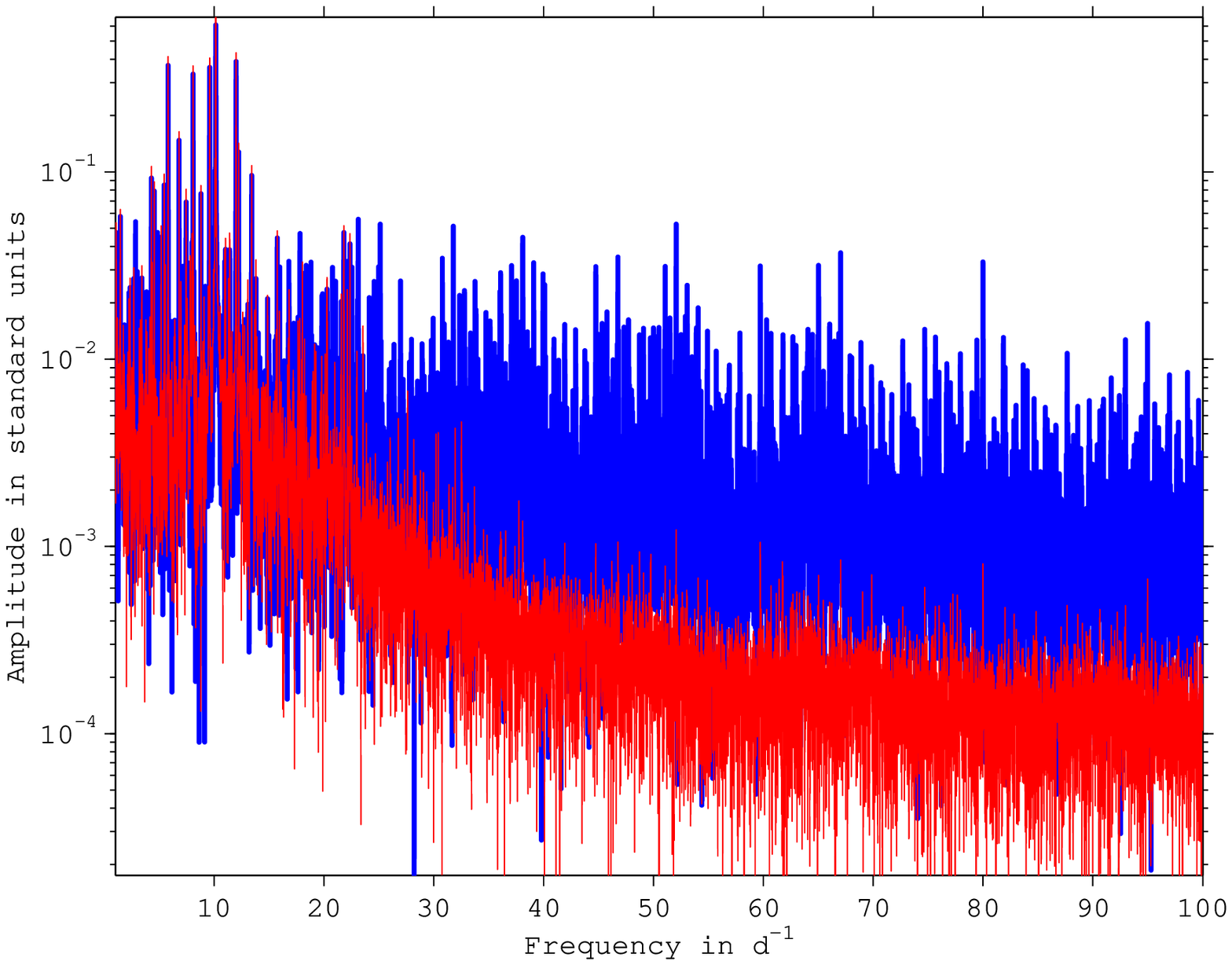}
  \includegraphics[width=8cm]{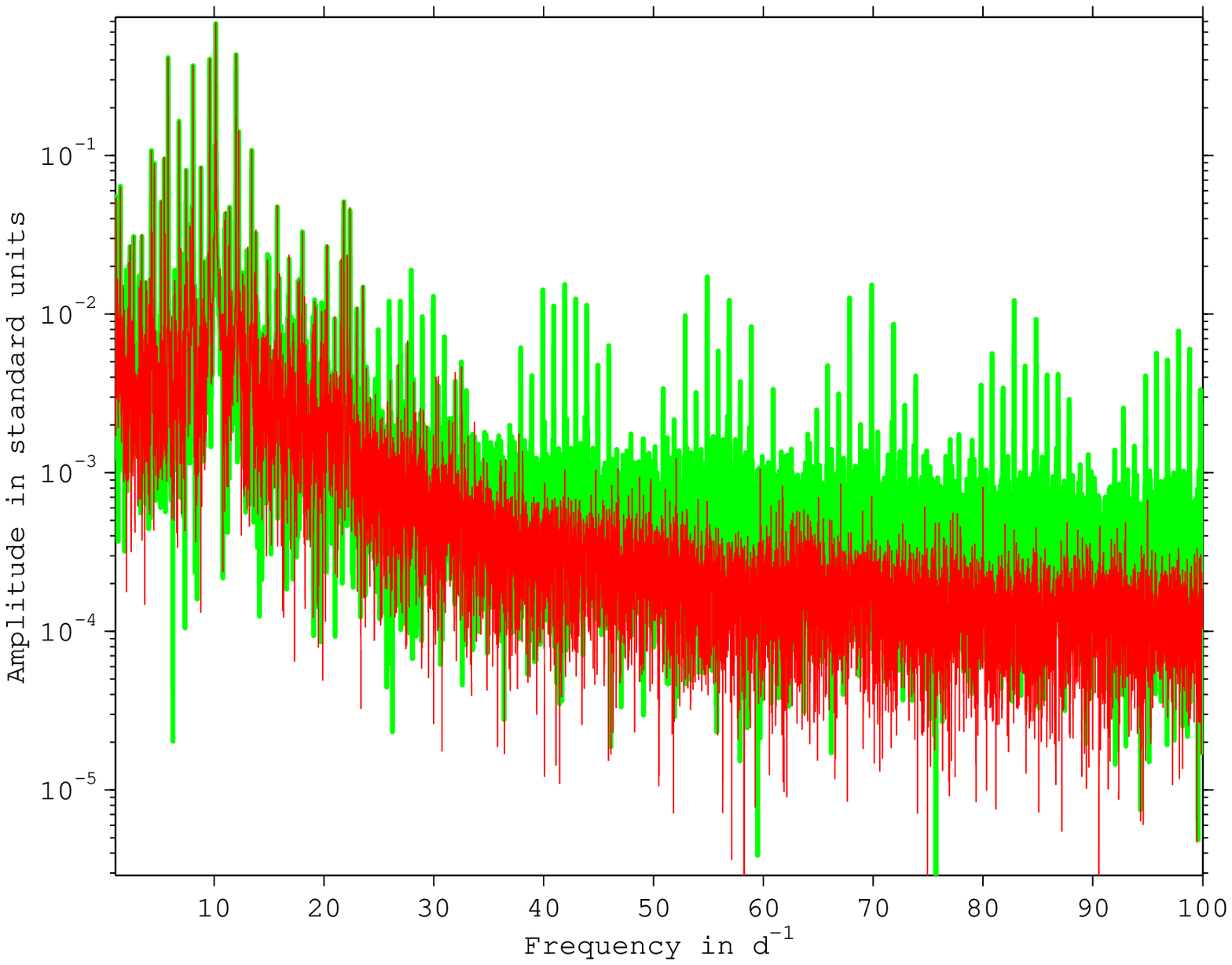}
  \caption{Power spectra of the light curves from HD 51722: upper panel shows gapped data in blue and ARMA interpolated data in red, lower panel shows linearly interpolated data in green and ARMA also in red.}
  \label{fig:ps8}
\end{figure}
\begin{figure}
  \centering
  \includegraphics[width=8cm]{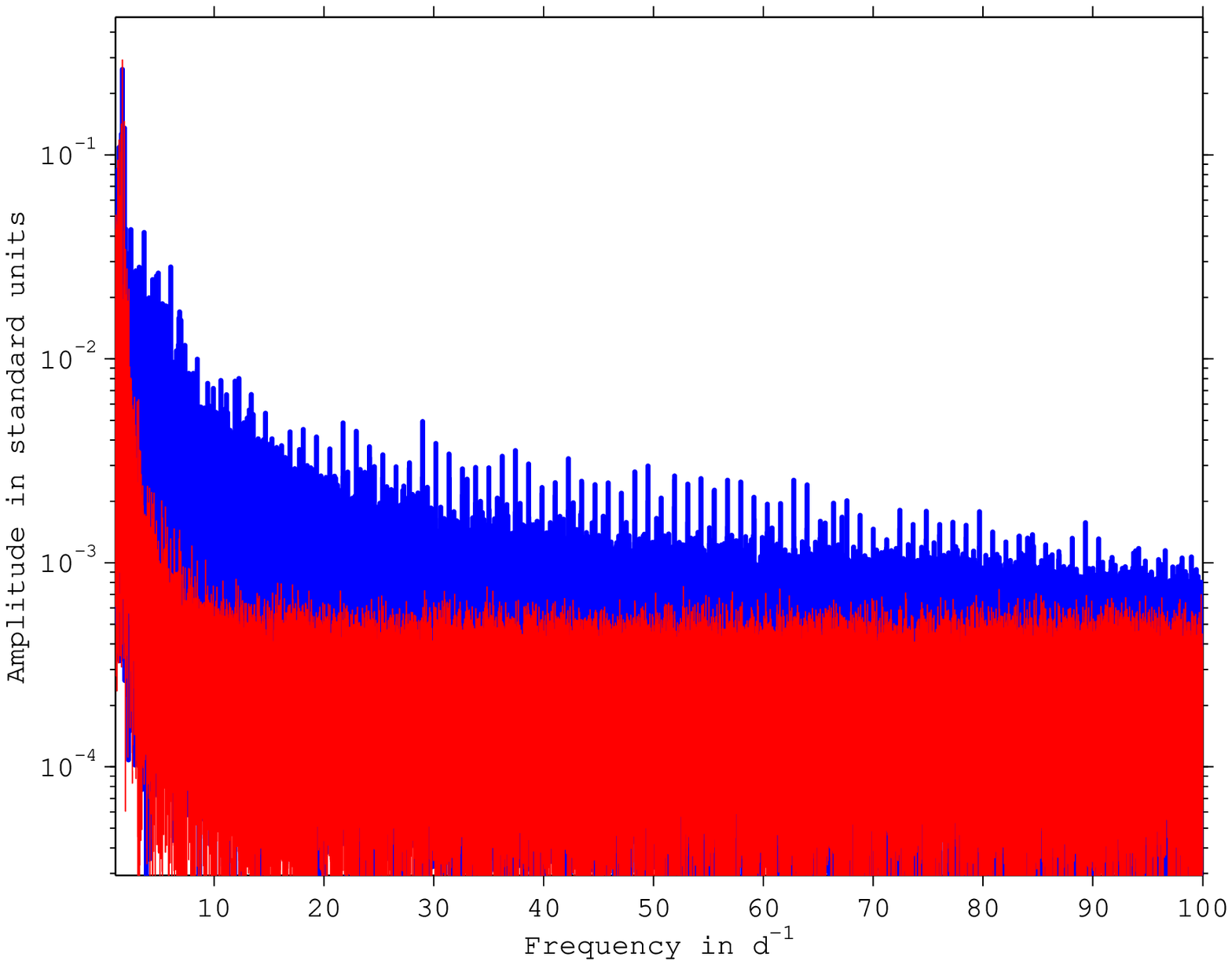}
  \includegraphics[width=8cm]{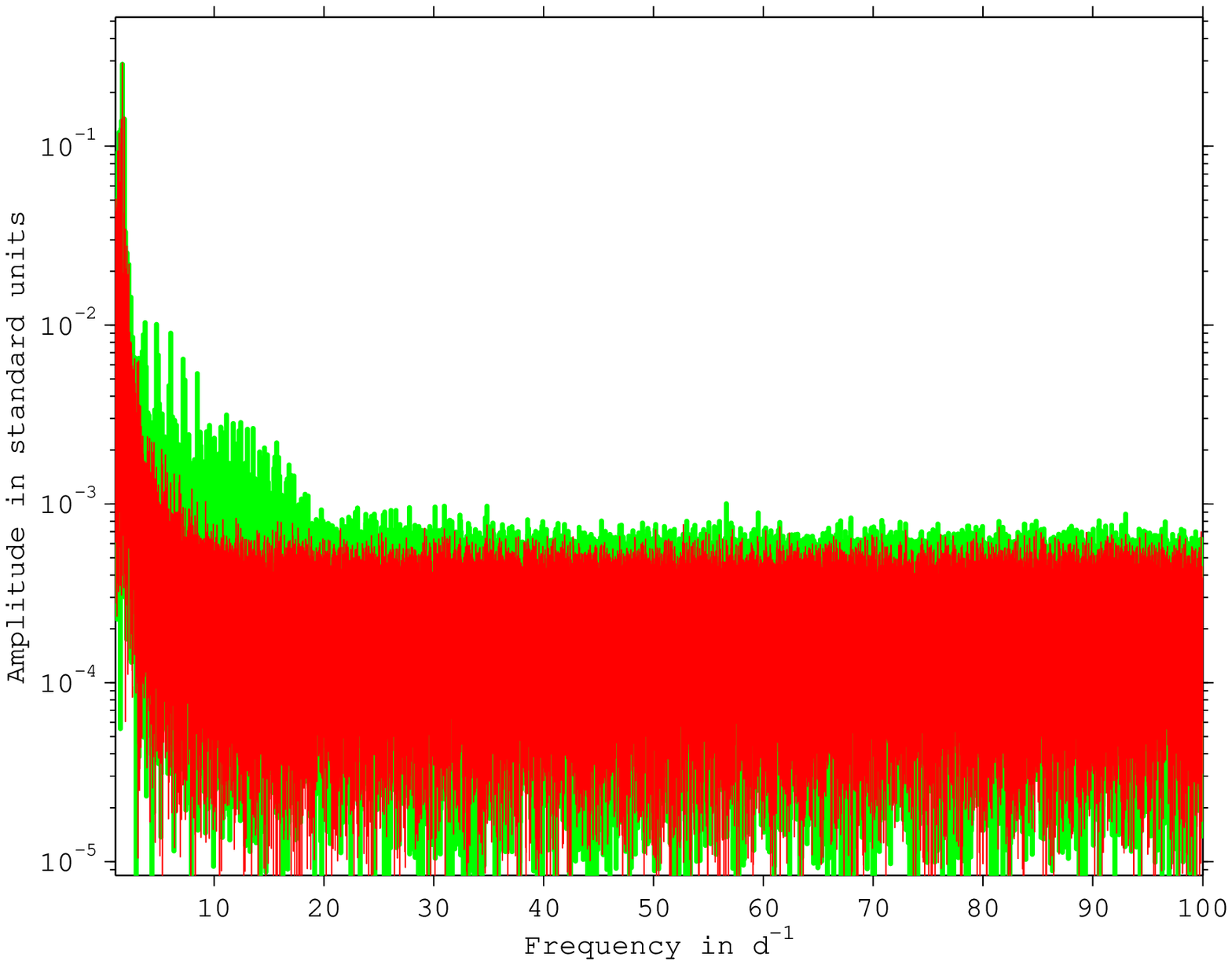}
  \caption{Power spectra of the light curves from HD 51359: upper panel shows gapped data in blue and ARMA interpolated data in red, lower panel shows linearly interpolated data in green and ARMA also in red.}
  \label{fig:ps9}
\end{figure}
\begin{figure}
  \centering
  \includegraphics[width=8cm]{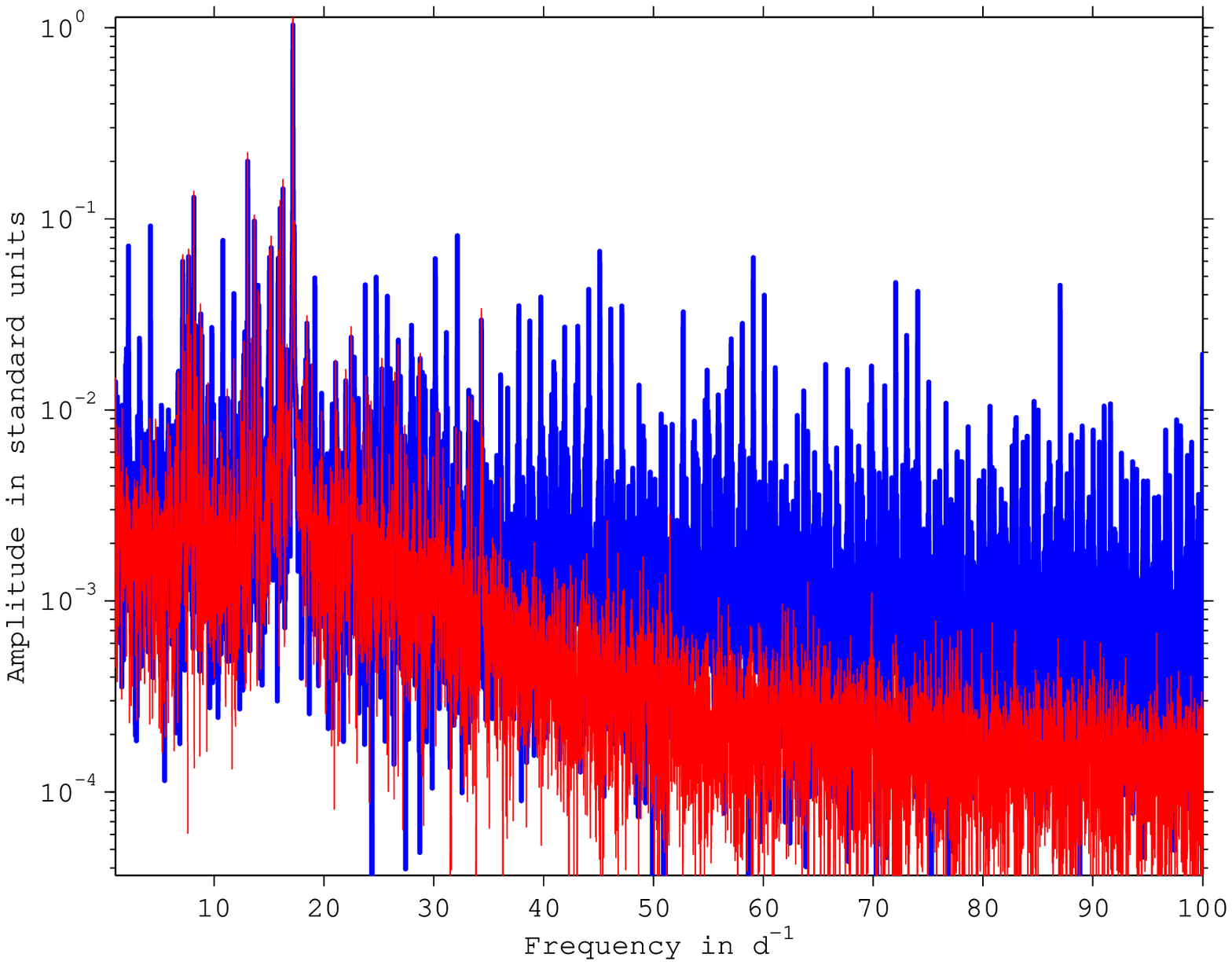}
  \includegraphics[width=8cm]{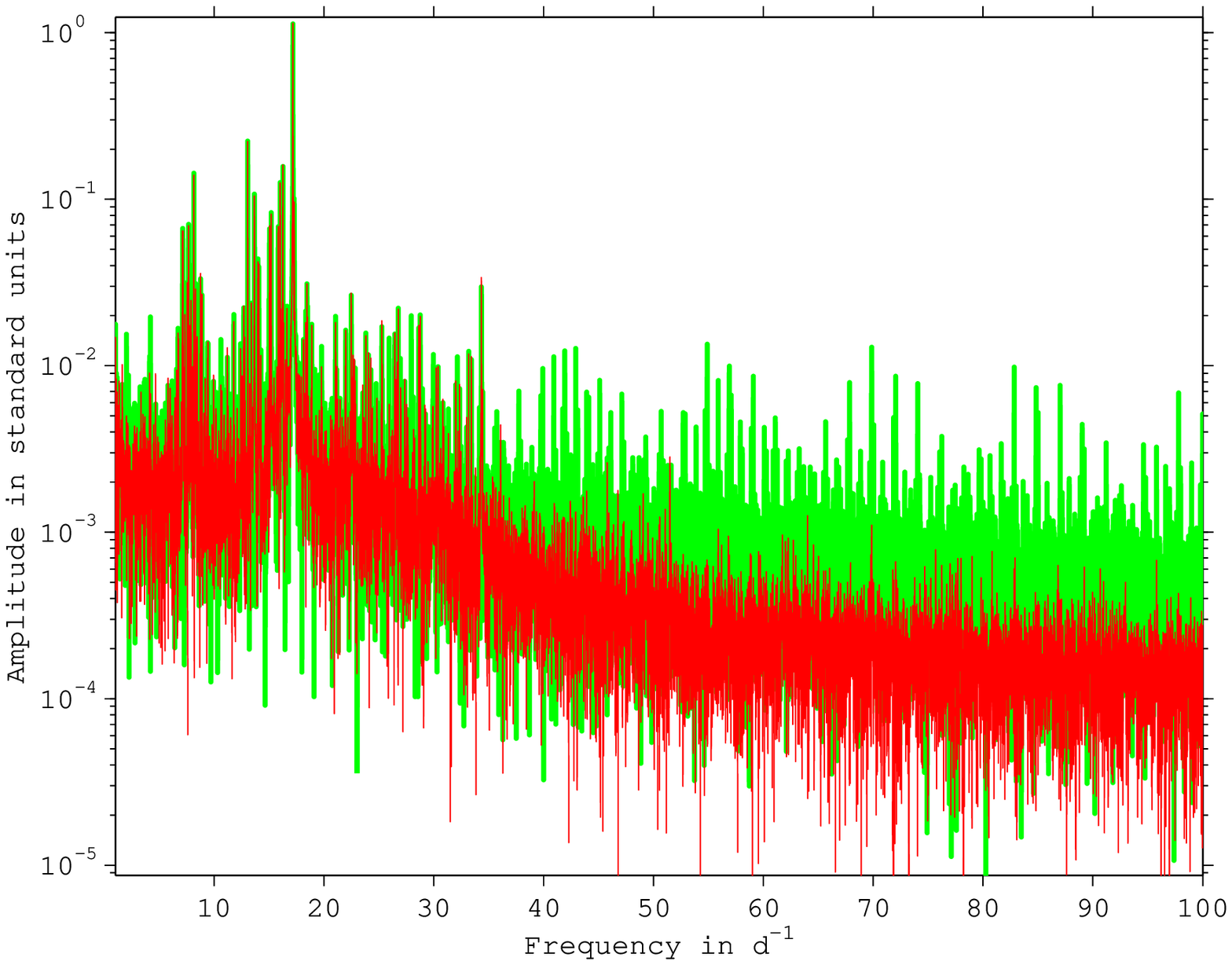}
  \caption{Power spectra of the light curves from HD 50870: upper panel shows gapped data in blue and ARMA interpolated data in red, lower panel shows linearly interpolated data in green and ARMA also in red.}
  \label{fig:ps10}
\end{figure}
\begin{figure}
  \centering
  \includegraphics[width=8cm]{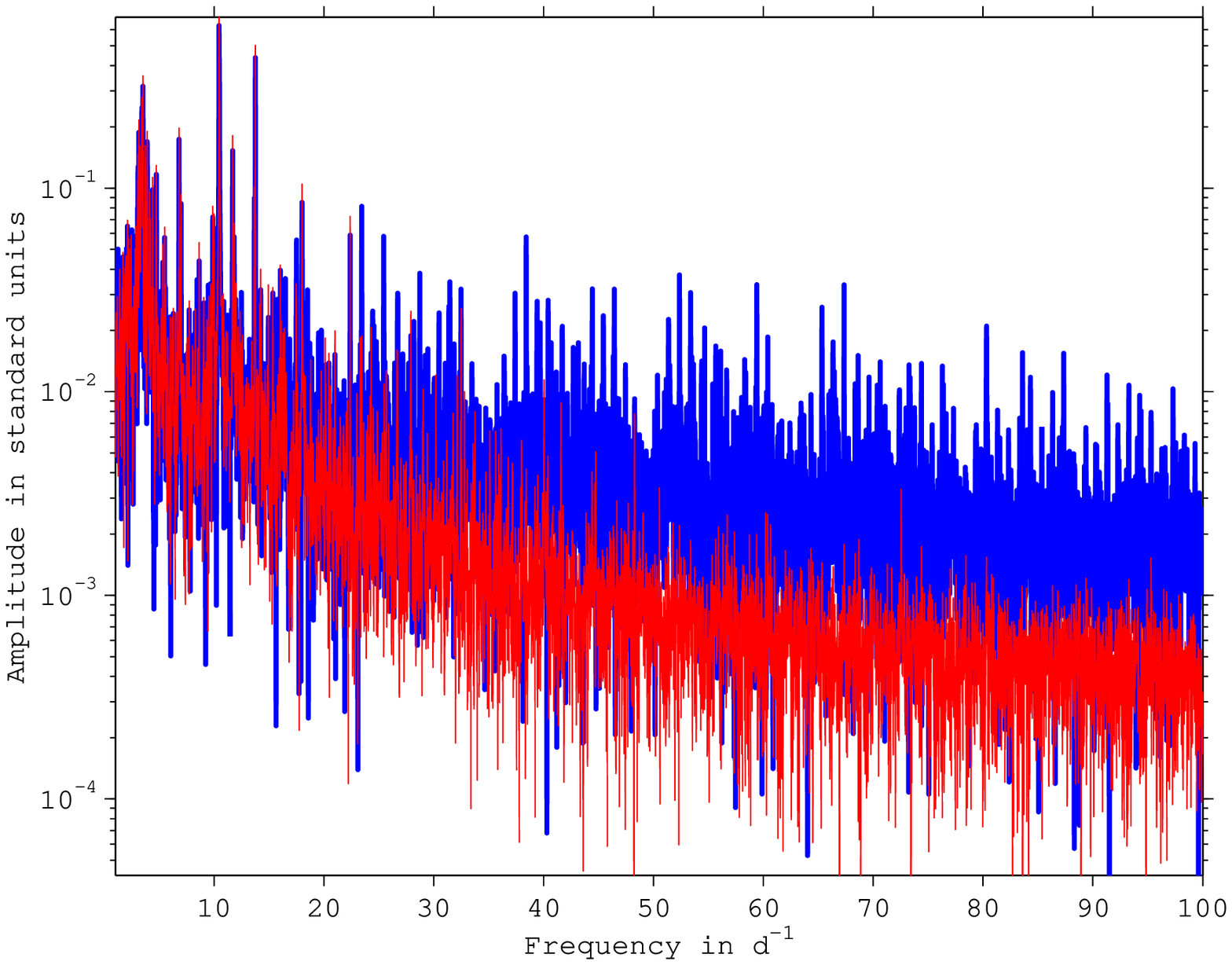}
  \includegraphics[width=8cm]{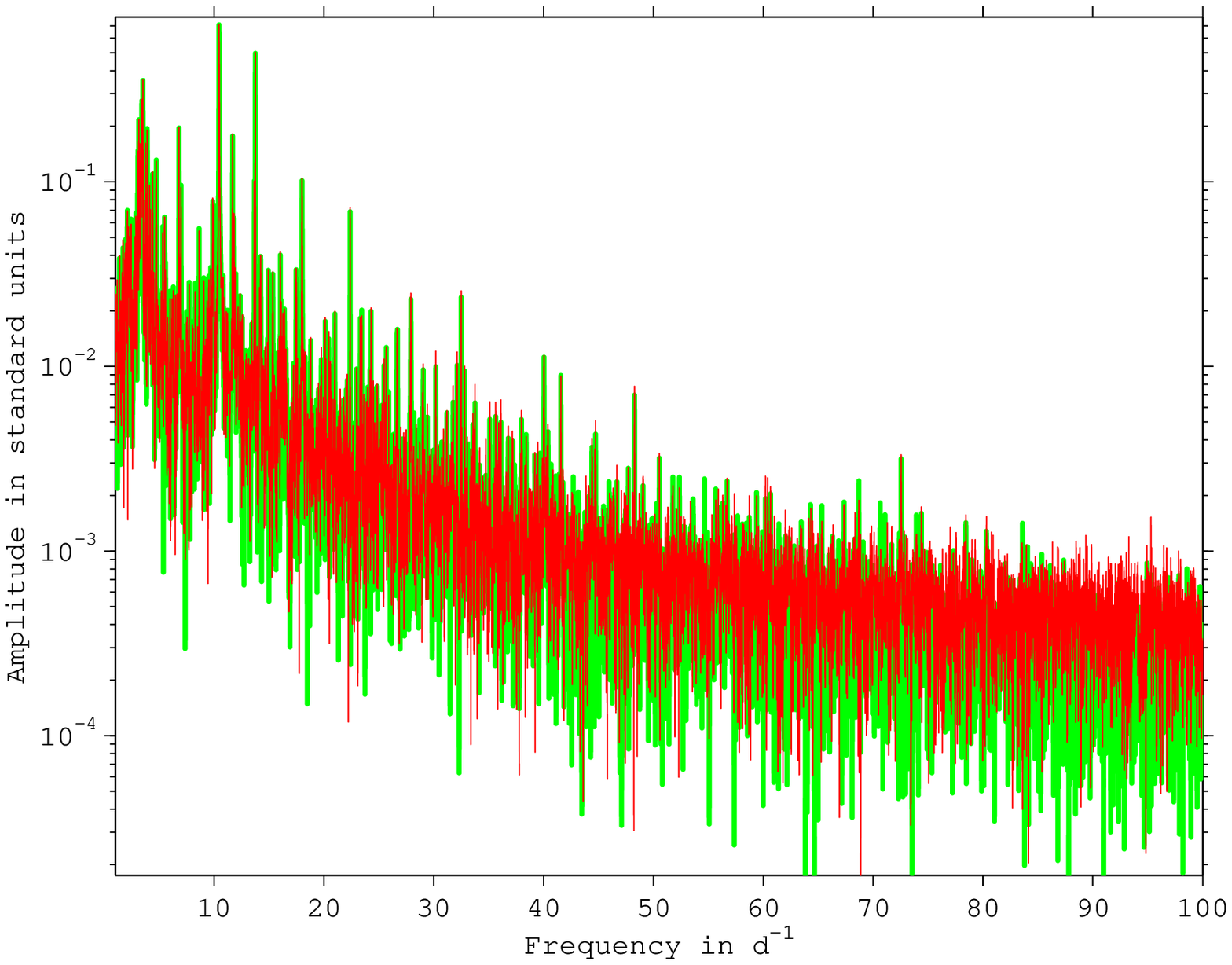}
  \caption{Power spectra of the light curves from HD 170699: upper panel shows gapped data in blue and ARMA interpolated data in red, lower panel shows linearly interpolated data in green and ARMA also in red.}
  \label{fig:ps11}
\end{figure}
\begin{figure}
  \centering
  \includegraphics[width=8cm]{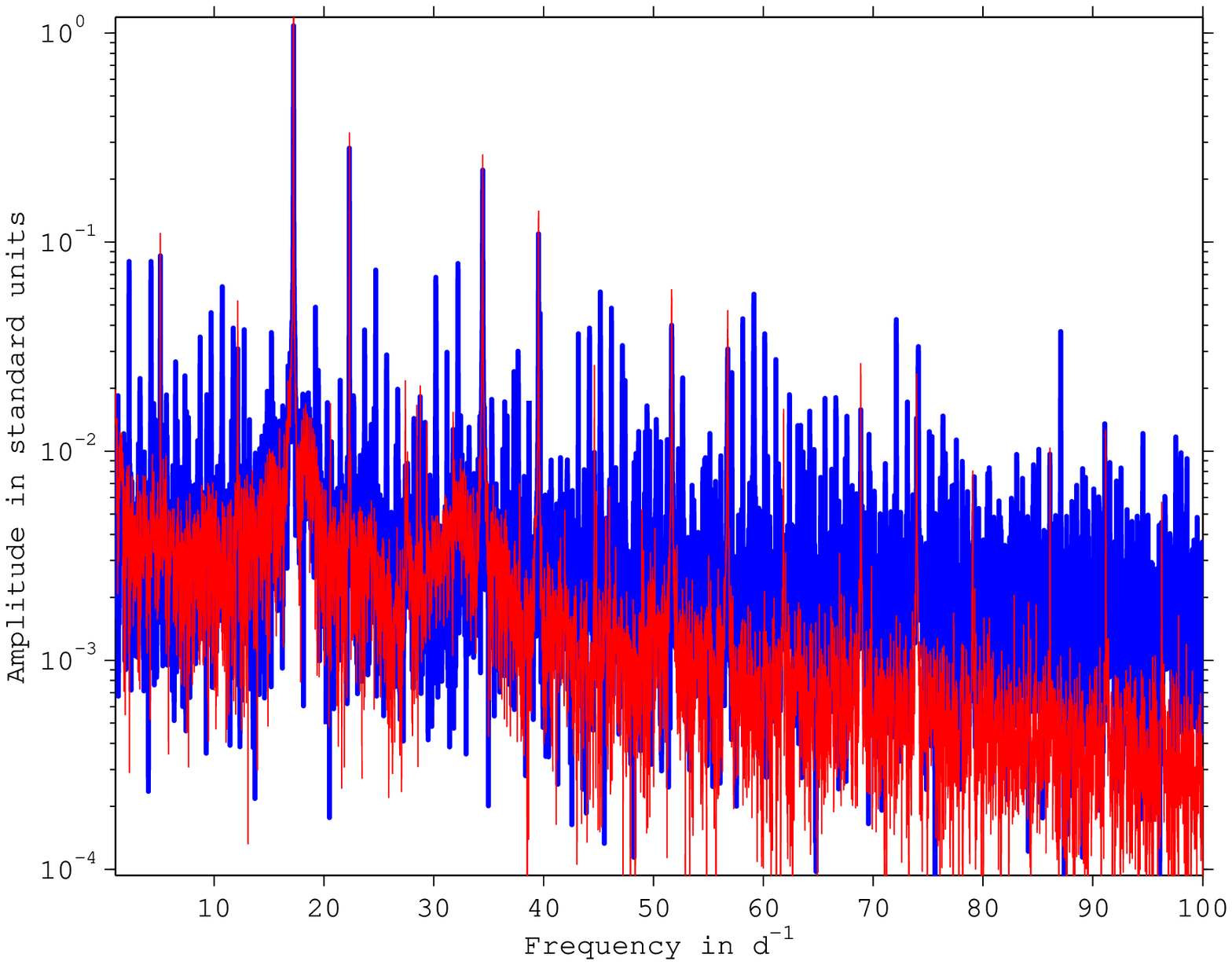}
  \includegraphics[width=8cm]{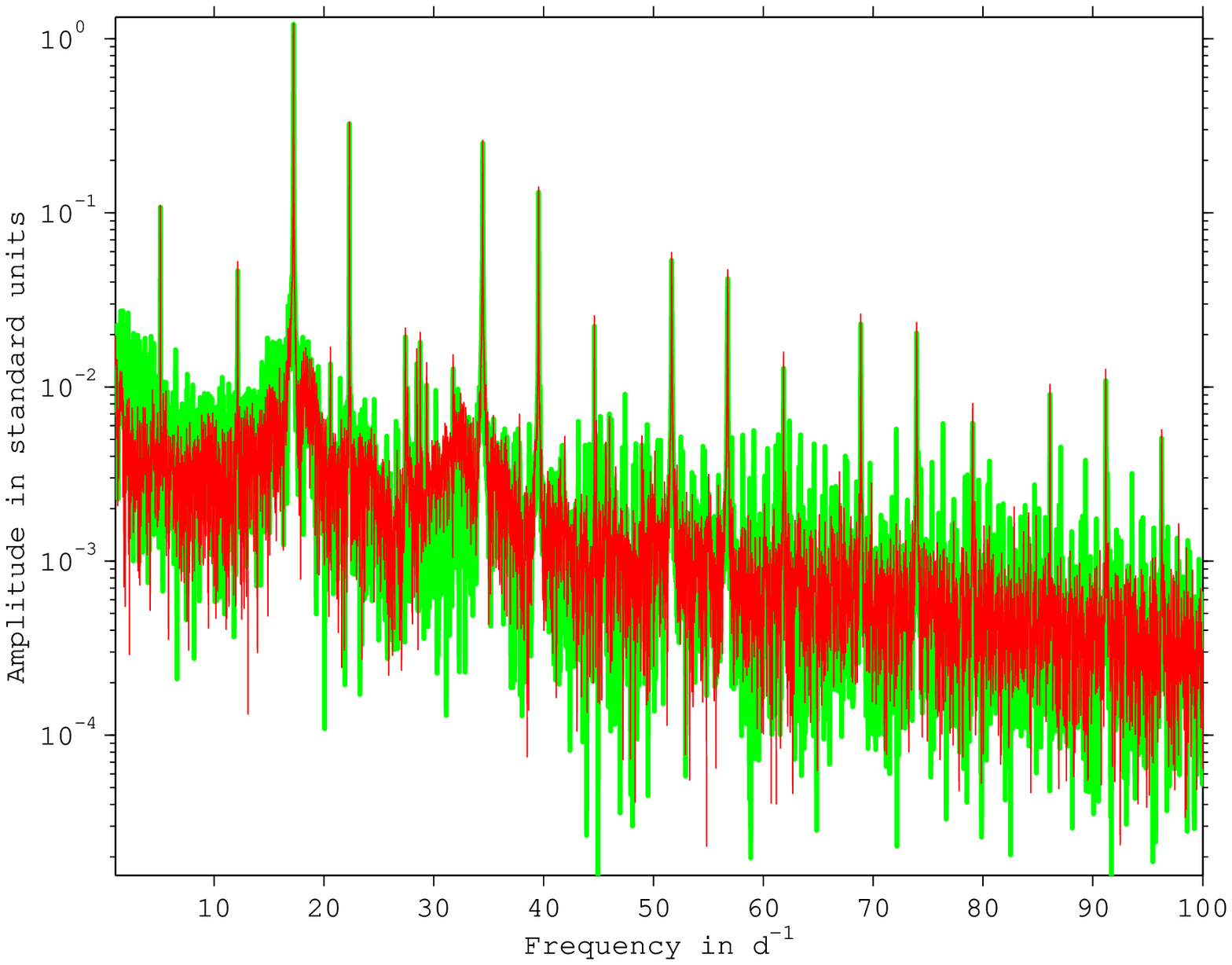}
  \caption{Power spectra of the light curves from GSC 00144-03031: upper panel shows gapped data in blue and ARMA interpolated data in red, lower panel shows linearly interpolated data in green and ARMA also in red.}
  \label{fig:ps12}
\end{figure}
\begin{figure}
  \centering
  \includegraphics[width=8cm]{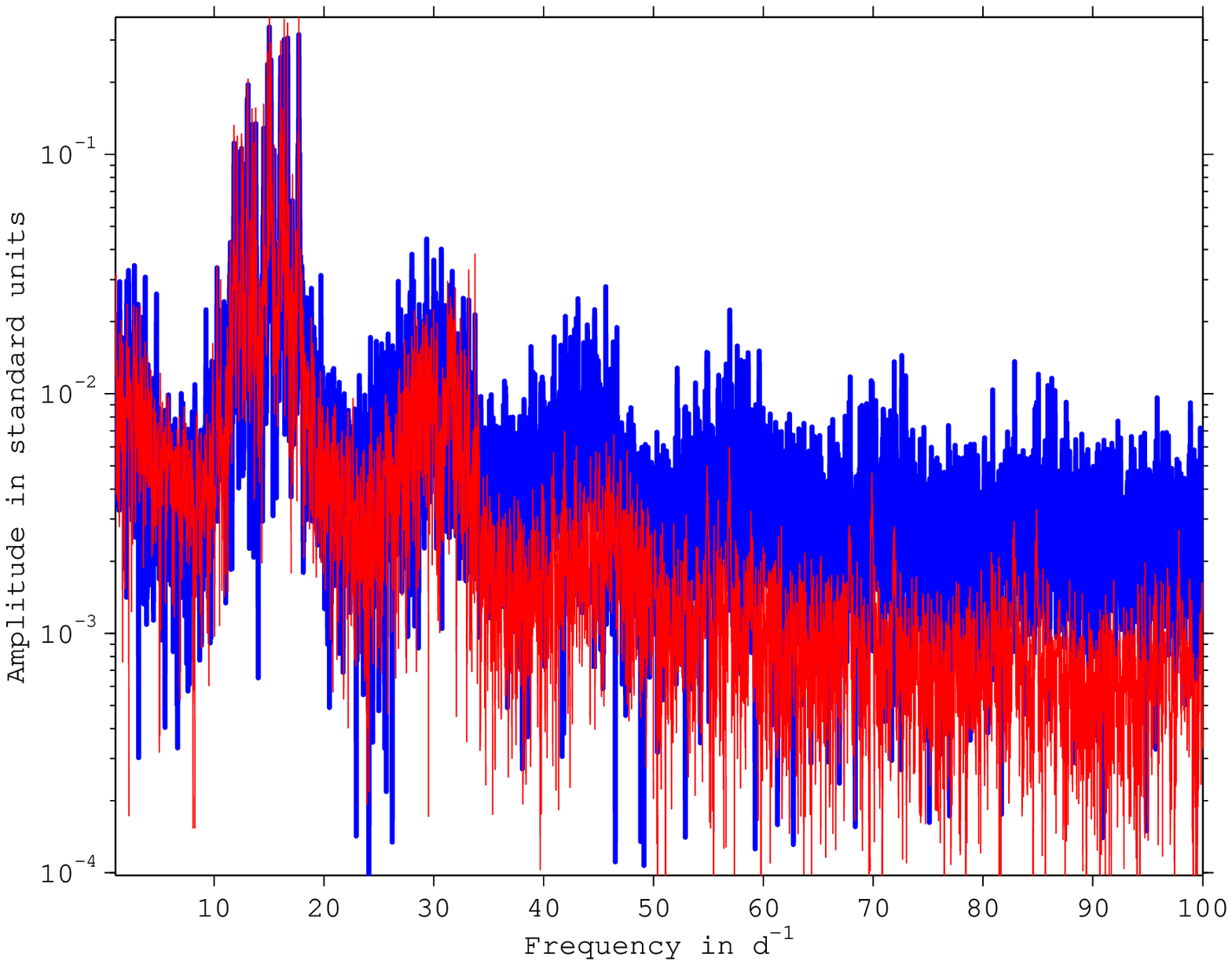}
  \includegraphics[width=8cm]{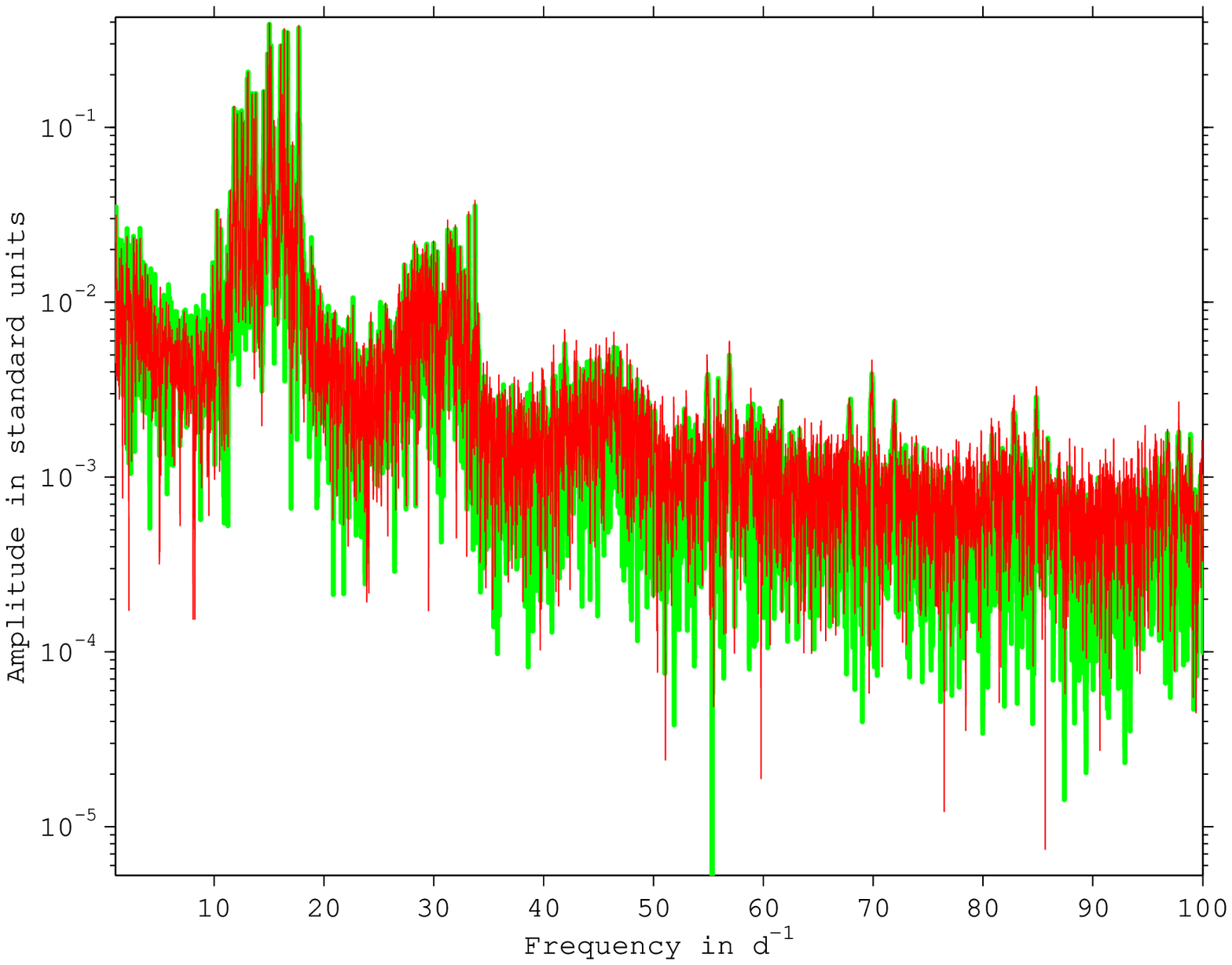}
  \caption{Power spectra of the light curves from HD 41641: upper panel shows gapped data in blue and ARMA interpolated data in red, lower panel shows linearly interpolated data in green and ARMA also in red.}
  \label{fig:ps13}
\end{figure}
\begin{figure}
  \centering
  \includegraphics[width=8cm]{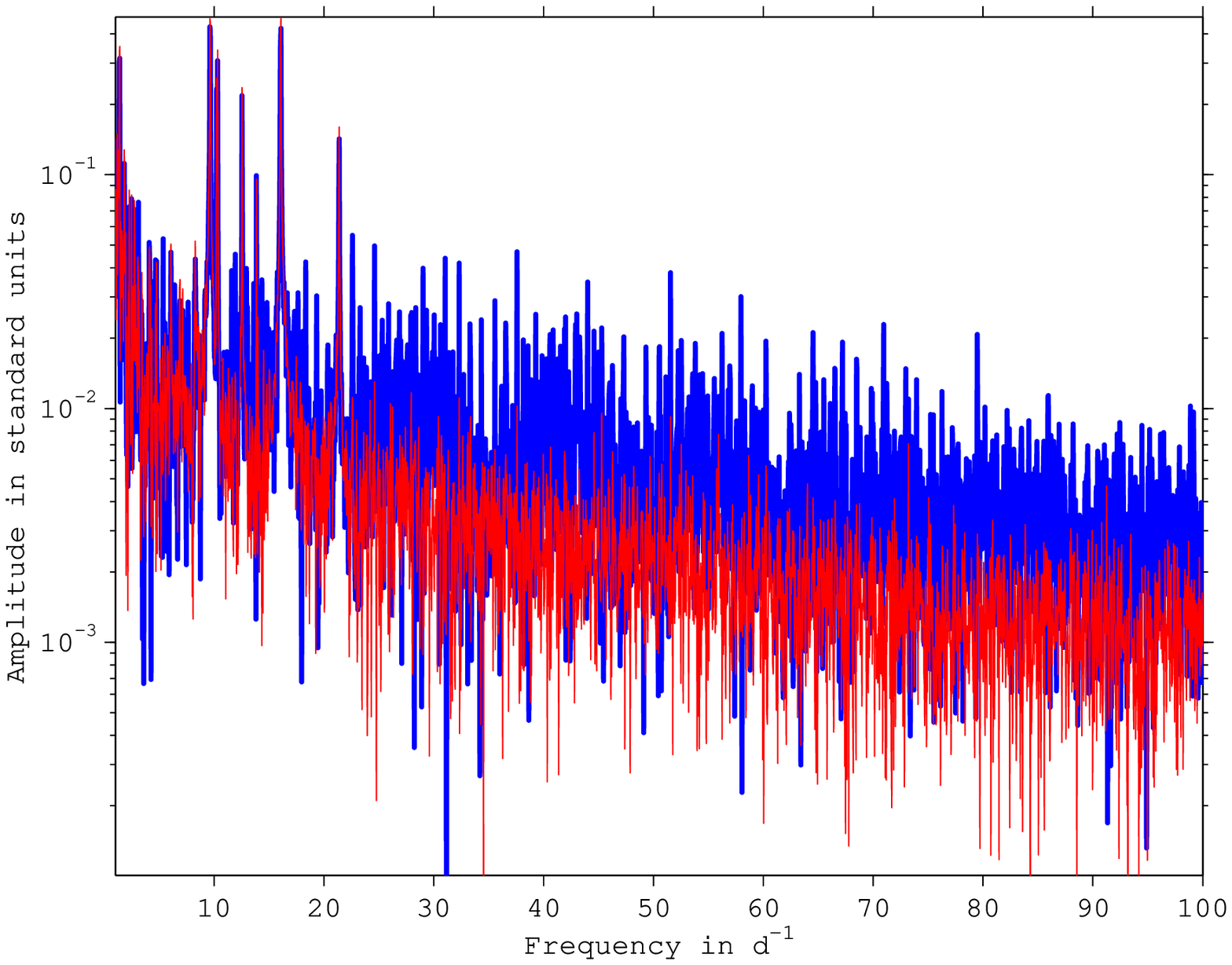}
  \includegraphics[width=8cm]{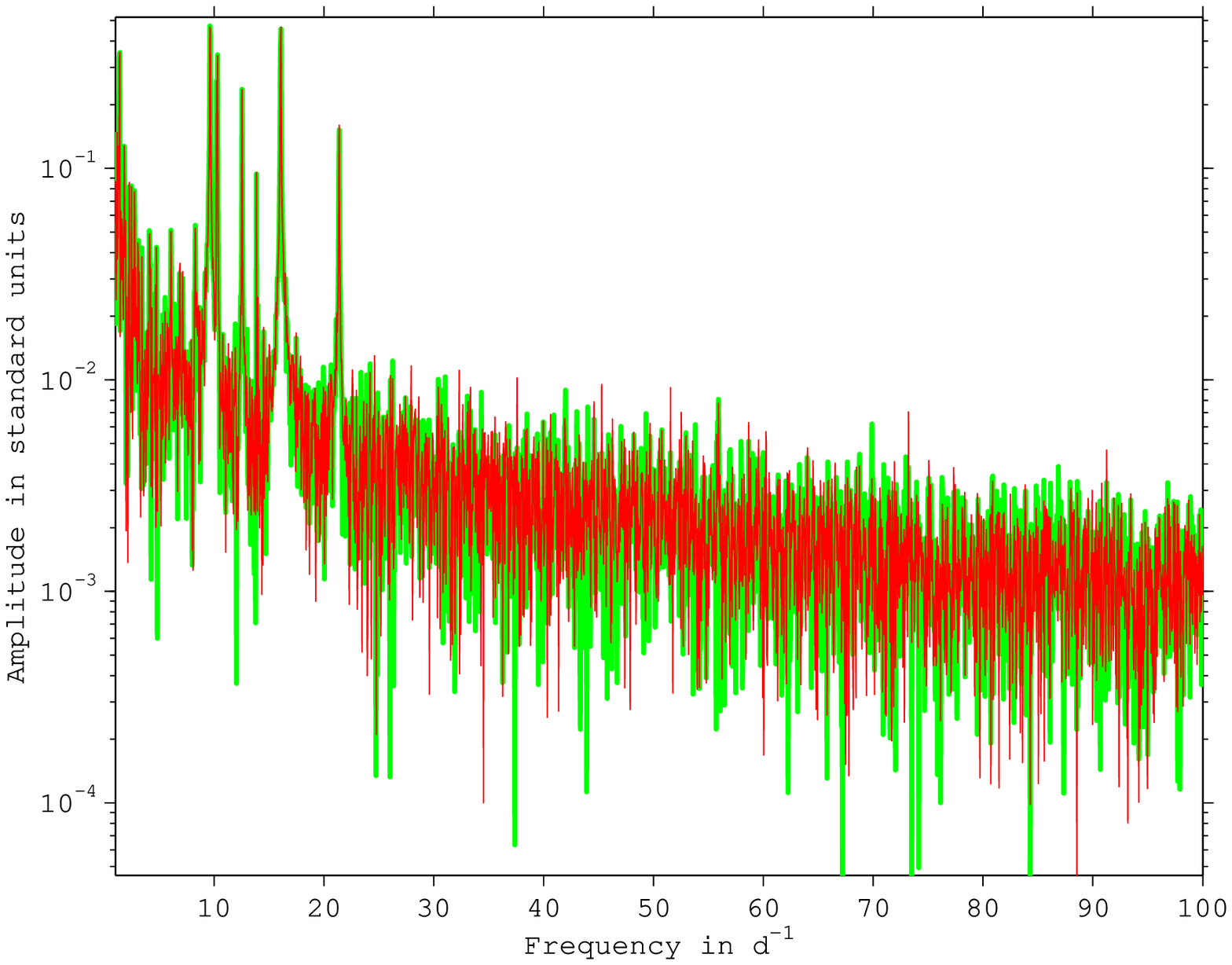}
  \caption{Power spectra of the light curves from HD 48784: upper panel shows gapped data in blue and ARMA interpolated data in red, lower panel shows linearly interpolated data in green and ARMA also in red.}
  \label{fig:ps14}
\end{figure}

\clearpage

\section[]{Histograms}
\begin{figure}[ht]
    \centering
   \includegraphics[width=8.5cm]{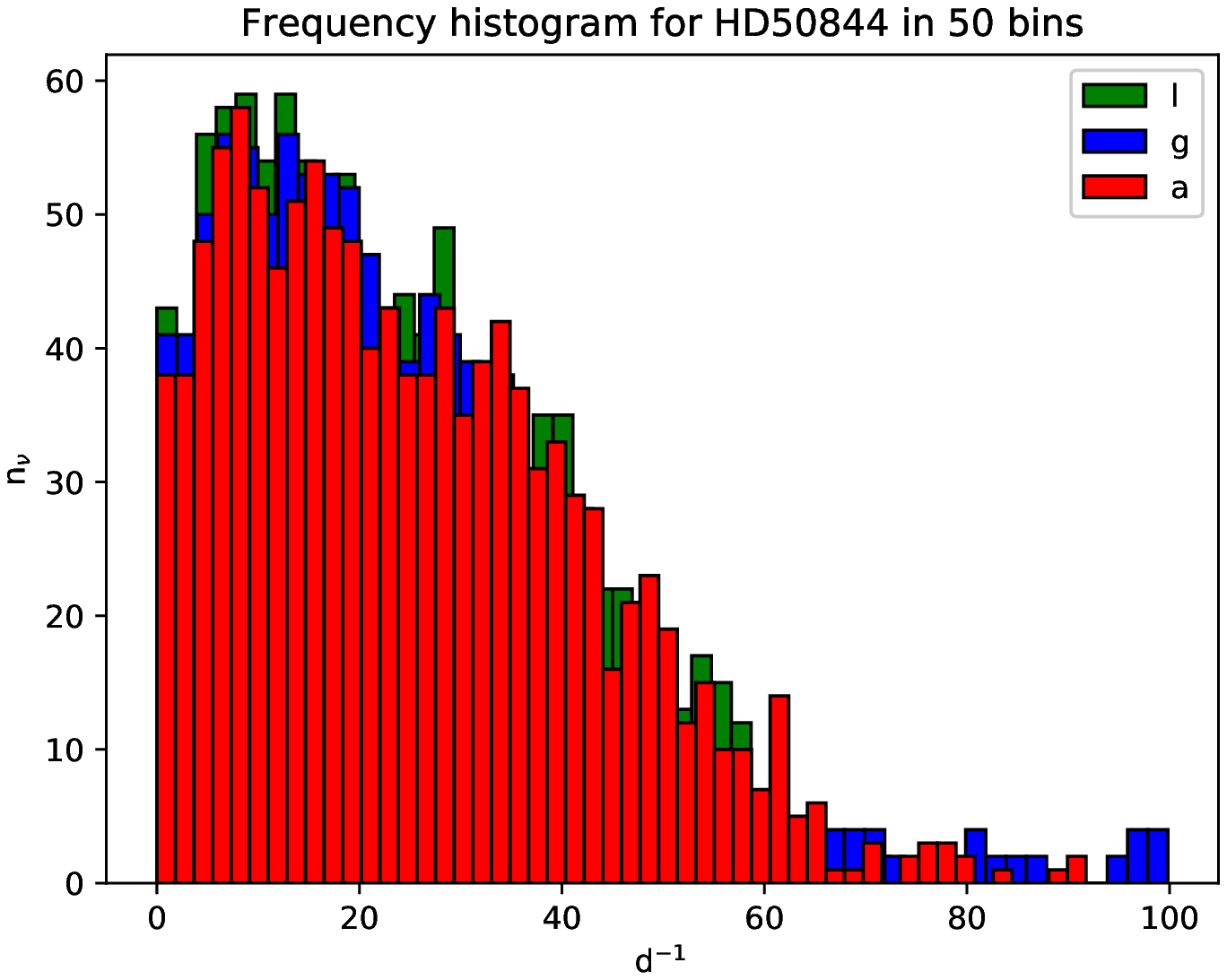}
   \caption{Histograms of detected frequencies in the light curves of HD 50844. blue bars correspond to gapped data, red bars to ARMA interpolated data, and green bars to linearly interpolated data. 
   }
   \label{fig:h1}
\end{figure}

\begin{figure}[ht]
   \centering
   \includegraphics[width=8.5cm]{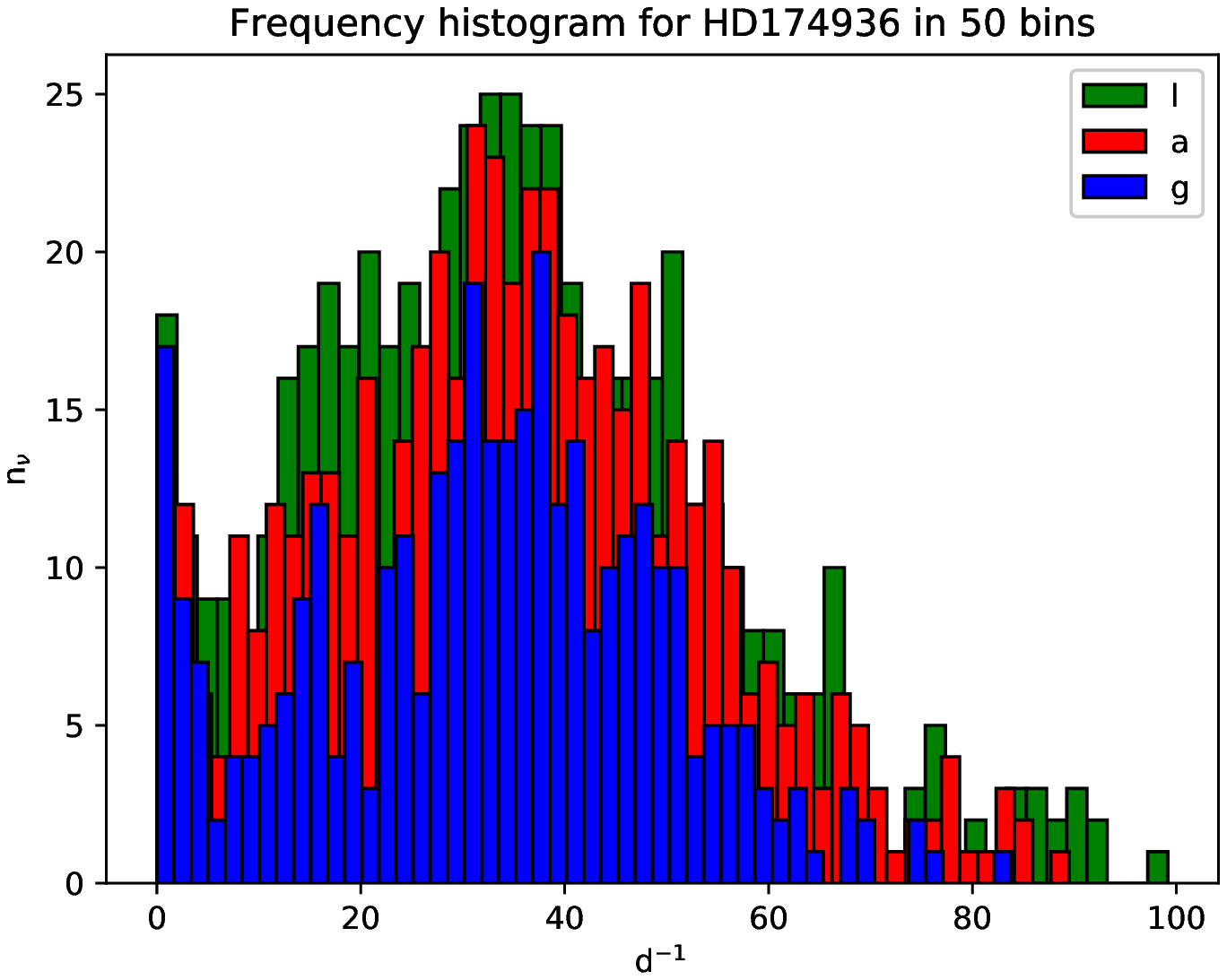}
   \caption{Histograms of detected frequencies in the light curves of HD 174936. blue bars correspond to gapped data, red bars to ARMA interpolated data, and green bars to linearly interpolated data. 
   }
   \label{fig:h2}
\end{figure}

\begin{figure}
   \centering
   \includegraphics[width=8.5cm]{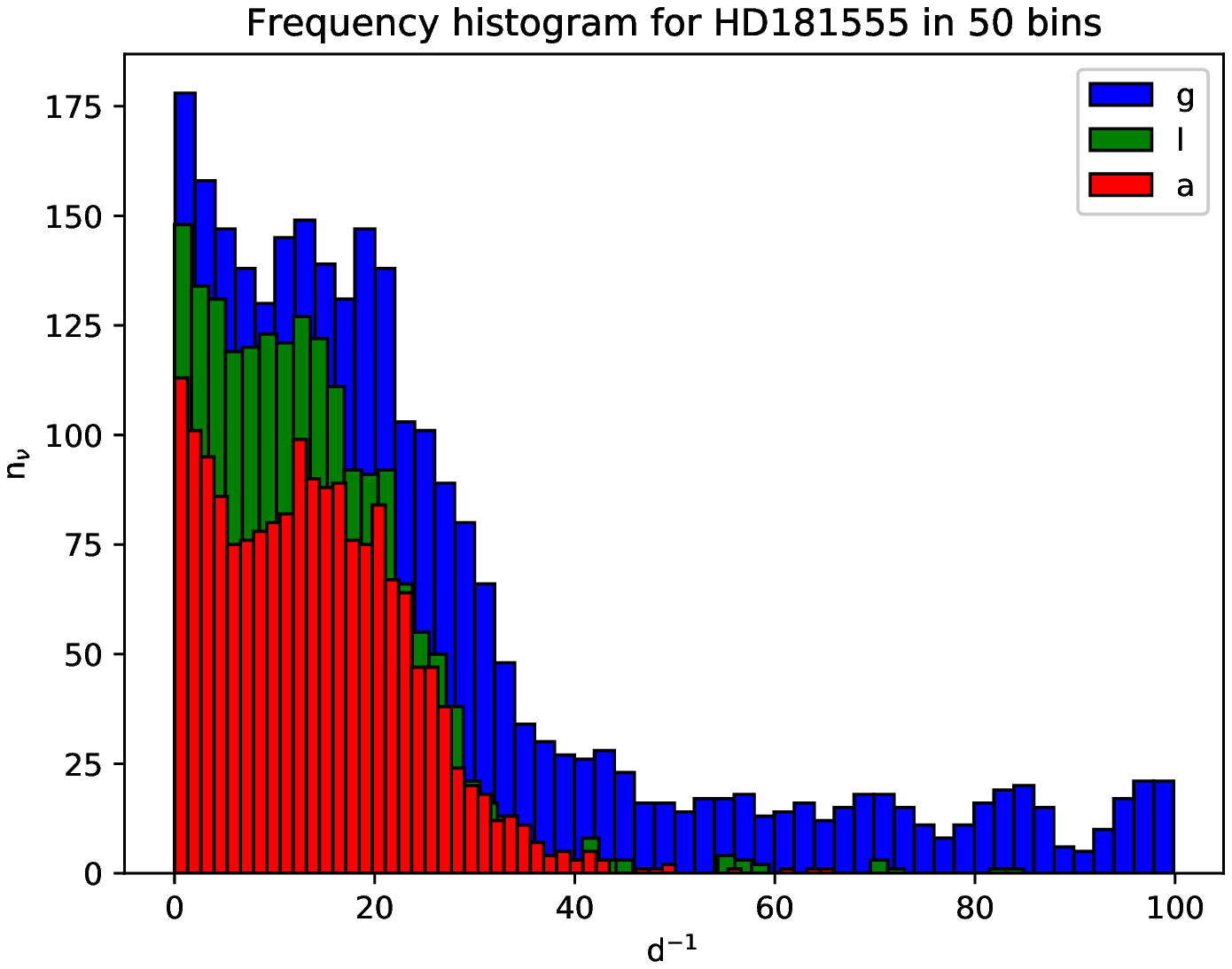}
   \caption{Histograms of detected frequencies in the light curves of HD 181555. blue bars correspond to gapped data, red bars to ARMA interpolated data, and green bars to linearly interpolated data.
   }
   \label{fig:h3}%
\end{figure}

\begin{figure}
    \centering
    \includegraphics[width=8.5cm]{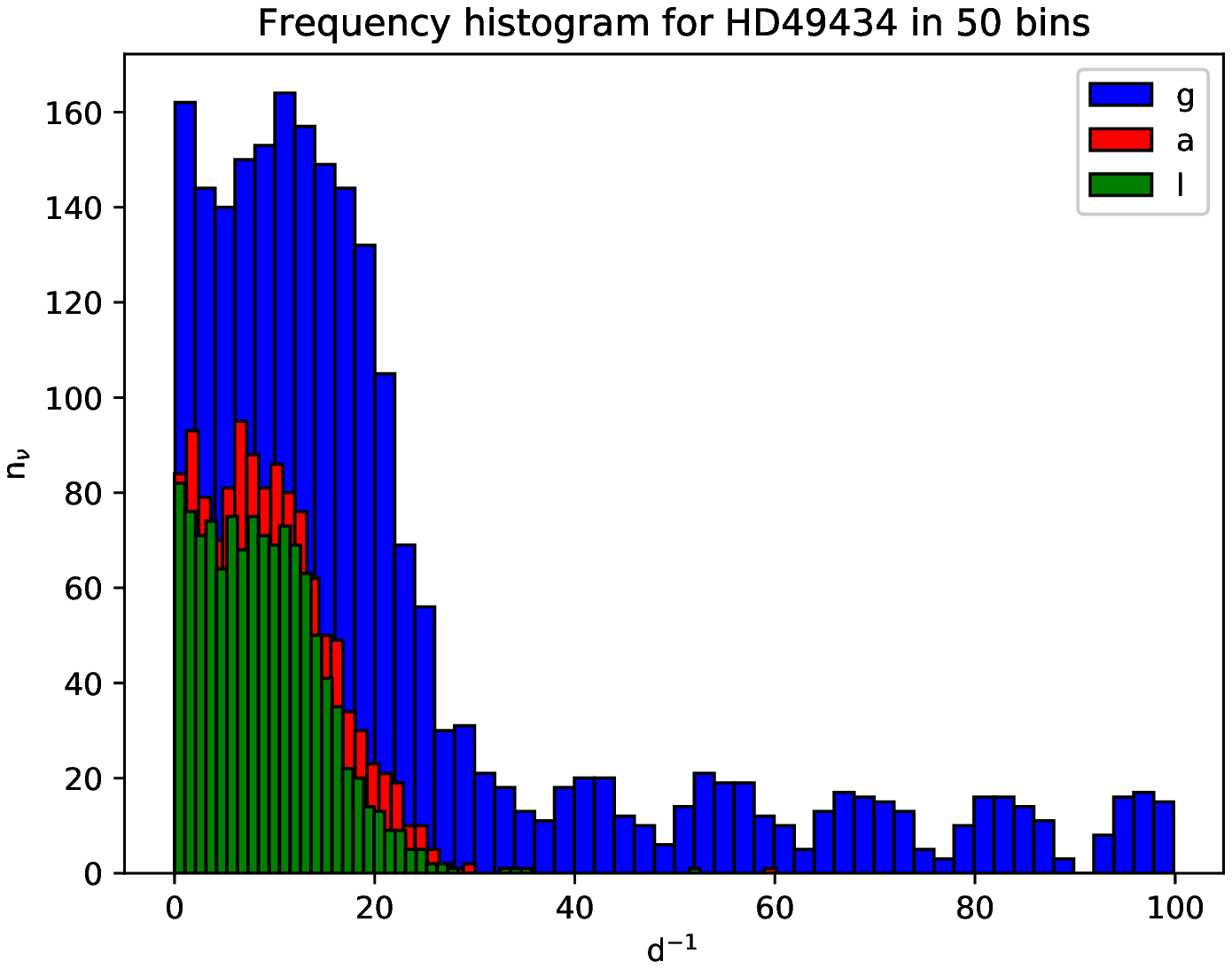}
    \caption{Histograms of detected frequencies in the light curves of HD 49434. blue bars correspond to gapped data, red bars to ARMA interpolated data, and green bars to linearly interpolated data. 
    }
    \label{fig:h4}
\end{figure}
 
\begin{figure}
    \centering
    \includegraphics[width=8.5cm]{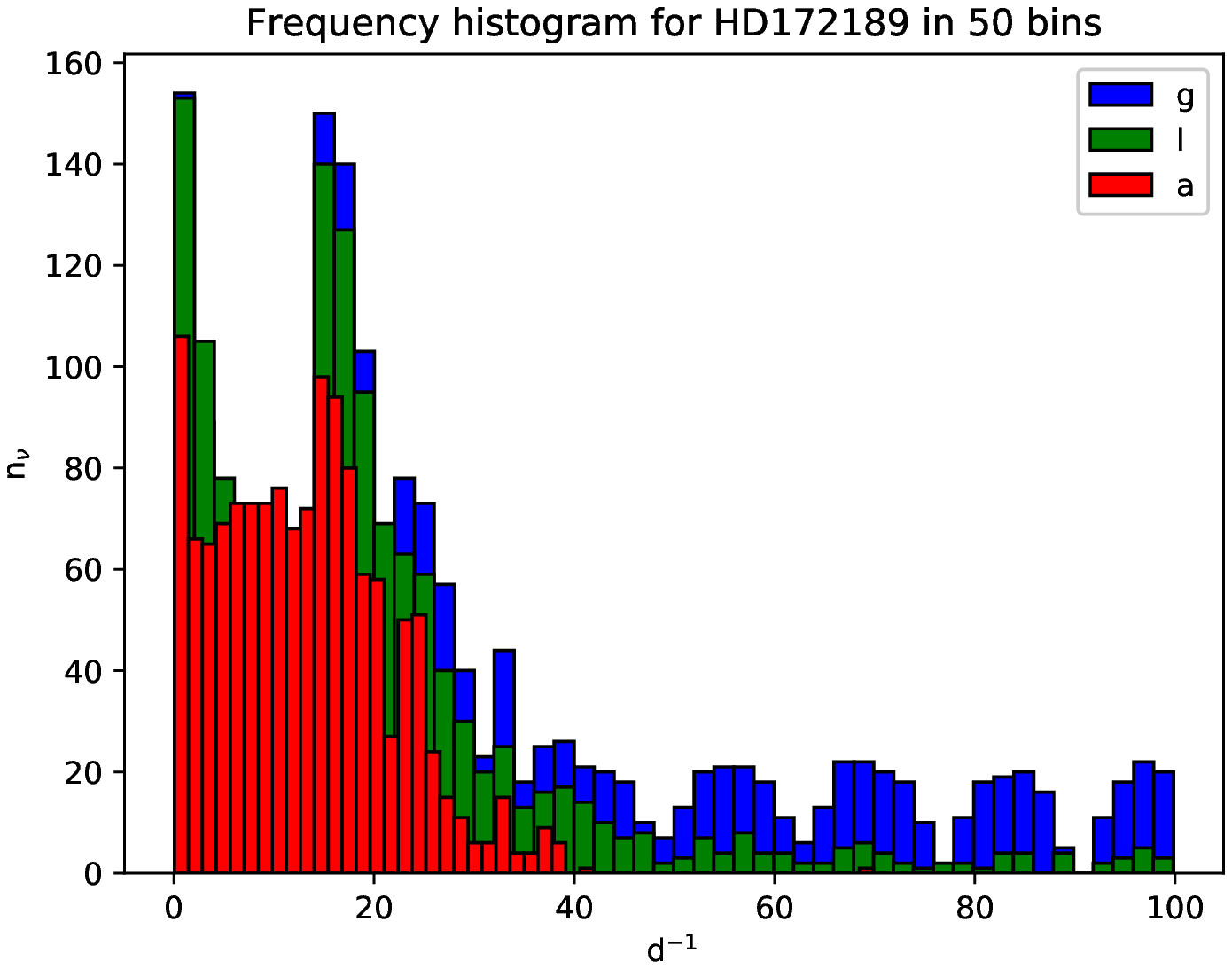}
    \caption{Histograms of detected frequencies in the light curves of HD 172189. blue bars correspond to gapped data, red bars to ARMA interpolated data, and green bars to linearly interpolated data.
    }
    \label{fig:h5}%
\end{figure}
 
\begin{figure}
    \centering
    \includegraphics[width=8.5cm]{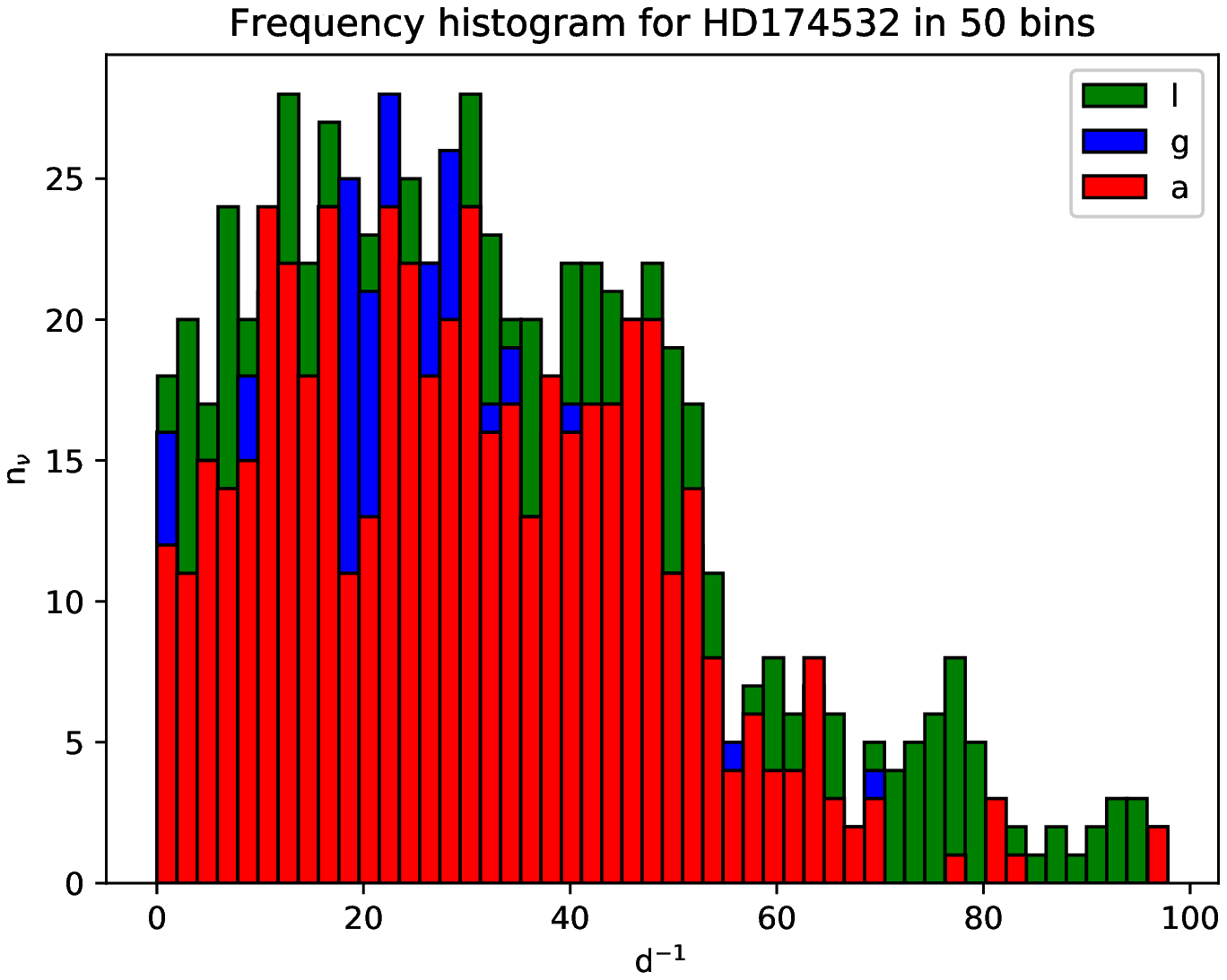}
    \caption{Histograms of detected frequencies in the light curves of HD 174532. blue bars correspond to gapped data, red bars to ARMA interpolated data, and green bars to linearly interpolated data. 
    }
   \label{fig:h6}%
\end{figure}
 
\begin{figure}
    \centering
    \includegraphics[width=8.5cm]{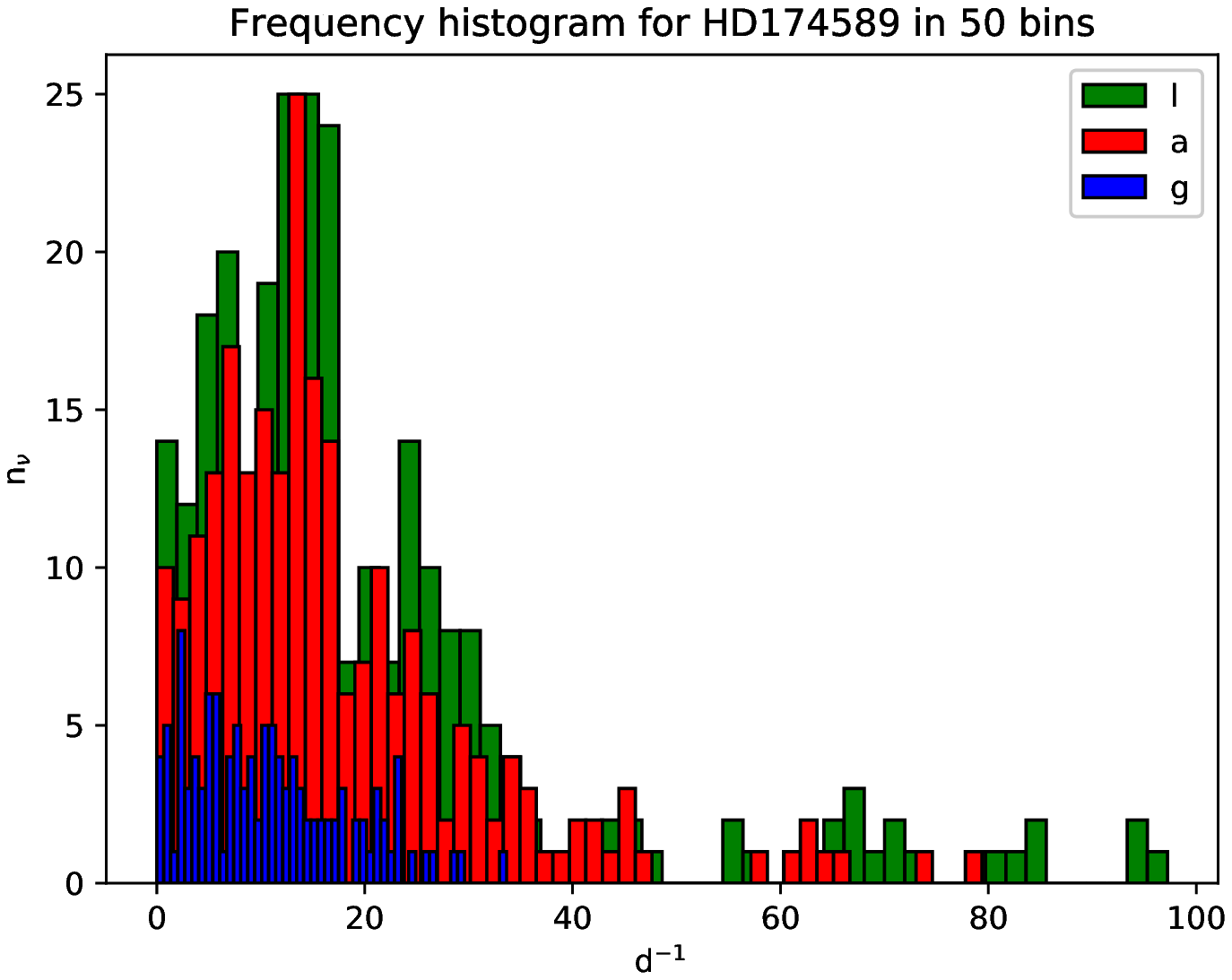}
    \caption{Histograms of detected frequencies in the light curves of HD 174589. blue bars correspond to gapped data, red bars to ARMA interpolated data, and green bars to linearly interpolated data. 
    }
    \label{fig:h7}%
\end{figure}
 
\begin{figure}
    \centering
    \includegraphics[width=8.5cm]{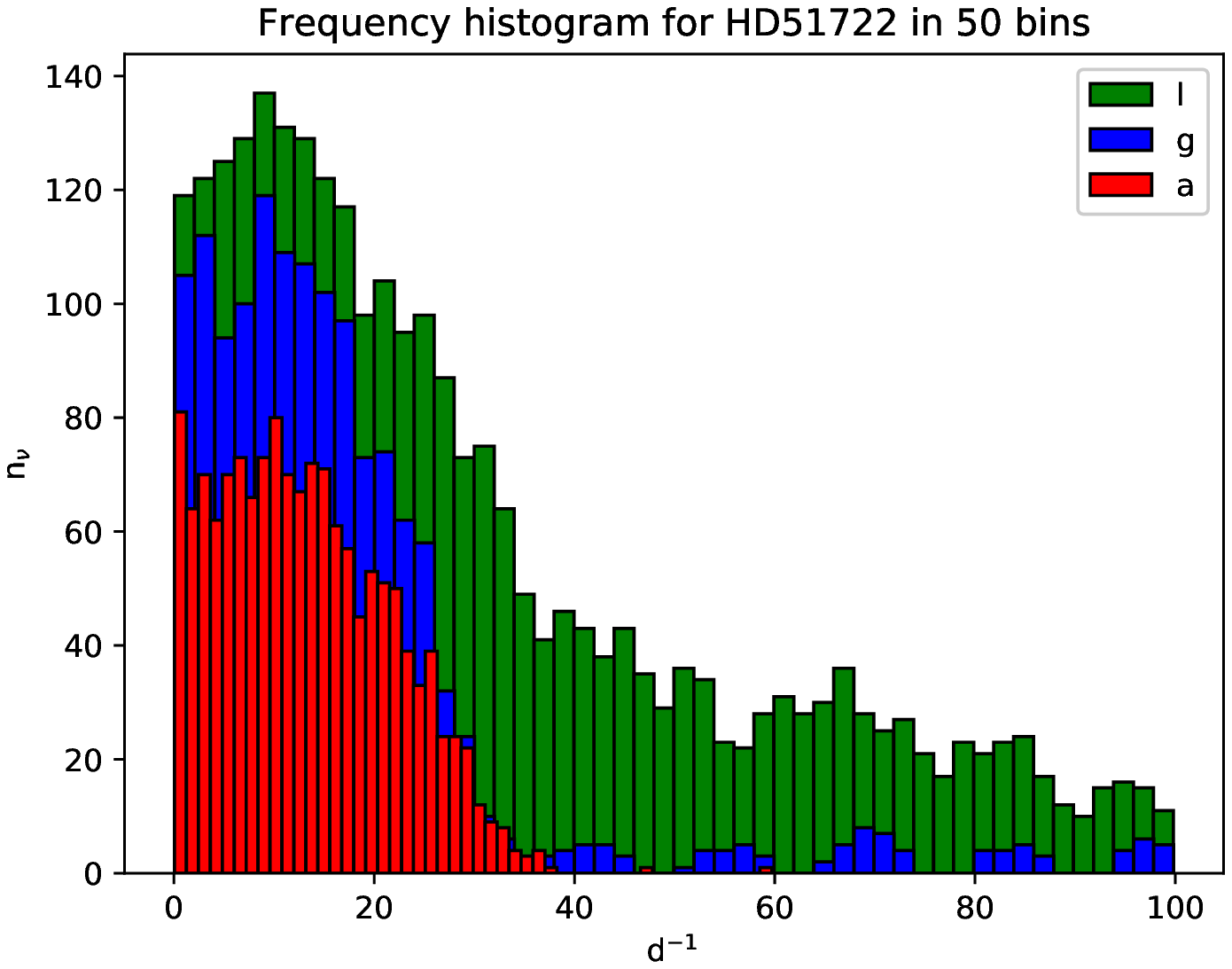}
    \caption{Histograms of detected frequencies in the light curves of HD 51722. blue bars correspond to gapped data, red bars to ARMA interpolated data, and green bars to linearly interpolated data.
    }
    \label{fig:h8}%
\end{figure}
 
\begin{figure}
   \centering
   \includegraphics[width=8.5cm]{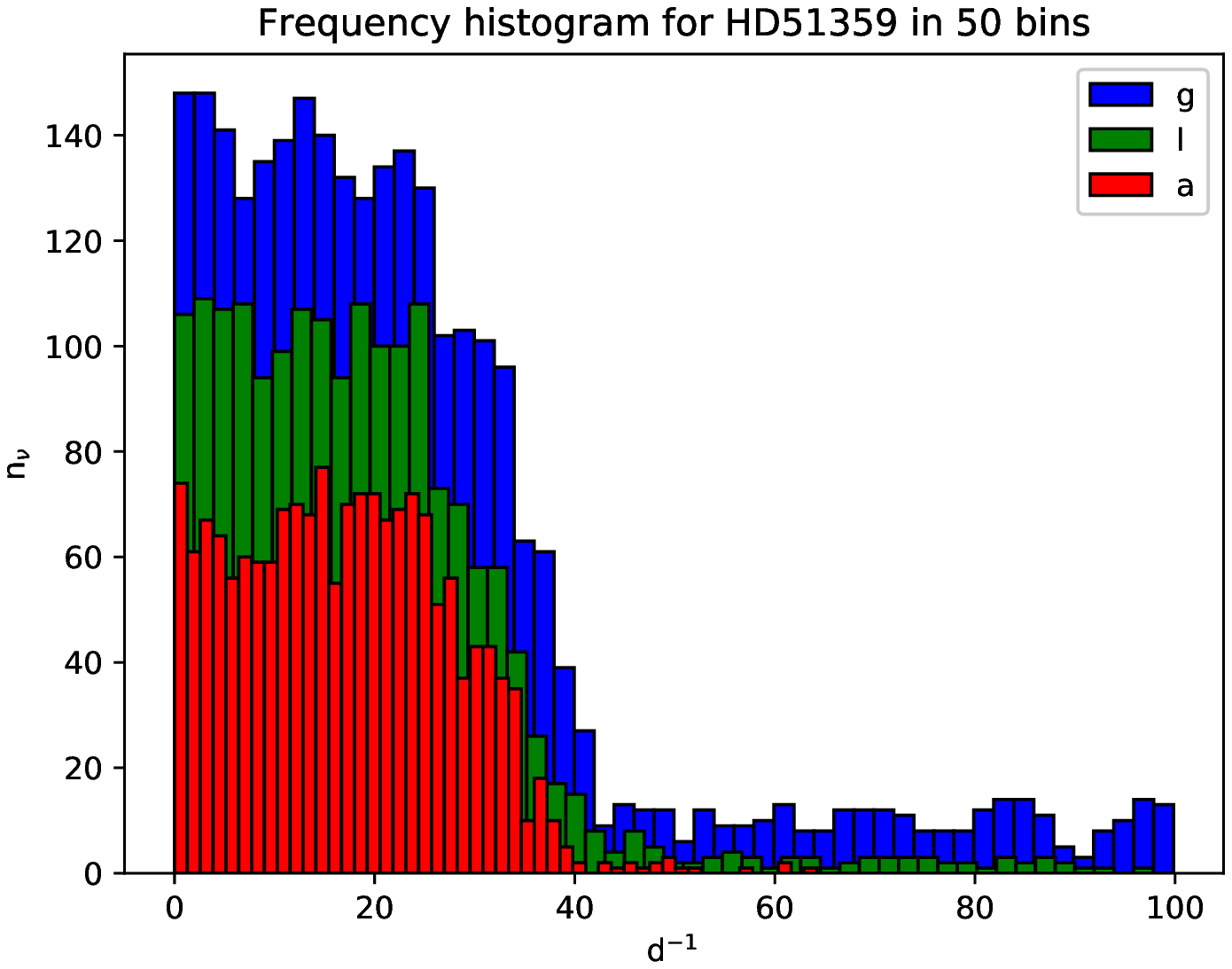}
   \caption{Histograms of detected frequencies in the light curves of HD 51359. blue bars correspond to gapped data, red bars to ARMA interpolated data, and green bars to linearly interpolated data. 
   }
   \label{fig:h9}%
\end{figure}
 
\begin{figure}
   \centering
    \includegraphics[width=8.5cm]{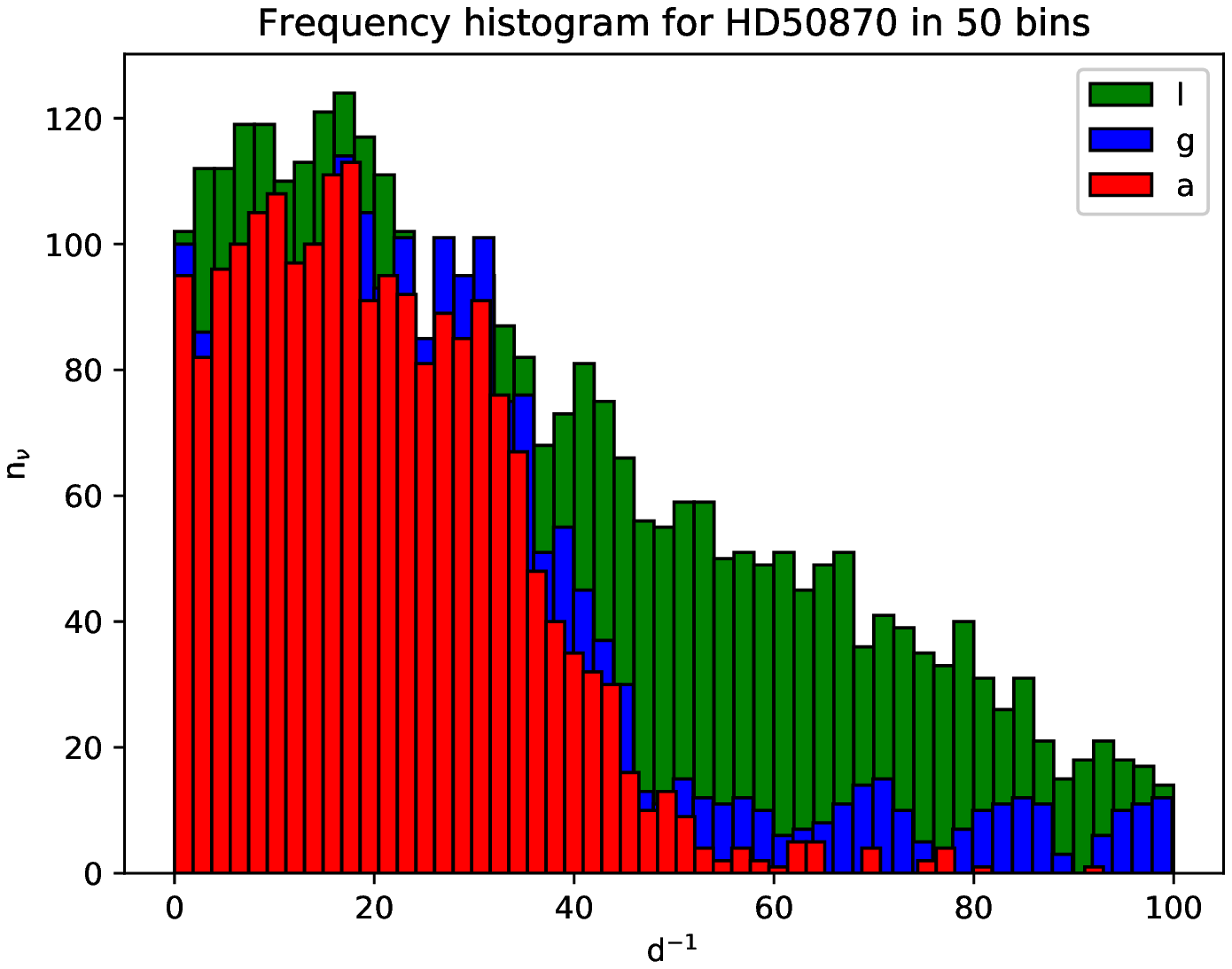}
    \caption{Histograms of detected frequencies in the light curves of HD 50870. blue bars correspond to gapped data, red bars to ARMA interpolated data, and green bars to linearly interpolated data.
    }
    \label{fig:h10}%
\end{figure}
 
\begin{figure}
   \centering
   \includegraphics[width=8.5cm]{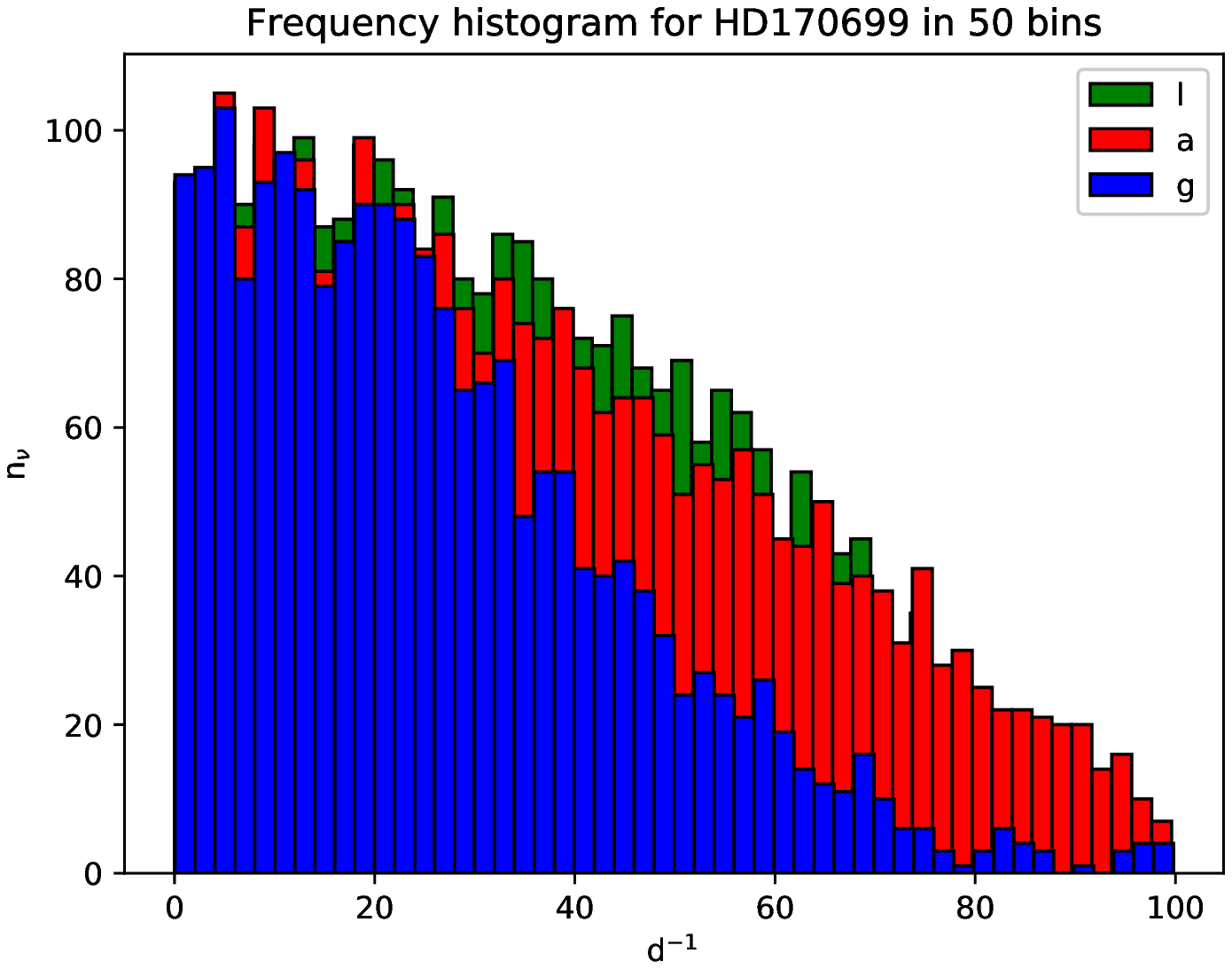}
   \caption{Histograms of detected frequencies in the light curves of HD 170699. blue bars correspond to gapped data, red bars to ARMA interpolated data, and green bars to linearly interpolated data.
   }
   \label{fig:h11}%
\end{figure}
 
\clearpage
 
\begin{figure}
    \centering
    \includegraphics[width=8.5cm]{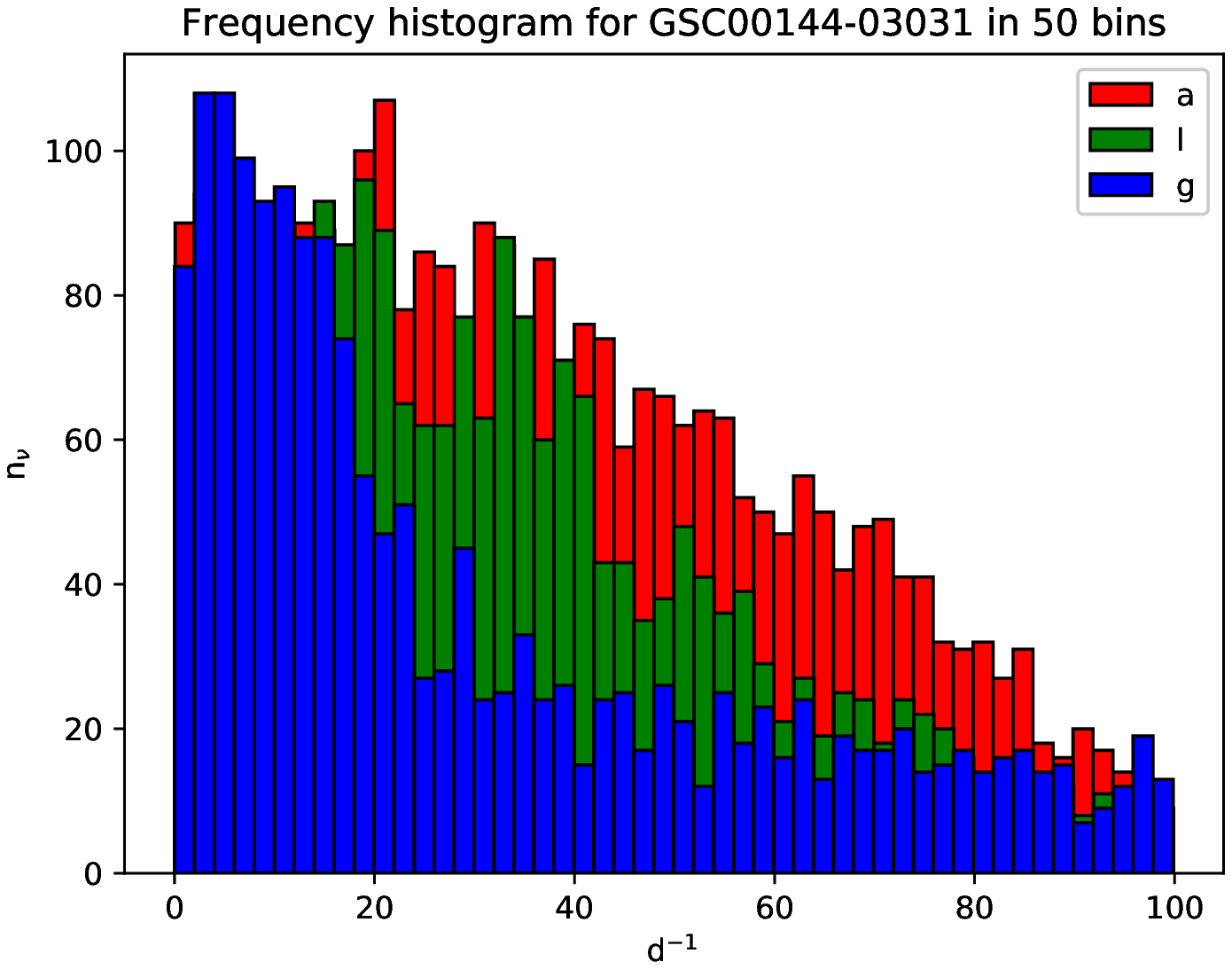}
    \caption{Histograms of detected frequencies in the light curves of GSC00144-03031. blue bars correspond to gapped data, red bars to ARMA interpolated data, and green bars to linearly interpolated data.
    }
    \label{fig:h12}%
\end{figure}
 
\begin{figure}
    \centering
    \includegraphics[width=8.5cm]{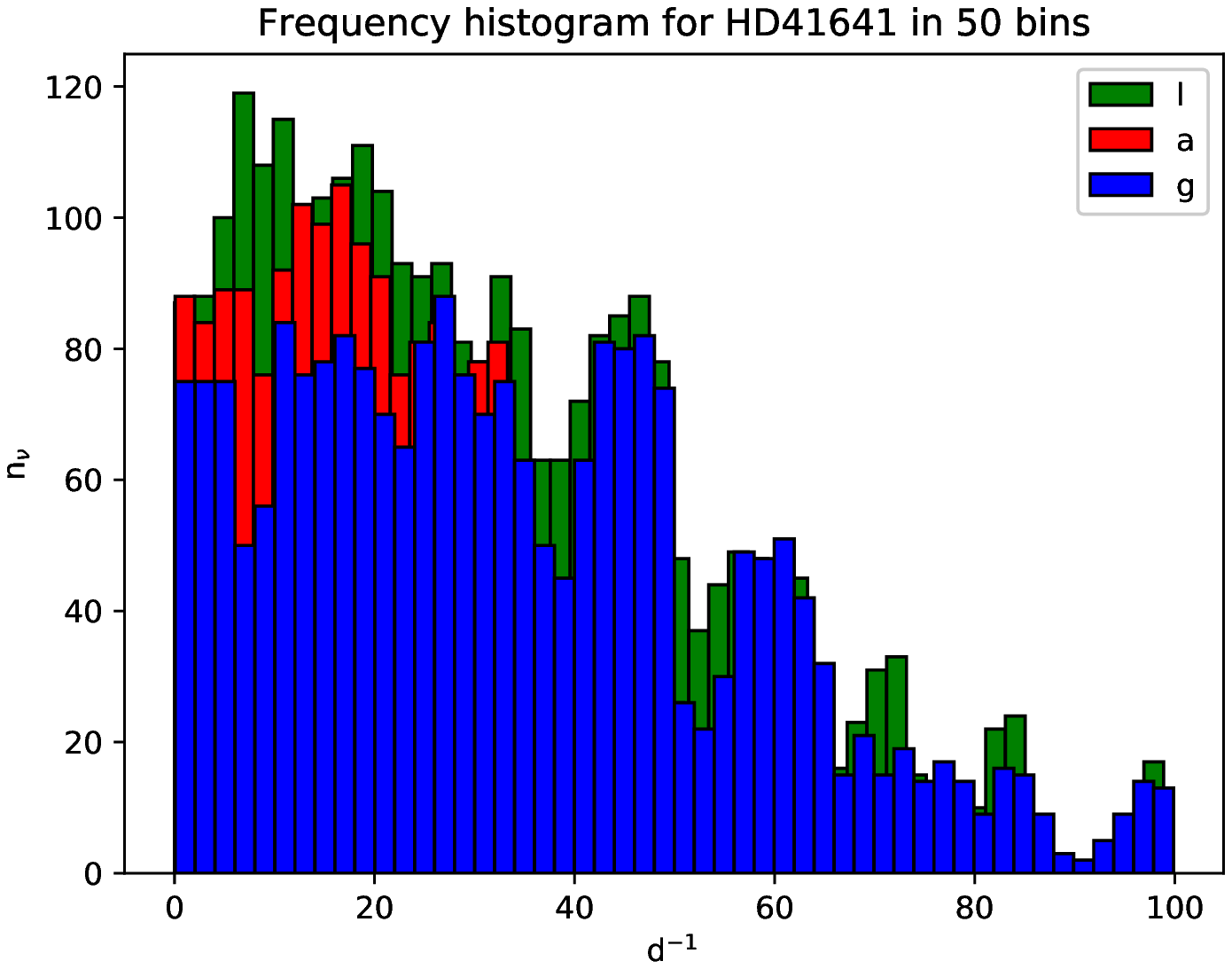}
    \caption{Histograms of detected frequencies in the light curves of HD 41641. blue bars correspond to gapped data, red bars to ARMA interpolated data, and green bars to linearly interpolated data. 
    }
    \label{fig:h13}%
\end{figure}
 
\begin{figure}
    \centering
    \includegraphics[width=8.5cm]{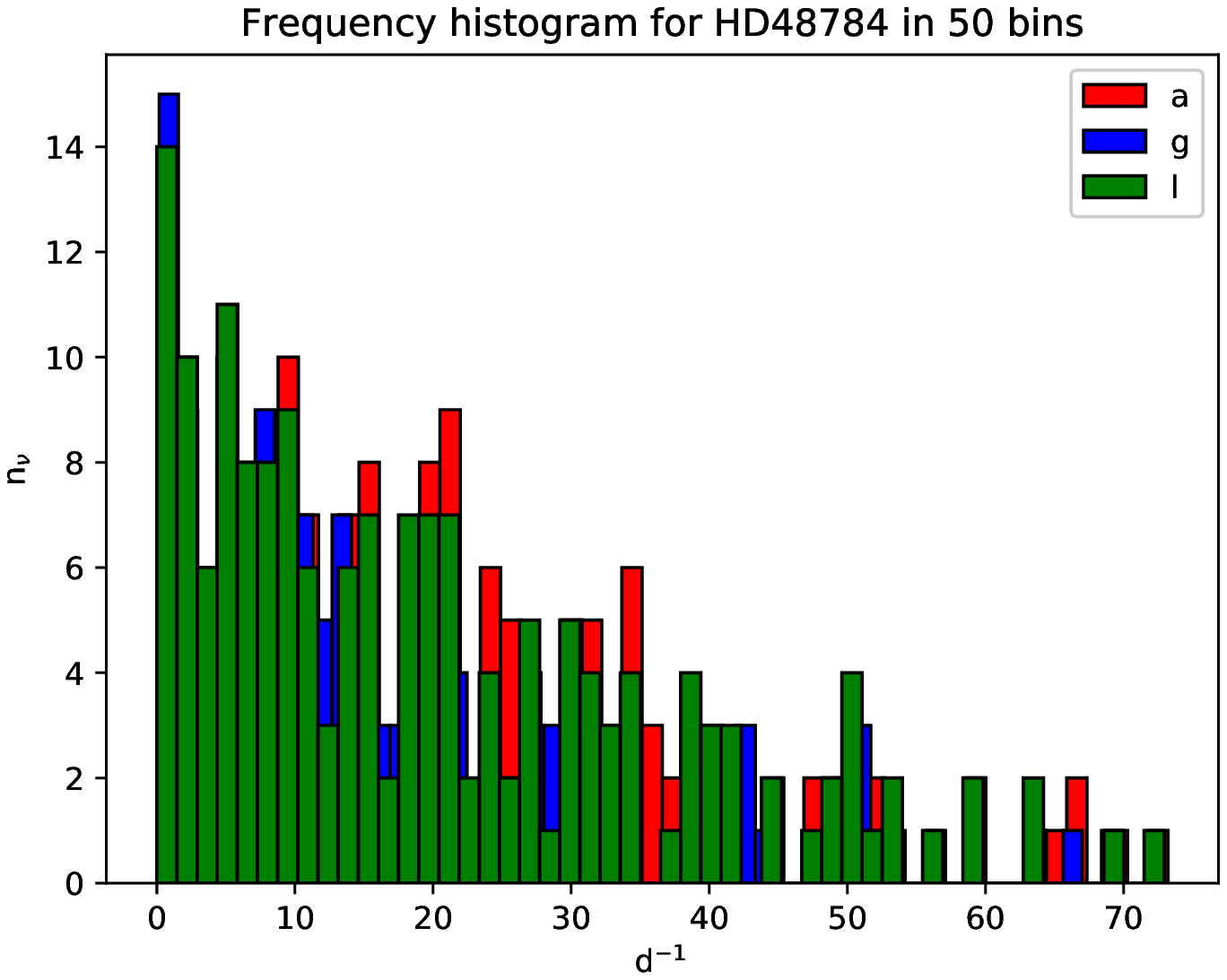}
    \caption{Histograms of detected frequencies in the light curves of HD 48784. blue bars correspond to gapped data, red bars to ARMA interpolated data, and green bars to linearly interpolated data. 
    }
    \label{fig:h14}%
\end{figure}

\clearpage

\end{document}